\documentclass[a4paper,11pt]{article}
\pdfoutput=1
\usepackage{jheppub}

\usepackage{amssymb,amsmath,amsfonts}
\usepackage[normalem]{ulem}
\usepackage[utf8x]{inputenc}
\usepackage{slashed}
\usepackage{graphicx}
\usepackage{tabularx}
\usepackage{here}
\usepackage{color}
\usepackage{csquotes} 
\usepackage{comment}
\usepackage{mathrsfs}
\usepackage{float}
\usepackage{ascmac}
\usepackage{multirow}
\usepackage{longtable}
\usepackage{bm}
\usepackage{ulem}
\usepackage{caption}
\usepackage[italicdiff]{physics}
\usepackage{tikz}
\usepackage[compat=1.1.0]{tikz-feynhand}
\newcommand{\com}{{\rm Com}}
\newcommand{\gr}{{\rm Gr}}
\newcommand{\conj}{{\rm Conj}}

\makeatletter
\newcommand*\rel@kern[1]{\kern#1\dimexpr\macc@kerna}
\newcommand*\widebar[1]{%
  \begingroup
  \def\mathaccent##1##2{%
    \rel@kern{0.8}%
    \overline{\rel@kern{-0.8}\macc@nucleus\rel@kern{0.2}}%
    \rel@kern{-0.2}%
  }%
  \macc@depth\@ne
  \let\math@bgroup\@empty \let\math@egroup\macc@set@skewchar
  \mathsurround\z@ \frozen@everymath{\mathgroup\macc@group\relax}%
  \macc@set@skewchar\relax
  \let\mathaccentV\macc@nested@a
  \macc@nested@a\relax111{#1}%
  \endgroup
}
\makeatother

\numberwithin{equation}{section}

\preprint{
\begin{minipage}{5cm}
\small
\flushright
EPHOU-26-02\\
KYUSHU-HET-356
\end{minipage}}

\title{
Residual group-like symmetries in selection rules without group actions
}

\author{Jun Dong$^{1}$,} 
\author{Tatsuo Kobayashi$^{1}$,} 
\author{Shuhei Miyamoto$^{1}$,} 
\author{Ryusei Nishida$^{1}$, \\ and } 
\author{Hajime Otsuka$^{2,3}$}
\affiliation{
$^1$Department of Physics, Hokkaido University, Sapporo 060-0810, Japan\\
$^2$Department of Physics, Kyushu University, 744 Motooka, Nishi-ku, Fukuoka 819-0395, Japan\\
$^3$Quantum and Spacetime Research Institute (QuaSR), Kyushu University, 744 Motooka, Nishi-ku, Fukuoka, 819-0395, Japan
}
\emailAdd{j-dong@particle.sci.hokudai.ac.jp}
\emailAdd{kobayashi@particle.sci.hokudai.ac.jp}
\emailAdd{s-miyamoto@particle.sci.hokudai.ac.jp}
\emailAdd{r-nishida@particle.sci.hokudai.ac.jp}
\emailAdd{otsuka.hajime@phys.kyushu-u.ac.jp}

\abstract{
We analyze loop-induced group-like symmetries in theories where fields are labeled by basis elements of a fusion algebra constructed from the conjugacy classes of finite groups. 
Although the fusion rules for conjugacy classes are in general violated at loop level, residual group-like symmetries, including both Abelian and non-Abelian ones, remain exact through a procedure referred to as ``groupification''. 
By examining various conjugacy classes of finite groups realized in heterotic string theory on non-Abelian orbifolds, we identify an approximate discrete symmetry that controls the magnitude of loop-induced couplings. 
As a result, most parameters appearing in non-invertible selection rules are natural in the sense of 't Hooft. 
Furthermore, we discuss anomalies of the groupification symmetry, which can impose additional constraints on models with non-invertible fusion rules. 
}

\makeatletter
\gdef\@fpheader{}
\makeatother

\begin{document}

\maketitle

\section{Introduction}
\label{sec:Intro}

Symmetries are important in particle physics as well as in other fields of physics. 
Conventionally, group-like symmetries have been used in order to constrain particle models. 
For example, in the case of Abelian groups, these symmetries lead to charge conservation laws that determine allowed coupling terms. 
Furthermore, such group-theoretical symmetries appear in four-dimensional low-energy effective field theory derived from string theory. 
However, not all stringy coupling selection rules can be understood only in terms of group theory, but certain stringy selection rules do not correspond to group-theoretical structures.

For example, in heterotic orbifold models, a string can be specified by its boundary condition, i.e., 
$X(\sigma=\pi)=gX(\sigma=0)=\theta X(\sigma = 0)+v$, where $\theta$ represents a discrete twist and $v$ denotes a discrete shift. 
Here, $g=(\theta,v)$ is a space group element, while $\theta$ is a point group element. 
The same boundary condition can also be satisfied by $hX$, where $h$ is another space group element, i.e.,
$hX(\sigma=\pi)=ghX(\sigma=0)$, which leads to $X(\sigma=\pi)=h^{-1}ghX(\sigma=0)$.
That implies that $g$ is equivalent to $h^{-1}gh$. 
Thus, each string state corresponds not to a space group element $g$, but rather to a conjugacy class $h^{-1}gh$ \cite{Dixon:1986jc,Dixon:1986qv,Hamidi:1986vh}. 
The associated coupling selection rules are governed by the multiplication rules of these conjugacy classes, which differ from those of individual group elements.
In group theory, the product of two elements is given by a single element. 
Under such multiplication rules, one can define an inverse element.
In contrast, the product of two conjugacy classes generally includes more than one conjugacy class. 
Hence, no inverse element can be defined under the multiplication rules of conjugacy classes, i.e., {\it non-invertible selection rules}. 
Such non-invertible selection rules lead to novel aspects when each field corresponds to a conjugacy class rather than to a group element.

In addition, in intersecting/magnetized D-brane models, each state also corresponds to a conjugacy class \cite{Kobayashi:2024yqq,Funakoshi:2024uvy}. 
In particular, the $\mathbb{Z}_2$ gauging of $\mathbb{Z}_N$ symmetries has been studied in the context of particle physics.
Interesting Yukawa textures were derived for the quark sector \cite{Kobayashi:2024cvp,Kobayashi:2025znw} and for the lepton sector~\cite{Kobayashi:2025ldi,Jiang:2025psz}.
These textures are also useful for addressing the strong CP problem without introducing an axion \cite{Liang:2025dkm,Kobayashi:2025thd,Kobayashi:2025rpx}. 
This analysis has also been extended to other gaugings, including Abelian and non-Abelian ones~\cite{Dong:2025jra} (see Refs.~\cite{Dong:2025jra,Kobayashi:2025wty,Qu:2026omn} for phenomenological applications).

Non-invertible selection rules exhibit another important property.
Allowed coupling terms at tree level are determined by the multiplication rules.
However, such coupling selection rules can be violated by loop effects \cite{Heckman:2024obe,Kaidi:2024wio,Funakoshi:2024uvy}. 
In principle, loop-induced couplings can be expressed in terms of tree-level couplings and are typically suppressed. 
This can lead to phenomenologically interesting implications \cite{Suzuki:2025oov,Kobayashi:2025cwx,Suzuki:2025bxg,Nomura:2025sod,Chen:2025awz,Okada:2025kfm}. 
Hence, it is important to study concretely which couplings can be generated through loop corrections and which remain forbidden even after accounting for loop corrections.

We study loop effects on non-invertible selection rules arising from the multiplication rules of conjugacy classes of discrete groups. 
As mentioned above, heterotic orbifold models provide one of the underlying frameworks leading to non-invertible selection rules, in particular, heterotic string theory on non-Abelian orbifolds $T^6/G$, where $G$ is a non-Abelian discrete group \cite{Inoue:1987ak,Inoue:1988ki,Inoue:1990ci,Konopka:2012gy,Fischer:2012qj,Fischer:2013qza,Funakoshi:2025lxs,Hernandez-Segura:2025sfr}.
Indeed, coupling selection rules in heterotic non-Abelian orbifold models were studied at tree level in Ref.~\cite{Kobayashi:2025ocp}. 
In this paper, we focus on point group selection rules with vanishing shift $v=0$.\footnote{See Refs.~\cite{Kobayashi:1990mc,Kobayashi:1991rp,Kobayashi:1995py,Kobayashi:2025ocp,Ramos-Sanchez:2018edc} for non-trivial aspects of space-group selection rules with $v \neq 0$ .}
See also Ref.~\cite{Dong:2025jra} for a discussion of the multiplication rules of conjugacy classes of discrete groups.
Specifically, we investigate loop effects across various discrete groups.

This paper is organized as follows. 
In Section~\ref{sec:loop}, we review non-invertible fusion rules for fields in the language of a hypergroup and discuss the loop effects on them. 
In Section~\ref{sec:groupification}, we discuss loop effects on non-invertible selection rules for fields labeled by conjugacy classes of various discrete groups. In particular, we list allowed 3-point couplings and residual group-like symmetries as well as approximate symmetries.
We also give a comment in string theory on orbifolds.
In Section~\ref{sec:anomaly}, the anomalies of these residual group-like symmetries are analyzed. 
We comment on loop effects on the selection rules in the context of heterotic Calabi-Yau compactifications in Section~\ref{sec:CY}. 
Finally, Section~\ref{sec:con} is devoted to the conclusions. 
Loop-induced 2-point couplings  are summarized in Appendices~\ref{app:2-point}.
In Appendix~\ref{app:Deta-6N}, we provide the multiplication rules of conjugacy classes of $\Delta(6N^2)$, focusing particularly on the $S_4$ and $\Delta(54)$ groups. 
Relations between Calabi-Yau selection rules and non-invertible selection rules are discussed in Appendix~\ref{app:CY}.

\section{Non-invertible selection rules and loop effects}
\label{sec:loop}
In this section, we give a brief review on non-invertible selection rules, mainly focusing on their loop effects \cite{Kaidi:2024wio}. Let us start with the definition of non-invertible selection rules considered in this paper.\par
Consider an algebra $\mathcal{A}$ with a set of finite basis elements $A =\{e, x, y, ... \}$, where $e$ is the unit element, which acts trivially on any element of $\mathcal{A}$. Multiplication rules between basis elements are defined as follows \cite{Kaidi:2024wio}:
\begin{align}
\label{eq:xy=NZ}
    xy = \sum_{z \in A} N^{z}_{xy} z,
\end{align}
where we focus on the case with $N^z_{xy}\in \mathbb{Z}_{\geq  0}$ which is the so-called fusion algebra and we assume $N^{z}_{xy}=N^{z}_{yx}$.
We introduce the symbol $\prec$ by defining $z \prec xy$ if $N^{z}_{xy} \ne 0$ and we also assume that for each $x \in A$, 
there exists a conjugate element $\bar{x}$, satisfying $x \bar{x} \succ e$.
Some elements $x$ may be self-conjugate, i.e.,  $xx \succ e$.
\par
Now we are ready to formulate a perturbative field theory without a group structure by associating a basis element $x_i$ with each field $\phi_i$.
Here and hereafter, we ignore Lorentz properties of fields such as scalars and spinors in order to focus on non-invertible selection rules.
We assume that the conjugate field of $\phi_i$ is labeled by $\bar{x_i}\in A$ and $n\mathrm{-point}$ bare interaction terms $\phi_{i_1} \phi_{i_2} \cdots \phi_{i_n}$ in Lagrangian must satisfy conditions $e \prec x_{i_1}x_{i_2}\cdots x_{i_n}$. 
These conditions imply that the allowed tree-level diagrams with external lines labeled by $x_1,x_2,...,x_n$ satisfy 
\begin{align}
    \label{eq:tree_selection_rule}
    e \prec x_1x_2\cdots x_n,\quad {\rm for\ any\ }n.
\end{align}
We next calculate loop effects in this framework.\par
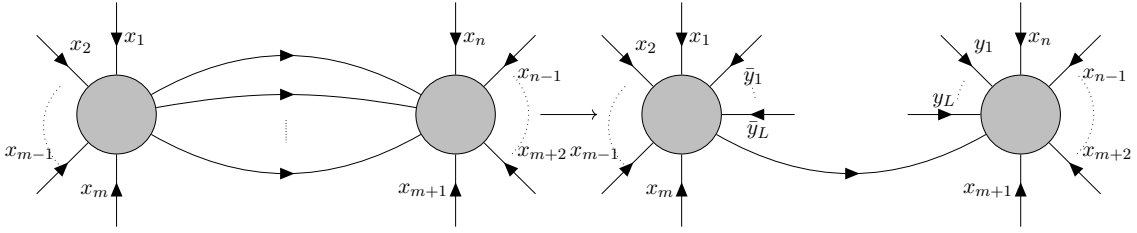
\begin{figure}[H]
  \centering
  \resizebox{\linewidth}{!}{
  \begin{tikzpicture}
  \begin{feynhand}  
    \vertex (i1) at (-1.414,1.414);
    \vertex (i2) at (0,2);
    \vertex (i3) at (0,-2);
    \vertex (i4) at (-1.414,-1.414);
    \vertex (i5) at (2,0);
    \vertex [grayblob] (a) [draw, text=black, circle] at (0,0) {{}};
    \propag [fermion] (i1) to [edge label = $x_2$] (a);
    \propag [fermion] (i2) to [edge label = $x_1$] (a);
    \propag [fermion] (i3) to [edge label = $x_{m}$] (a);
    \propag [fermion] (i4) to [edge label = $x_{m-1}$] (a);
    \draw [dotted] (-1,0.5) arc (135:225:1);
    \draw [dotted] (7,-0.7) arc (-45:45:1);
    \vertex [grayblob] (b) [draw, text=black, circle] at (6,0) {{}};
    \vertex (i6) at (4,0);
    \vertex (i7) at (6+1.414,1.414);
    \vertex (i8) at (6,2);
    \vertex (i9) at (6,-2);
    \vertex (i10) at (6+1.414,-1.414);
    \propag [fermion] (i7) to [edge label = $x_{n-1}$] (b);
    \propag [fermion] (i8) to [edge label = $x_n$] (b);
    \propag [fermion] (i9) to [edge label = $x_{m+1}$] (b);
    \propag [fermion] (i10) to [edge label' = $x_{m+2}$] (b);

    \propag [fermion] (a) to [out=-30, in=210,looseness=1] (b);
    \propag[fermion] (a) to [in=150, out=30, looseness=1] (b);
    \propag [fermion] (a) to [out=10, in=170,looseness=1] (b);
    \draw[densely dotted] (3,-0.1) -- (3,-0.5);

    \draw[->] (7.5,0)--(8.5,0);

  \vertex (j1) at (10-1.414,1.414);
    \vertex (j2) at (10,2);
    \vertex (j3) at (10,-2);
    \vertex (j4) at (10-1.414,-1.414);
    \vertex (j5) at (12,0);
    \vertex [grayblob] (c) [draw, text=black, circle] at (10,0) {{}};
    \propag [fermion] (j1) to [edge label = $x_2$] (c);
    \propag [fermion] (j2) to [edge label = $x_1$] (c);
    \propag [fermion] (j3) to [edge label = $x_{m}$] (c);
    \propag [fermion] (j4) to [edge label = $x_{m-1}$] (c);
    \draw [dotted] (9,0.5) arc (135:225:1);
    \draw [dotted] (17,-0.7) arc (-45:45:1);
    \vertex [grayblob] (d) [draw, text=black, circle] at (16,0) {{}};
    \vertex (j6) at (14,0);
    \vertex (j7) at (16+1.414,1.414);
    \vertex (j8) at (16,2);
    \vertex (j9) at (16,-2);
    \vertex (j10) at (16+1.414,-1.414);
    \propag [fermion] (j7) to [edge label = $x_{n-1}$] (d);
    \propag [fermion] (j8) to [edge label = $x_n$] (d);
    \propag [fermion] (j9) to [edge label = $x_{m+1}$] (d);
    \propag [fermion] (j10) to [edge label' = $x_{m+2}$] (d);

\vertex (j11) at (10+1.414,1.414);
\propag [fermion] (j11) to [edge label = $\bar{y}_{1}$] (c);
\vertex (j12) at (10+2,0);
\propag [fermion] (j12) to [edge label = $\bar{y}_{L}$] (c);
    \propag [fermion] (c) to [out=-30, in=210,looseness=1] (d);
    \draw [dotted] (11.3,0.3) arc (10:30:1);
\vertex (j13) at (16-1.414,1.414);
\propag [fermion] (j13) to [edge label = $y_{1}$] (d);
\draw [dotted] (15,0.6) arc (150:170:1);
\vertex (j14) at (16-2,0);
\propag [fermion] (j14) to [edge label = $y_{L}$] (d);
  \end{feynhand}
  \end{tikzpicture}}
  \caption{A loop diagram is shown in the left panel. The internal lines are cut in the right.}
    \label{fig:cut_loop_diagram}
\end{figure}

Let us consider $L$-loop diagrams with external lines labeled by $x_1,x_2,..., x_n \in A$, and non-vanishing $n$-point couplings arise according to the following calculation (see Figure \ref{fig:cut_loop_diagram}). 
An $L$-loop diagram can be converted into a tree diagram with $n+2L$ external lines by cutting its $L$ internal propagators. 
We denote the $2L$ additional external lines resulting from the cut by $y_1,y_2,...,y_L, \bar{y}_1,\bar{y}_2,...,\bar{y}_L$. Note that $y_i$ is accompanied by its conjugate $\bar{y}_i$. 

The obtained tree diagram with $n+2L$ external lines satisfies the tree-level selection rule discussed above:
\begin{align}
    e\prec x_1x_2\cdots x_n(y_1\bar{y}_1)(y_2\bar{y}_2)\cdots (y_L\bar{y}_L).
\end{align}
This condition implies that there exist basis elements $\bar{z}_i \prec y_i\bar{y}_i$ such that 
\begin{align}
    e\prec x_1x_2\cdots x_n \bar{z}_1 \bar{z}_2\cdots \bar{z}_L.
\end{align}
The existence of $\bar{z}_i$ further implies that there exists $w \prec z_1z_2\cdots z_L$ satisfying
\begin{align}
    w \prec x_1x_2\cdots x_n.
\end{align}
Here, we define 
\begin{align}
&    \com(A) :=  \{z\ |\ z\prec y\bar{y}\ {\rm for\ some\ } y\in A\},  \\
 &  \com(A)^L := \{w\ |\ w\prec z_1z_2\cdots z_L\ {\rm for\ some\ }z_1,z_2,...,z_L\in \com(A) \}. \notag
\end{align}
Then, the above discussion shows that $n$-point couplings at $L$-loop level are non-zero only if the following $L$-loop selection rule is satisfied \cite{Kaidi:2024wio}:
\begin{align}
    w \prec x_1x_2\cdots x_n,\quad {\rm for\ some\ }w\in \com(A)^L.
\end{align}
Here and in what follows, we simply assume that the theory contains dynamical particles labeled by all conjugacy classes unless otherwise specified. 
Later, we comment on the case in which dynamical particles are restricted to a subset of conjugacy classes.

Moreover, the selection rules at all loop orders can be simplified by a procedure referred to as {\it groupification}. 
To this end, we introduce the following equivalence relation on any pair of elements $x,y\in A$: 
\begin{align}
  x\sim y \Leftrightarrow {\rm There\ exists\ }w\in \com(A)^{\infty}\ {\rm such\ that\ }x\prec wy,
\end{align}
and define the quotient set as $\gr[A] := A/\sim$. 
To endow $\gr[A]$ with a group structure, we further define the product: 
\begin{align}
  [x]\cdot[y] = [z]\quad {\rm if\ and\ only\ if\ } N^{z}_{xy}\neq 0.
\end{align}
Note that when the multiplication rules (\ref{eq:xy=NZ}) are commutative, i.e., $N^z_{xy}=N^z_{yz}$, $\gr[A]$ forms an Abelian group whose unit element is obviously $[e]$.\par
As a result, the selection rule reduces to
\begin{align}
  [e] = [x_1][x_2]\cdots [x_n],
\end{align}
which takes the form of a group-like selection rules.

\section{Group-like symmetries through groupification}
\label{sec:groupification}

Here, we study loop effects on non-invertible selection rules by examining concrete examples.
For example, selection rules in heterotic string theory on $T^6/S_3$ and $T^6/T_7$ as well as $T^2/S_3$ were studied in Ref.~\cite{Kobayashi:2025ocp}. Furthermore, in Ref.~\cite{Dong:2025jra}, the multiplication rules of conjugacy classes of various discrete groups such as $D_N$, $\Delta(3N^2)$, $\Delta(6N^2)$ were studied.\footnote{The names of discrete groups in this paper follow Refs.~\cite{Ishimori:2010au,Kobayashi:2022moq}.}

The multiplication rules (\ref{eq:xy=NZ}) of conjugacy classes are commutative, i.e., $N^z_{xy}=N^z_{yx}$.
Let $[a]$ and $[b]$ denote the conjugacy classes including the elements $a$ and $b$, respectively. 
The product of conjugacy class $[a]\cdot[b]$ contains the element $ab$.
It is straightforward to see that
\begin{align}
    ab=b(b^{-1}ab),
\end{align}
where the right-hand side is included in the product conjugacy class $[b]\cdot[a]$.
Thus, we find $[a]\cdot[b]=[b]\cdot[a]$. 
Therefore, the groupification of the conjugacy class algebra yields an Abelian group.

\subsection{$D_N$}
\label{sec:DN}
First, we discuss the dihedral group $D_N \cong \mathbb{Z}_N \rtimes \mathbb{Z}_2$, whose order is $2N$. Taking $a$ as the $\mathbb{Z}_N$ generator and $b$ as the $\mathbb{Z}_2$ generator, they satisfy the following algebraic relations:
\begin{align}
  a^N = b^2 = e,\quad bab = a^{-1},
\end{align}
where $e$ denotes the identity.
Note that arbitrary elements in $D_N$ can be expressed as $a^mb^n,\ m=0,1,...,N-1,\ n= 0,1$. 

\subsubsection{Multiplication rules}

Here, we give a brief review on conjugacy classes of $D_N$ and their multiplication rules \cite{Dong:2025jra}.
We obtain conjugacy classes of $D_N$ as follows \cite{Ishimori:2010au,Kobayashi:2022moq}, 
\begin{itemize}
  \item When $N$ is even, 
  \begin{align}
    &C_1 = \{e\}, \notag\\
    &C^1_2=\{a,a^{-1}\},\notag\\
    &C^2_2=\{a^2,a^{-2}\},\notag\\
    &\vdots\notag\\
    &C^{\frac{N-2}{2}}_2 = \{a^{\frac{N-2}{2}},a^{-\frac{N-2}{2}}\},\notag\\
    &C^{\frac{N}{2}}_1 = \{ a^{\frac{N}{2}} \},\notag\\
    &B_1 = \{b,a^2b,...,a^{N-2}b\},\notag\\
    &B_2 = \{ ab,a^3b,...,a^{N-1}b \}.
  \end{align}
  \item When $N$ is odd,
  \begin{align}
    &C_1 = \{e\}, \notag\\
    &C^1_2=\{a,a^{-1}\},\notag\\
    &C^2_2=\{a^2,a^{-2}\},\notag\\
    &\vdots\notag\\
    &C^{\frac{N-1}{2}}_2 = \{a^{\frac{N-1}{2}},a^{-\frac{N-1}{2}}\},\notag\\
    &B_1 = \{b,ab,a^2b,...,a^{N-1}b\}.
  \end{align}
\end{itemize}
We now turn to multiplication rules of the conjugacy classes of $D_N$. In general, multiplication and fusion coefficients between two conjugacy classes of finite group $G$ are defined by
\begin{align}
\label{eq:def_conjugacy_class_multiplication}
  C_g\cdot C_h = \sum_{i\in R}N^i_{gh}C_i,
\end{align}
with
\begin{align}
\label{eq:def_conjugacy_class_fusion_coefficient}
  N^{i}_{gh} = \frac{|\{ (g',h')\in C_g\times C_h\ |\ g'h'\in C_i \}|}{|C_i|},
\end{align}
where $R$ is a set of representatives for conjugacy classes of $G$. We denote by $\conj(G)$ the fusion algebra constructed from the conjugacy classes of $G$. Then we can calculate the multiplication rules of $D_N$ as listed below. In what follows, we omit the products between $C_1$ and other classes because they are trivial.
\begin{itemize}
  \item When $N$ is even,
  \begin{align}
    &C^{k}_2\cdot C^l_2 = C^{k+l}_2+C^{k-l}_{2},\notag\\
    &C^{k}_2\cdot C^{\frac{N}{2}}_1 = C^{k+\frac{N}{2}}_2,\notag\\
    &C^{k}_2\cdot B_1 = 
    \left\{ \,
    \begin{aligned}
    & 2B_1,\ {\rm if}\ k{\rm\ is\ even }\\
    & 2B_2,\ {\rm if}\ k{\rm\ is\ odd }\\
    \end{aligned}
\right.
,\notag\\
&C^{k}_2\cdot B_2 = 
    \left\{ \,
    \begin{aligned}
    & 2B_2,\ {\rm if}\ k{\rm\ is\ even }\\
    & 2B_1,\ {\rm if}\ k{\rm\ is\ odd }\\
    \end{aligned}
\right.
,\notag\\
    &C^{\frac{N}{2}}_1\cdot C^{\frac{N}{2}}_1 = C_1,\notag\\
    &C^{\frac{N}{2}}_1\cdot B_1 = 
    \left\{ \,
    \begin{aligned}
    & B_1,\ {\rm if}\ N/2{\rm\ is\ even }\\
    & B_2,\ {\rm if}\ N/2{\rm\ is\ odd }\\
    \end{aligned}
\right.
,\notag\\
&C^{\frac{N}{2}}_1\cdot B_2 = 
    \left\{ \,
    \begin{aligned}
    & B_2,\ {\rm if}\ N/2{\rm\ is\ even }\\
    & B_1,\ {\rm if}\ N/2{\rm\ is\ odd }\\
    \end{aligned}
\right.
,\notag\\
&B_1\cdot B_1 = B_2\cdot B_2 =  
    \left\{ \,
    \begin{aligned}
    & \frac{N}{2}C_1+\frac{N}{2}C^2_2+\frac{N}{2}C^4_2+\cdots +\frac{N}{2}C^{\frac{N-4}{2}}_2+\frac{N}{2}C^{\frac{N}{2}}_1,\ {\rm if}\ N/2{\rm\ is\ even }\\
    & \frac{N}{2}C_1+\frac{N}{2}C^2_2+\frac{N}{2}C^4_2+\cdots +\frac{N}{2}C^{\frac{N-2}{2}}_2,\ {\rm if}\ N/2{\rm\ is\ odd }\\
    \end{aligned}
\right.
,\notag\\
&B_1\cdot B_2 = 
\left\{ \,
    \begin{aligned}
    & \frac{N}{2}C^1_2+\frac{N}{2}C^3_2+\cdots + \frac{N}{2}C^{\frac{N-2}{2}}_2,\ {\rm if}\ N/2{\rm\ is\ even }\\
    & \frac{N}{2}C^1_2+\frac{N}{2}C^3_2+\cdots + \frac{N}{2}C^{\frac{N-4}{2}}_2+\frac{N}{2}C^{\frac{N}{2}}_1,\ {\rm if}\ N/2{\rm\ is\ odd }\\
    \end{aligned}
\right.,
  \end{align}
  where $k,l = 1,2,...,\frac{N-2}{2}$ and note that when $k\pm l$ becomes $0$ or $N/2$, we replace $C^{k\pm l}_2$ by $2C_1,2C^{\frac{N}{2}}_2$ respectively.
  \item When $N$ is odd,
  \begin{align}
    &C^{k}_2\cdot C^l_2 = C^{k+l}_2+C^{k-l}_{2},\notag\\
    &C^{k}_2\cdot B_1 = 2B_1,\notag\\
    &B_1\cdot B_1 = NC_1+NC^1_2+NC^2_2+\cdots + NC^{\frac{N-1}{2}}_2,
  \end{align}
  where $k,l = 1,2,...,\frac{N-1}{2}$ and note that when $k\pm l$ becomes $0$, we replace $C^{k\pm l}_2$ by $2C_1$.
\end{itemize}
When $N$ is even, it turns out that there is a $\mathbb{Z}_2$ symmetry: $C^{{\rm even}}_2\rightarrow C^{{\rm even}}_2$ and $C^{{\rm odd}}_2\rightarrow -C^{{\rm odd}}_2$, as seen in the following multiplication rules: 
\begin{align}
\label{eq:DN_symmetry}
  &C^{{\rm even}}_2\cdot C^{{\rm even}}_2 = C^{{\rm even}}_2 + C^{{\rm even}}_2,\notag\\
  &C^{{\rm even}}_2\cdot C^{{\rm odd}}_2 = C^{{\rm odd}}_2 + C^{{\rm odd}}_2,\\
  &C^{{\rm odd}}_2\cdot C^{{\rm odd}}_2 = C^{{\rm even}}_2+C^{{\rm even}}_2.\notag
\end{align}
In addition, the multiplication rules have the permutation symmetry between $B_1$ and $B_2$, i.e., $S_2$, which is outer automorphism of $D_N$ with $N=$ even. 
Although $S_2$ is isomorphic to $\mathbb{Z}_2$, we denote this permutation symmetry by $S_2$ to distinguish other $\mathbb{Z}_2$, where each conjugacy class carries a definite charge.

\subsubsection{Loop effects}

In this section, we analyze loop effects. 
As discussed in Section \ref{sec:loop}, the sets $\com(A)^L$ and $\gr(A)$ determine the structure of loop corrections. 
The former $\com(A)^L$ characterizes loop-induced violations of the tree-level selection rules, while the latter $\gr(A)$ encodes the residual symmetry that remains unbroken by loop effects. 
From the multiplication rules of $\conj(D_N)$ presented in the previous section, we observe that all conjugacy classes of $D_N$ are self conjugate. Therefore, all elements in $\com(\conj(D_N))$ appear on the right-hand side of the product of a conjugacy class with itself. 
It follows that when $N$ is even, the conjugacy classes included in $\com(\conj(D_N))$ consist of even powers of $a$ including the identity $e$; 
on the other hand, when $N$ is odd, 
the conjugacy classes in $\com(\conj(D_N))$ contain $a^n$ for any $n$ as well as the identity $e$, i.e., 
\begin{align}
    \com(\conj(D_N)) = 
    \left\{ \,
    \begin{aligned}
    & \{ C_1, C^n_2 | n\in 2\mathbb{Z}\}\,\, {\rm when\ }N{\rm \ is\ even}\\
    &  \{ C_1, C^n_2 | n\in \mathbb{Z}\}\,\, {\rm when\ }N{\rm \ is\ odd}\\
    \end{aligned}
\right.
.
\end{align}
Note that if $N/2$ is even, $C^{N/2}_1$ is also included in $\com(\conj(D_N))$. 

Moreover, we can derive the groupification of $D_N$ from the above multiplication rules by focusing on the products between the elements in $\com(\conj(D_N))$ and the remaining conjugacy classes. 
We find that when $N$ is even, 
\begin{align}
    \gr[\conj(D_N)] =  \{ [C_1],[C^{{\rm odd}}_2], [B_1], [B_2] \}\cong \mathbb{Z}_2\times \mathbb{Z}_2,
\end{align}
where we define
\begin{gather}
    [C_1] = \com(\conj(D_N)),\quad [C^{{\rm odd}}_2] = \{ C^n_2|n\in 2\mathbb{Z}+1 \},\notag\\
    [B_1]=\{ B_1 \},\quad [B_2] = \{ B_2 \}.
\end{gather}
Accordingly, the classes $[C_1]$, $[C^{{\rm odd}}_2]$, $[B_1]$, and  $[B_2]$ carry (even, even), (even,odd), (odd,even), and (odd,odd) charges of $\mathbb{Z}_2\times \mathbb{Z}_2$, respectively. 
In addition, we can impose the permutation symmetry $S_2$ between $B_1$ and $B_2$.

On the other hand, when $N$ is odd,
\begin{align}
    \gr[\conj(D_N)] =  \{ [C_1],[B_1]\}\cong \mathbb{Z}_2,
\end{align}
where we define
\begin{align}
    [C_1] = \com(\conj(D_N)),\quad [B_1]=\{ B_1 \}.
\end{align}
In this case, $[C_1]$, and $[B_1]$ carry
$\mathbb{Z}_2$ even and odd charges, respectively.

\subsubsection{Example: $S_3$}

Let us consider the simplest example, $S_3\cong D_3$. 
$S_3$ is the smallest non-Abelian discrete group.
For example, $S_3$ orbifolded heterotic models were studied in Refs.~\cite{Inoue:1987ak,Inoue:1988ki,Inoue:1990ci,Konopka:2012gy,Fischer:2012qj,Fischer:2013qza,Funakoshi:2025lxs,Hernandez-Segura:2025sfr,Kobayashi:2025ocp}.
The conjugacy classes $S_3$ are given by
\begin{align}
  C_1=\{e\},\quad C^{1}_2 = \{a,a^2\},\quad B_1 = \{b,ab,a^2b\}.
\end{align}
They obey the multiplication rules summarized in Table~\ref{tab:S3}.
\begin{table}[H]
    \centering
    \caption{Multiplication rules for conjugacy classes of $S_3$.}
    \label{tab:S3}
    \begin{tabular}{|c||c|c|c|}
    \hline
    & $C_1$ & $C^1_2$ & $B_1$\\
    \hline\hline
    $C_1$ & $C_1$& $C^1_2$&$B_1$\\
    \hline
    $C^1_2$ &$C^1_2$ &$2C_1+C^1_2$ &$2B_1$\\
    \hline
    $B_1$ & $B_1$&$2B_1$ &$3C_1+3C^1_2$\\
    \hline
    \end{tabular}
\end{table}
From these results, we obtain the subset $\com (\conj(S_3))$ as
\begin{align}
  \com(\conj(S_3)) = \{ C_1,C^1_2 \},
\end{align}
and we can also compute the groupification of $\conj(S_3)$ as follows:
\begin{align}
  \gr[\conj(S_3)] = \{[C_1],[B_1] \} \cong \mathbb{Z}_2,
\end{align}
where we define
\begin{align}
  [C_1] = \{C_1,C^1_2\},\quad [B_1]=\{B_1\}.
\end{align}
It turns out that $[C_1]$, and $[B_1]$ have
$\mathbb{Z}_2$ even and odd charges, respectively.

Now, let us study loop effects more explicitly. 
In order to simplify the notation for fields and their couplings, we introduce the following shorthand notation:
\begin{align}
  &0 = C_1=\{ e \},\quad \dot{1} = C^1_2=\{ a,a^2 \},\quad \dot{2} = B_1=\{ b,ab,a^2b \}.
\end{align}
We denote self-conjugate elements by dotted number, while the conjugate element of $k$ is written as $\bar{k}$ and $0$ represents the unit element.
The notation $k$ and $\bar{k}$ will be used later. 
Using this notation, we label the corresponding fields as $\phi_0,\phi_{\dot 1}, \phi_{\dot 2}$ and their couplings as $\lambda_{000}$, $\lambda_{00\dot{1}}$, etc.

Let us focus on three point couplings.\footnote{Any tree-level string amplitudes can be decomposed into combinations of three string interactions, i.e., pants diagrams and also loop amplitudes can be reduced to such building blocks, although we work within an effective field theory framework.}
The selection rules imposed by $\conj(S_3)$ allow the 3-point couplings 
shown in the second row of Table \ref{tab:3_point_coupling_S3} at tree level. 
Specifically, the following couplings are non-vanishing:
\begin{align}
    \lambda^{(0)}_{000}, \quad \lambda^{(0)}_{0\dot{1}\dot{1}}, \quad \lambda^{(0)}_{0\dot{2
    }\dot{2}}, \quad \lambda^{(0)}_{\dot{1} \dot{1} \dot{1}}, \quad \lambda^{(0)}_
    {\dot{1}\dot{2
    } \dot{2} }.
\end{align}
We can define an approximate $\mathbb{Z}'_2$ symmetry such that $\phi_0$ and $\phi_{\dot{2}}$, corresponding to $C_1$ and $B_1$, carry $\mathbb{Z}'_2$ even charge, while $\phi_{\dot{1}}$, corresponding to $C^1_2$, carries odd charge.
However, the couplings $\lambda^{(0)}_{\dot{1} \dot{1} \dot{1}}$ and $\lambda^{(0)}_{\dot{1}\dot{2  } \dot{2} }$ violate this $\mathbb{Z}'_2$ symmetry. 
Note that $C_2^1$ belongs to $\com(\conj(S_3))$.
In the spurion terminology~\cite{Suzuki:2025bxg,Suzuki:2025kxz}, we can regard $\lambda^{(0)}_{\dot{1} \dot{1} \dot{1}}$ and $\lambda^{(0)}_{\dot{1}\dot{2  } \dot{2} }$ as carrying $\mathbb{Z}'_2$ odd charge, while the other couplings are $\mathbb{Z}'_2$ even.

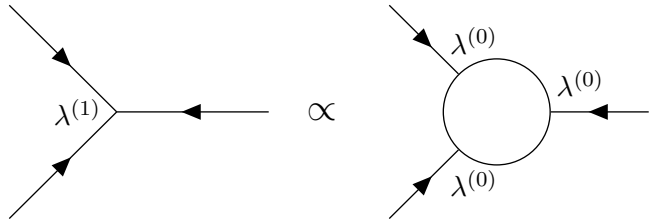
\begin{figure}[H]
  \centering
  \begin{tikzpicture}
  \begin{feynhand}  
    \vertex (i1) at (-1.414,1.414);
    \vertex (i2) at (2,0);
    \vertex (i3) at (-1.414,-1.414);
    \vertex [ringblob] (a) [draw, text=black, circle] at (0,0) {{}};
    \propag [fermion] (i1) to  (a);
    \propag [fermion] (i2) to  (a);
    \propag [fermion] (i3) to  (a);
    \vertex (v_1) at (1.1,0) {};
\node[above=2pt] at (v_1) {$\lambda^{(0)}$};
\vertex (v_2) at (-0.3,0.6) {};
\node[above=1pt] at (v_2) {$\lambda^{(0)}$};
\vertex (v_3) at (-0.3,-0.6) {};
\node[below=1pt] at (v_3) {$\lambda^{(0)}$};
    \vertex (i4) at (-5-1.414,1.414);
    \vertex (i5) at (-5+2,0);
    \vertex (i6) at (-5-1.414,-1.414);
    \vertex (v_3) at (-5,0);
    \node[left=2pt] at (v_3) {$\lambda^{(1)}$};
    \propag [fermion] (i4) to  (v_3);
    \propag [fermion] (i5) to  (v_3);
    \propag [fermion] (i6) to  (v_3);
    \node at (-2.3, 0) {\Large{$\propto$}};
  \end{feynhand}
  \end{tikzpicture}
  \caption{Loop-induced couplings.}
    \label{fig:3-point_1_loop_diagram}
\end{figure}

We now examine loop effects on the 3-point couplings.
For example, the loop diagram shown in Figure \ref{fig:3-point_1_loop_diagram} generates a new coupling $\lambda^{(1)}_{0 0 \dot{1}}$ of the form: 
\begin{align}
\label{eq:new-coupling-S3}
{
\lambda^{(1)}_{ 0 0 \dot{1}
}
}&\propto
\lambda^{(0)}_{ 0 \dot{1} \dot{1} }\lambda^{(0)}_{ 0 \dot{1} \dot{1} }\lambda^{(0)}_{ \dot{1} \dot{1} \dot{1} }
+
\lambda^{(0)}_{ 0 \dot{2} \dot{2} }\lambda^{(0)}_{ 0 \dot{2} \dot{2} }\lambda^{(0)}_{ \dot{1} \dot{2} \dot{2} }.
\end{align}
This corresponds to the third row in Table \ref{tab:3_point_coupling_S3} and violates the tree-level selection rules. 
The 3-point couplings other than those shown in Table \ref{tab:3_point_coupling_S3} are forbidden by $\gr[\conj(S_3)] = \{[C_1],[B_1] \} \cong \mathbb{Z}_2$. 
We can see the correspondence of $\mathbb{Z}'_2$ charges between the left and right hand sides in Eq.~(\ref{eq:new-coupling-S3}). 
When both $\lambda^{(0)}_{\dot{1} \dot{1} \dot{1}}$ and $\lambda^{(0)}_{\dot{1}\dot{2  } \dot{2} }$ are small, the induced coupling $\lambda^{(1)}_{0 0 \dot{1}}$ is also suppressed. 
This structure, controlled by the approximate $\mathbb{Z}'_2$ symmetry, is interesting. 
In the limit $\lambda^{(0)}_{\dot{1} \dot{1} \dot{1}}, \lambda^{(0)}_{\dot{1}\dot{2  } \dot{2} } \to 0$, the loop-induced coupling 
$\lambda^{(1)}_{0 0 \dot{1}}$ vanishes as well, and the tree-level selection rules are preserved. 
Indeed, in this limit, the approximate $\mathbb{Z}'_2$ symmetry becomes exact. 
Hence, most of the parameters discussed in the non-invertible selection rules are natural in the sense of ’t Hooft.

Table~\ref{tab:3_point_coupling_S3} lists the $3$-point couplings that are allowed at tree level and $1$-loop level. 

\begin{table}[H]
  \centering
  \caption{Allowed $3$-point couplings obtained from $\conj(S_3)$.}
  \label{tab:3_point_coupling_S3}
  \begin{tabular}{|c|c|c|}
  \hline
      &  3-point coupling&types\\
      \hline
     tree & $C_1C_1C_1$, $C^1_2C^1_2C_1$, $B_1B_1C_1$, $C^1_2C^1_2C^1_2$, $C^1_2B_1B_1$&5\\
     \hline
     1-loop & $C_1C_1C^1_2$&1(6)\\
     \hline
  \end{tabular}
\end{table}

Note that the coupling $\lambda^{(0)}_{00\dot{1}}$ is forbidden at tree level, because of the property of the unit, i.e., $\phi_0 \phi_{\dot{1}}\sim \phi_{\dot{1}}$.
We can generalize this result and find 
the tree-level couplings $\lambda^{(0)}_{00{k}}$ is forbidden by $\conj(G)$, where $\phi_{{k}}$ corresponds to any except the unit.
These couplings may be induced by loop effects.

We have examined the diagram shown in Figure \ref{fig:3-point_1_loop_diagram}.
Similarly, we can discuss other diagrams and find the same behaviors on the approximate symmetry and $\gr[\conj(S_3)]$.
See Appendix~\ref{app:2-point} for 2-point couplings.

\subsubsection{Example: $D_4$}

As another concrete example, we discuss the conjugacy classes of $D_4$.
$D_4$ orbifolded heterotic models were studied, e.g., in Refs.~\cite{Fischer:2012qj,Fischer:2013qza,Hernandez-Segura:2025sfr}.
The conjugacy classes of $D_4$ are listed as follows,
\begin{align}
  C_1&=\{e\},\quad C^{1}_2 = \{a,a^3\},\quad C^2_1=\{a^2\},\notag\\
  &B_1 = \{b,a^2b\},\quad B_2 = \{ab,a^3b\}.
\end{align}
They obey the multiplication rules shown in Table~\ref{tab:D4}.
\begin{table}[H]
    \centering
    \caption{Multiplication rules for conjugacy classes of $D_4$.}
    \label{tab:D4}
    \begin{tabular}{|c||c|c|c|c|c|}
    \hline
         &  $C_1$ & $C^{1}_{2}$ & $C^{2}_1$ & $B_1$ & $B_2$ \\
         \hline\hline
         $C_1$&  $ C_1 $& $ C^{1}_2 $ & $C^{2}_1  $ & $ B_1 $ & $ B_2 $   \\
         \hline
         $C^{1}_{2}$&  $C^{1}_{2}$  & $ 2C_1+2C^{2}_1 $ & $ C^{1}_2 $ & $ 2B_2 $ &  $ 2B_1 $  \\
         \hline
         $C^{2}_1$&  $C^{2}_1$ & $ C^{1}_2 $ & $ C_1 $ & $ B_1 $ &  $ B_2 $ \\
         \hline
         $B_1$&  $B_1$  & $ 2B_2 $ & $ B_1 $ & $ 2C_1+2C^{2}_1 $ & $2C^{1}_2$  \\
         \hline
         $B_2$&  $B_2$  & $ 2B_1 $ & $ B_2 $ & $2C^{1}_2  $ & $ 2C_1+2C^{2}_1 $ \\
         \hline
    \end{tabular}
\end{table}
From these results, we can obtain the subset $\com(\conj(D_4))$ as
\begin{align}
  \com(\conj(D_4)) = \{ C_1,C^2_1 \},
\end{align}
and we can also compute the groupification of $\conj(D_4)$ as
\begin{align}
  \gr[\conj(D_4)] = \{ [C_1], [C^1_2], [B_1], [B_2] \} \cong \mathbb{Z}_2\times \mathbb{Z}_2,
\end{align}
where we define
\begin{align}
  [C_1] = \{C_1,C^2_1\},\quad [C^1_2] = \{C^1_2\},\quad [B_1]=\{B_1\},\quad [B_2] = \{B_2\}.
\end{align}
It turns out that the classes $[C_1]$, $[C^{1}_2]$, $[B_1]$, and  $[B_2]$ carry (even, even), (even,odd), (odd,even), and (odd,odd) charges of $\gr (\conj (D_4)) \cong\mathbb{Z}_2\times \mathbb{Z}_2$, respectively. 
Note that the multiplication rules have the permutation symmetry $S_2$ between $B_1$ and $B_2$.

Now, let us examine the loop effects more explicitly, in analogy with the $S_3$ case.
We introduce the following new notation:
\begin{align}
  &0 =C_1= \{ e \},\quad \dot{1} =C_2^1= \{ a,a^3 \},\quad \dot{2} =C_1^2= \{ a^2 \},\notag\\
  &\dot{3} =B_1= \{ b, a^2b \},\quad \dot{4} =B_2= \{ ab, a^3b \}.
\end{align}
Using this notation, we denote the corresponding fields, $\phi_0$, $\phi_{\dot{1}}, \cdots$, and their couplings by, e.g., $\lambda_{00\dot{1}}$. 
The selection rules imposed by $\conj(D_4)$ allow the 3-point couplings listed in the second row of Table \ref{tab:3_point_coupling_D4}, which correspond to the following couplings:
\begin{align}
        \lambda^{(0)}_{000}, \quad \lambda^{(0)}_{0\dot{1}\dot{1}}, \quad
        \lambda^{(0)}_{\dot{1}\dot{1}\dot{2}}, \quad
        \lambda^{(0)}_{0\dot{2
    }\dot{2}},\quad 
    \lambda^{(0)}_{0\dot{3
    }\dot{3}},\quad 
    \lambda^{(0)}_{\dot{2}\dot{3
    }\dot{3}},\quad 
    \lambda^{(0)}_{0\dot{4
    }\dot{4}},\quad 
    \lambda^{(0)}_{\dot{2}\dot{4
    }\dot{4}},\quad 
    \lambda^{(0)}_{\dot{1} \dot{3} \dot{4}},
\end{align}
while the others are forbidden at tree level.
We can define an approximate $\mathbb{Z}_2'$ symmetry such that 
$\phi_{\dot{2}}$, corresponding to $C^2_1$, carries $\mathbb{Z}'_2$ odd charge, while the others are $\mathbb{Z}'_2$ even.
However, the couplings $\lambda^{(0)}_{\dot{1} \dot{1} \dot{2}}$, $\lambda^{(0)}_{\dot{2} \dot{3} \dot{3}}$  and 
$\lambda^{(0)}_{\dot{2} \dot{4} \dot{4}}$ violate this $\mathbb{Z}'_2$ symmetry. 
Note that $C_1^2$ belongs to $\com(\conj(D_4))$.
In the spurion language, these couplings are $\mathbb{Z}'_2$ odd, while the other couplings are $\mathbb{Z}'_2$ even. 

The loop diagrams induce the following couplings:
\begin{align}
\label{eq:loop-D4}
{
\lambda^{(1)}_{ 0 0 \dot{2}
}
}&\propto
\lambda^{(0)}_{ 0 \dot{1} \dot{1} }\lambda^{(0)}_{ 0 \dot{1} \dot{1} }\lambda^{(0)}_{ \dot{1} \dot{1} \dot{2} }
+
\lambda^{(0)}_{ 0 \dot{3} \dot{3} }\lambda^{(0)}_{ 0 \dot{3} \dot{3} }\lambda^{(0)}_{ \dot{2} \dot{3} \dot{3} }
+
\lambda^{(0)}_{ 0 \dot{4} \dot{4} }\lambda^{(0)}_{ 0 \dot{4} \dot{4} }\lambda^{(0)}_{ \dot{2} \dot{4} \dot{4} }, \notag \\
{
\lambda^{(1)}_{ \dot{2} \dot{2} \dot{2}
}
}&\propto
\lambda^{(0)}_{ \dot{1} \dot{1} \dot{2} }\lambda^{(0)}_{ \dot{1} \dot{1} \dot{2} }\lambda^{(0)}_{ \dot{1} \dot{1} \dot{2} }
+
\lambda^{(0)}_{ \dot{2} \dot{3} \dot{3} }\lambda^{(0)}_{ \dot{2} \dot{3} \dot{3} }\lambda^{(0)}_{ \dot{2} \dot{3} \dot{3} }
+
\lambda^{(0)}_{ \dot{2} \dot{4} \dot{4} }\lambda^{(0)}_{ \dot{2} \dot{4} \dot{4} }\lambda^{(0)}_{ \dot{2} \dot{4} \dot{4} }.
\end{align}
These couplings are shown in Table \ref{tab:3_point_coupling_D4} and violate the tree-level selection rules.
However, they are consistent with $\gr (\conj(D_4))$, which forbids the other three-point couplings. 
We can verify the correspondence of $\mathbb{Z}'_2$ charges between the left and right hand sides in Eq.~(\ref{eq:loop-D4}). 
When all of $\lambda^{(0)}_{\dot{1} \dot{1} \dot{2}}$, $\lambda^{(0)}_{\dot{2} \dot{3} \dot{3}}$  and 
$\lambda^{(0)}_{\dot{2} \dot{4} \dot{4}}$ are small, the induced couplings $\lambda_{00 \dot{2}}^{(1)} $ and $\lambda_{\dot{2} \dot{2} \dot{2}}^{(1)}$ are also small.
In the limit $\lambda^{(0)}_{\dot{1} \dot{1} \dot{2}}, \lambda^{(0)}_{\dot{2} \dot{3} \dot{3}}, 
\lambda^{(0)}_{\dot{2} \dot{4} \dot{4}} \to 0$, 
the loop-induced couplings $\lambda_{00 \dot{2}}^{(1)} $ and $\lambda_{\dot{2} \dot{2} \dot{2}}^{(1)}$ vanish, and the tree-level selection rules are not violated. 
Indeed, in this limit, the approximate $\mathbb{Z}'_2$ becomes exact.

Note that the multiplication rules have the permutation symmetry $S_2$.
We can impose this symmetry $S_2$ between $\phi_{\dot{3}}$ and $\phi_{\dot{4}}$ in a field theory by requiring 
$\lambda^{(0)}_{0\dot{3} \dot{3}}=\lambda^{(0)}_{0\dot{4} \dot{4}}$.
When we combine $S_2$ with $\gr[\conj(D_4)]\cong\mathbb{Z}_2 \times \mathbb{Z}_2$, the remaining symmetry can be enhanced to $\mathbb{Z}_2\times D_4$.
When 
$\lambda^{(0)}_{0\dot{3} \dot{3}} \neq \lambda^{(0)}_{0\dot{4} \dot{4}}$, 
the symmetry is not enhanced from $\gr[\conj(D_4)]\cong\mathbb{Z}_2 \times \mathbb{Z}_2$. 
This argument can be generalized to $D_N$ with $N=$ even, and one can realize the non-Abelian group-like symmetry $\mathbb{Z}_2\times D_4$ when $\lambda^{(0)}_{0\dot{3} \dot{3}}=\lambda^{(0)}_{0\dot{4} \dot{4}}$.

\begin{table}[H]
  \centering
  \caption{Allowed $3$-point couplings obtained from $\conj(D_4)$}
  \label{tab:3_point_coupling_D4}
  \resizebox{\linewidth}{!}{
  \begin{tabular}{|c|c|c|}
  \hline
      &  3-point coupling&types\\
      \hline
     tree & $C_1C_1C_1$,$C^1_2C^1_2C_1$,$C^1_2C^1_2C^2_1$,$C^2_1C^2_1C_1$,$B_1B_1C_1$,$B_1B_1C^2_1$,$B_2B_2C_1$,$B_2B_2C^2_1$,$C^1_2B_1B_2$&9\\
     \hline
     1-loop & $C^2_1C^2_1C^2_1$, $C_1C_1C^2_1$&2(11)\\
     \hline
  \end{tabular}}
\end{table}

\subsection{$T_N$}
\label{sec:TN}
We now move on to a different class of finite groups, represented as $T_N \cong \mathbb{Z}_N \rtimes \mathbb{Z}_3$, whose order is $3N$. Putting $a$ as the $\mathbb{Z}_N$ generator and $b$ as the $\mathbb{Z}_3$ generator, $a$ and $b$ satisfy the following algebraic relations:
\begin{gather}
    a^N = b^3 = e,\quad b^{-1}ab = a^m\,\  (m\neq 0) .
\end{gather}
Note that an arbitrary element of $T_N$ can be expressed in the form $a^mb^n,\ m,n =0,1,...,N-1,\ n=0,1,2$. We focus on the case $m\geq 2$ since for $m=1$, the group reduces to a direct product rather than a semidirect product. The possible combinations of $(N,m)$ are restricted by consistency with the above algebraic relations:
\begin{align}
  (N,m) = (7,2),(7,4),(9,4),(9,7),(13,3),...
\end{align}
For example, $T_7$ orbifolded heterotic models were studied in Refs.\cite{Fischer:2012qj,Fischer:2013qza,Hernandez-Segura:2025sfr,Kobayashi:2025ocp}.

\subsubsection{Multiplication rules}
In the remainder of this section, we assume that $N$ is a prime number. The conjugacy classes of $T_N$ are then given by
\begin{align}
  &C_1 = \{ e \},\notag\\
  &C^k_3 = \{ a^k,a^{km},a^{km^2} \},\notag\\
  &C^1_N = \{ b,ba,...,ba^{N-1} \},\notag\\
  &C^2_N = \{ b^2,b^2a,...,b^2a^{N-1} \}.
\end{align}
The multiplication rules among these conjugacy classes of $T_N$ are calculated as follows,
\begin{align}
  &C^k_3\cdot C^l_3 = C^{k+l}_3+C^{k+lm}_3+C^{k+lm^2}_3,\notag\\
  &C^k_3\cdot C^1_N = 3C^1_N,\notag\\
  &C^k_3\cdot C^2_N = 3C^2_N,\notag\\
  &C^1_N\cdot C^1_N = NC^2_N,\notag\\
  &C^2_N\cdot C^2_N = NC^1_N,\notag\\
  &C^1_N\cdot C^2_N = NC_1 + \sum_k NC^k_3.
\end{align}
 As in the previous section, we omit the products between $C_1$ and other classes because they are trivial.
Note that when $k+l,k+lm,k+lm^2 \equiv 0\pmod N$, the corresponding classes $C^{k+l}_3,C^{k+lm}_3,C^{k+lm^2}_3$ should be replaced by $3C_1$ respectively.

\subsubsection{Loop effects}
The above rules imply that the $C^{k}_3$ and $C^{-k}_3$ are conjugate each other and the same holds for $C^1_N$ and $C^2_N$. It follows that $\com(\conj(\Delta(T_N)))$ contains all conjugacy except for $C^1_N,C^2_N$. Moreover, we can calculate the groupification of $T_N$ from the above multiplication rules, focusing on the products between the elements in $\com(\conj(D_N))$ and other conjugacy classes. 
Thus, we find that
\begin{align}
    \gr[\conj(T_N)] =  \{ [C_1],[C^1_N],
[C^2_{N}]\}\cong\mathbb{Z}_3 ,
\end{align}
where we define
\begin{gather}
    [C_1] = \com(\conj(T_N)),\quad [C^1_N]=\{ C^1_N \},\quad [C^2_N] = \{ C^2_N \}.
\end{gather}
Their $\mathbb{Z}_3$ charges are given by
\begin{align}
    &[C_1]: 0,\quad
    [C_{N}^1] : 1,\quad [C_{N}^2] : 2.
\end{align}

\subsubsection{Example: $T_7$}
Let us consider a concrete example, $T_7\cong \mathbb{Z}_7\rtimes \mathbb{Z}_3$ conjugacy classes, which are given by 
\begin{gather}
  C_1 = \{e\},\quad C^1_3 = \{ a,a^2,a^4 \},\quad C^3_3 = \{ a^3,a^5,a^6 \},\notag\\
  C^1_7 = \{ b,ba,...,ba^6 \},\quad C^2_7 = \{ b^2,b^2a,...,b^2a^6 \}.
\end{gather}
They obey the multiplication rules shown in Table~\ref{tab:T7}.
Note that the multiplication rules have the permutation symmetry $S_2$ between $C_7^1$ and $C_7^2$ as well as permutation between $C_3^1$ and $C_3^3$. That is the outer automorphism.

\begin{table}[H]
  \centering
  \caption{Multiplication rules for conjugacy classes of $T_7$.}
    \label{tab:T7}
  \resizebox{\textwidth}{!}{
  \begin{tabular}{|c||c|c|c|c|c|}
  \hline
     & $C_1$ & $C^1_3$ & $C^3_3$& $C^{1}_7$&$C^{2}_7$ \\
     \hline\hline
     $C_1$&$C_1$ & $C^1_3$ & $C^3_3$&$C^{1}_7$ &$C^{2}_7$ \\
     \hline
     $C^1_3$&$C^1_3$&$C^1_3+2C^3_3$   &$3C_1+C^1_3+C^3_3$&$3C^{1}_7$ &$3C^{2}_7$  \\
     \hline
     $C^3_3$&$C^3_3$  &$3C_1+C^1_3+C^3_3$  &$2C^1_3+C^3_3$& $3C^{1}_7$&$3C^{2}_7$  \\
     \hline
     $C^{1}_7$&$C^{1}_7$&$3C^{1}_7$&$3C^{1}_7$&$7C^{2}_7$&$7C_1+7C^{1}_3+7C^{3}_3$\\
     \hline
$C^{2}_7$&$C^{2}_7$&$3C^{2}_7$&$3C^{2}_7$&$7C_1+7C^{1}_3+7C^{3}_3$&$7C^{1}_7$\\
     \hline
  \end{tabular}
  }
\end{table}
From the above results, we can obtain the subset $\com(\conj(T_7))$ as
\begin{align}
  \com(\conj(T_7)) = \{ C_1,C^1_3,C^3_3 \},
\end{align}
and we can also compute the groupification of $\conj(T_7)$ as follows:
\begin{align}
  \gr(\conj(T_7)) = \{ [C_1],[C^1_7],[C^2_7] \}\cong \mathbb{Z}_3,
\end{align}
where we define
\begin{align}
  [C_1] = \{ C_1,C^1_3,C^3_3 \},\quad [C^1_7]=\{ C^1_7 \},\quad [C^2_7] = \{ C^2_7 \}.
\end{align}
Hence, the three equivalence classes $[C_1],[C^1_7]$ and $[C^2_7]$ form a $\mathbb{Z}_3$ structure. 

Now, let us study this case more explicitly.
We introduce the following new notation:
\begin{align}
  &0 = C_1=\{ e \},\quad 1 =C_3^1= \{ a,a^2,a^4 \},\quad \bar{1} =  C_3^3\{ a^3, a^5,a^6 \},\notag\\
  &2 = C_7^1=\{ b, ba,\cdots,ba^6 \},\quad \bar{2} =C_7^2= \{ b^2, b^2a,\cdots, ba^6 \}.
\end{align}
As mentioned in the previous section, 
we denote self-conjugate elements by dotted number and conjugate element of $k$ by $\bar{k}$ and $0$ is the unit element.
By using this notation, we denote the corresponding fields by $\phi_0$, $\phi_{{1}}, \cdots$, and their couplings by, e.g., $\lambda_{001}$.
The selection rules imposed by $\conj(T_7)$ allow the tree-level 3-point couplings shown in Table \ref{tab:3_point_coupling_T7}, which correspond to the following couplings:
\begin{align}
        \lambda^{(0)}_{000}, \quad \lambda^{(0)}_{111}, \quad 
        \lambda^{(0)}_{\bar{1}\bar{1}\bar{1}}, \quad
        \lambda^{(0)}_{222}, \quad 
        \lambda^{(0)}_{\bar{2}\bar{2}\bar{2}}, \quad
        \lambda^{(0)}_{1 1 \bar{1}}, \quad \lambda^{(0)}_
    {\bar{1}\bar{1} 1 },\quad 
        \lambda^{(0)}_{01\bar{1}},\quad
        \lambda^{(0)}_{02\bar{2}},\quad
        \lambda^{(0)}_{12\bar{2}},\quad
        \lambda^{(0)}_{\bar{1}2\bar{2}},
\end{align}
and the others are forbidden at  tree level.
We can define an approximate $\mathbb{Z}_3'$ symmetry such that 
$\phi_1$ and $\phi_{\bar 1}$, corresponding to $C_3^1$ and $C_3^3$, carry $\mathbb{Z}'_3$ charges, 1 and 2, respectively, while the others have charge 0.
However, the following couplings:
\begin{gather}
    \lambda_{11\bar{1}},\quad 
    \lambda_{12\bar{2}},\\
   \lambda_{1\bar{1}\bar{1}},
 \quad \lambda_{\bar{1}2\bar{2}},
\end{gather}
violate the approximate $\mathbb{Z}_3'$ symmetry.
Note that $C_3^1$ and $C_3^3$ belong to $\com(\conj(T_7))$.
In the spurion language, the former set of couplings carries $\mathbb{Z}'_3$ charge 1, while the latter set carries $\mathbb{Z}_3'$ charge 2.

The loop diagrams induce the following couplings:
\begin{align}
\label{eq:loop-T7}
{
\lambda^{(1)}_{ 0 0 1
}
}&\propto
\lambda^{(0)}_{ 0 1 \bar{1} }\lambda^{(0)}_{ 0 1 \bar{1} }\lambda^{(0)}_{ 1 1 \bar{1} }
+
\lambda^{(0)}_{ 0 2 \bar{2} }\lambda^{(0)}_{ 0 2 \bar{2} }\lambda^{(0)}_{ 1 2 \bar{2} },  \notag \\
{
\lambda^{(1)}_{ 0 0 \bar{1}
}
}&\propto
\lambda^{(0)}_{ 0 1 \bar{1} }\lambda^{(0)}_{ 0 1 \bar{1} }\lambda^{(0)}_{ 1 \bar{1} \bar{1} }
+
\lambda^{(0)}_{ 0 2 \bar{2} }\lambda^{(0)}_{ 0 2 \bar{2} }\lambda^{(0)}_{ \bar{1} 2 \bar{2} }, \notag \\
{
\lambda^{(1)}_{ 0 1 1
}
}&\propto
\lambda^{(0)}_{ 0 1 \bar{1} }\lambda^{(0)}_{ 1 1 1 }\lambda^{(0)}_{ 1 \bar{1} \bar{1} }
+
\lambda^{(0)}_{ 0 1 \bar{1} }\lambda^{(0)}_{ 1 1 \bar{1} }\lambda^{(0)}_{ 1 1 \bar{1} }
+
\lambda^{(0)}_{ 0 2 \bar{2} }\lambda^{(0)}_{ 1 2 \bar{2} }\lambda^{(0)}_{ 1 2 \bar{2} }, \notag \\
{
\lambda^{(1)}_{ 0 \bar{1} \bar{1}
}
}&\propto
\lambda^{(0)}_{ 0 1 \bar{1} }\lambda^{(0)}_{ 1 1 \bar{1} }\lambda^{(0)}_{ \bar{1} \bar{1} \bar{1} }
+
\lambda^{(0)}_{ 0 1 \bar{1} }\lambda^{(0)}_{ 1 \bar{1} \bar{1} }\lambda^{(0)}_{ 1 \bar{1} \bar{1} }
+
\lambda^{(0)}_{ 0 2 \bar{2} }\lambda^{(0)}_{ \bar{1} 2 \bar{2} }\lambda^{(0)}_{ \bar{1} 2 \bar{2} }.
\end{align}
These couplings are shown in Table \ref{tab:3_point_coupling_T7} and violate the tree-level selection rules. 
Nevertheless, all the couplings shown in Table \ref{tab:3_point_coupling_T7} are consistent with $\gr(\conj(T_7))=\mathbb{Z}_3$, which forbids the other three-point couplings. 
We can verify the correspondence of $\mathbb{Z}'_3$ charges between left and right sides in Eq.~(\ref{eq:loop-T7}). 
In particular, when all spurionic couplings carrying $\mathbb{Z}'_3 $ charges 1 and 2 vanish, the tree-level selection rules are not violated.
Indeed, this limit, the approximate $\mathbb{Z}'_3$ becomes exact.

Since the multiplication rules have the permutation symmetry $S_2$, let us impose this $S_2$ symmetry between $\phi_{2}$ and $\phi_{\bar 2}$ in a field theory by requiring $\lambda^{(0)}_{222}=\lambda^{(0)}_{\bar 2 \bar 2  \bar 2}$. 
When we combine $S_2$ with $\gr[\conj(T_7)]\cong \mathbb{Z}_3$, the remaining symmetry can be enhanced to $S_3$.
When $\lambda^{(0)}_{222} \neq \lambda^{(0)}_{\bar 2 \bar 2  \bar 2}$, 
the symmetry is not enhanced from $\gr[\conj(T_7)]\cong \mathbb{Z}_3$. 
This argument can be generalized to $T_N$ for the other $N$, and one can realize the non-Abelian group-like symmetry $S_3$ when $\lambda^{(0)}_{222} = \lambda^{(0)}_{\bar 2 \bar 2  \bar 2}$.

We can construct the theory, where the conjugates of $\phi_2$ and $\phi_{\bar 2}$ are related to each other, i.e., $\phi_2=\bar \phi_{\bar 2}$ and $\bar \phi_2=\phi_{\bar 2}$. 
The hermiticity of Lagrangian requires $\lambda^{(0)}_{222}=(\lambda^{(0)}_{\bar 2 \bar 2  \bar 2})^*$ and $\lambda^{(0)}_{0 2\bar 2}=$ real, etc.
If we require CP invariance further, the couplings are constrained as $\lambda^{(0)}_{222}=(\lambda^{(0)}_{\bar 2 \bar 2  \bar 2})^*=$ real.
In this theory, the above $S_2$ permutation symmetry is realized as CP symmetry, as discussed in Ref.~\cite{Kobayashi:2025wty}. 

Alternatively, we can construct another theory, where some fields $\phi_2$ ($\phi_{\bar 2}$) are not directly related to their charge conjugates $\bar \phi_{\bar{2}}$ ($\bar \phi_{2}$). 
If $\lambda^{(0)}_{222}=\lambda^{(0)}_{\bar 2 \bar 2  \bar 2}$ holds in this theory, 
there appears the permutation symmetry between independent fields $\phi_2$ and $\phi_{\bar 2}$. 
That is the flavor symmetry i.e., $S_3$.
This theory can be CP symmetric, and CP symmetry can be enhanced to generalized CP symmetry (for more details, see, Section~\ref{sec:non-Abelian}.).

\begin{longtable}[H]{|c|p{12cm}|c|}
\caption{Allowed $3$-point couplings obtained from $\conj(T_7)$.} \label{tab:3_point_coupling_T7} \\
\hline
& \multicolumn{1}{c|}{3-point coupling}& types\\
\hline
\endhead
tree & $C_1C_1C_1$, 
$C^1_3C^1_3C^1_3$,
$C^3_3C^3_3C^3_3$,
$C^1_7C^1_7C^1_7$,
$C^2_7C^2_7C^2_7$,
$C^1_3C^1_3C^3_3$,
$C^3_3C^3_3C^1_3$,
$C_1C^1_3C^3_3$,
$C_1C^1_7C^2_7$,
$C^1_3C^1_7C^2_7$,
$C^3_3C^1_7C^2_7$
&11\\
\hline
1-loop & $C_1C_1C^1_3$,
$C_1C_1C^3_3$,
$C^1_3C^1_3C_1$,
$C^3_3C^3_3C_1$
&4(15)\\
\hline
\end{longtable}

Note that the couplings $\lambda^{(0)}_{00{1}}$ and $\lambda^{(0)}_{00 \bar{1}}$  are forbidden at  tree level, because of the property of the unit, i.e., $\phi_0 \phi_{{1}}\sim \phi_{{1}}$.
In addition, the couplings $\lambda^{(0)}_{01{1}}$ and $\lambda^{(0)}_{0\bar 1 \bar{1}}$  are forbidden at tree level.
That is because that $\phi_1\phi_{1}$ does not include $\phi_0$.
If $\phi_1\phi_{1}$ includes $\phi_0$, $\phi_1$ must be self-conjugate.
We can generalize this result and find the tree-level couplings 
$\lambda~{(0)}_{1kk}$ are forbidden by $\conj(G)$, where $\phi_k$ is not self-conjugate.
These couplings may be induced by loop effects.

\subsection{$\Delta(3N^2)$}
\label{sec:Delta3N^2}
Next, we discuss different classes of finite groups, represented as $\Delta(3N^2) \cong (\mathbb{Z}_N\times \mathbb{Z}'_N) \rtimes \mathbb{Z}_3$, 
which has order $3N^2$. Putting $a$ as the $\mathbb{Z}_N$ generator, $a'$ as the $\mathbb{Z}'_N$ generator, and $b$ as the $\mathbb{Z}_3$ generator, they satisfy the following algebraic relations:
\begin{gather}
    a^N = a'^N= e,\quad aa' = a'a,\notag\\
    b^3 = e,\quad bab^{-1} = a^{-1}a'^{-1},\quad ba'b^{-1} = a,
\end{gather}
where $e$ denotes the identity. 
Note that arbitrary elements in $\Delta(3N^2)$ can be  expressed as $b^ka^ma'^n,\ k= 0,1,2,\ m,n =0,1,...,N-1$.

\subsubsection{Multiplication rules}

We obtain conjugacy classes of $\Delta(3N^2)$ as follows  \cite{Ishimori:2010au,Kobayashi:2022moq},
\begin{itemize}
  \item When $N/3$ is not an integer,
  \begin{align}
    &C_1 = \{ e \},\notag\\
    &C^{(k,l)}_3 = \{ a^ka'^l,a^{-k+l}a'^{-k},a^{-l}a'^{k-l} \},\ (k,l)\neq (0,0),\notag\\
    &B^1_{N^2} = \{ ba^la'^m\ |\ l,m = 0,1,...,N-1 \},\notag\\
    &B^2_{N^2} = \{ b^2a^la'^m\ |\ l,m = 0,1,...,N-1 \}.
  \end{align}
  \item When $N/3$ is an integer,
  \begin{align}
    &C_1 = \{ e \},\notag\\
    &C^{k}_1 = \{ a^ka'^{-k} \},\quad k = \frac{N}{3},\frac{2N}{3},\notag\\
    &C^{(k,l)}_3 = \{ a^ka'^l,a^{-k+l}a'^{-k},a^{-l}a'^{k-l} \},\ (k,l)\neq (0,0),(N/3,2N/3),(2N/3,N/3),\notag\\
    &B^{(1,p)}_{N^2/3} = \{ ba^{p-n-3m}a'^n\ |\ m=0,1,...,\frac{N-3}{3},n=0,1,...,N-1 \},\quad p = 0,1,2,\notag\\
    &B^{(2,p)}_{N^2/3} = \{ b^2a^{p-n-3m}a'^n\ |\ m=0,1,...,\frac{N-3}{3},n=0,1,...,N-1 \},\quad p = 0,1,2.\notag
  \end{align}
\end{itemize}
The multiplication rules for the conjugacy classes of $\Delta(3N^2)$ are calculated as follows \cite{Dong:2025jra}. As in the previous cases, we omit the products between $C_1$ and other classes because they are trivial.
\begin{itemize}
  \item When $N/3$ is not an integer,
  \begin{align}
  \label{eq:multi-Delta-3N}
    &C^{(k,l)}_3\cdot C^{(m,n)}_3 = C^{(k+m,l+n)}_3+C^{(k-m+n,l-m)}_3+C^{(k-n,l+m-n)}_3,\notag\\
  &C^{(k,l)}_3\cdot B^1_{N^2} = 3B^1_{N^2},\notag\\
  &C^{(k,l)}_3\cdot B^2_{N^2} = 3B^2_{N^2},\notag\\
  &B^{1}_{N^2}\cdot B^{1}_{N^2} = N^2B^2_{N^2},\notag\\
  &B^{2}_{N^2}\cdot B^{2}_{N^2} = N^2B^1_{N^2},\notag\\
  &B^1_{N^2}\cdot B^2_{N^2} = N^2C_1+\sum_{(k,l)\neq(0,0)}N^2C^{(k,l)}_3.
  \end{align}
  Note that when $(k+m,l+n),(k-m+n,l-m),(k-n,l+m-n)$ become $(0,0)$, we replace $C^{(k+m,l+n)}_3,C^{(k-m+n,l-m)}_3,C^{(k-n,l+m-n)}_3$ by $3C_1$ respectively.
  \item When $N/3$ is an integer,
  \begin{align}
    &C^k_1\cdot C^{l}_1 = 
    \left\{ \,
    \begin{aligned}
    & C^{2k}_1\,\ {\rm if}\ k=l\\
    & C_1\,\ {\rm if}\ k\neq l\\
    \end{aligned}
\right.
,\notag\\
    &C^{k}_1\cdot C^{(l,m)}_3 = C^{(k+l,-k+m)}_3,\notag\\
    &C^{k}_1\cdot B^{(1,p)}_{N^2/3} = B^{(1,p)}_{N^2/3},\notag\\
    &C^{k}_1\cdot B^{(2,p)}_{N^2/3} = B^{(2,p)}_{N^2/3},\notag\\
    &C^{(k,l)}_3\cdot C^{(m,n)}_3 = C^{(k+m,l+n)}_3+C^{(k-m+n,l-m)}_3+C^{(k-n,l+m-n)}_3,\notag\\
    &C^{(k,l)}_3\cdot B^{(1,p)}_{N^2/3} = 3B^{(1,p')}_{N^2/3},\quad p+k+l\equiv p'\pmod 3,\notag\\
    &C^{(k,l)}_3\cdot B^{(2,p)}_{N^2/3} = 3B^{(2,p')}_{N^2/3},\quad p+k+l\equiv p'\pmod 3,\notag\\
    &B^{(1,p)}_{N^2/3}\cdot B^{(1,q)}_{N^2/3} = \frac{N^2}{3}B^{(2,r)}_{N^2/3},\quad p+q\equiv r\pmod 3,\notag\\
    &B^{(2,p)}_{N^2/3}\cdot B^{(2,q)}_{N^2/3} = \frac{N^2}{3}B^{(1,r)}_{N^2/3},\quad p+q\equiv r\pmod 3,\notag\\
    &B^{(1,p)}_{N^2/3}\cdot B^{(2,q)}_{N^2/3} = \left\{ \,
    \begin{aligned}
    & \frac{N^2}{3}C_1+\frac{N^2}{3}C^{N/3}_1+\frac{N^2}{3}C^{2N/3}_1\\&+\frac{N^2}{3}\sum_{(k,l),k+l=0\bmod3}C^{(k,l)}_3 ,\ {\rm if}\ p+q\equiv 0\pmod 3\\
    & \frac{N^2}{3}\sum_{(k,l),k+l=p+q\bmod3}C^{(k,l)}_3 ,\ {\rm otherwise.} \\
    \end{aligned}
\right.
  \end{align}
  Note that when $(k+m,l+n),(k-m+n,l-m),(k-n,l+m-n)$ become $(0,0)$ or $(N/3,2N/3)$ or $(2N/3,N/3)$, we replace $C^{(k+m,l+n)}_3,C^{(k-m+n,l-m)}_3,C^{(k-n,l+m-n)}_3$ by $3C_1,3C^{N/3}_1,3C^{2N/3}_1$ respectively.
\end{itemize}

\subsubsection{Loop effects}

The above multiplication rules imply that when $N/3$ is not an integer, the classes $C^{(k,l)}$ and $C^{(-k,-l)}_3$ are conjugate to each other. The same statement applies for $B_1$ and $B_2$. It follows that $\com(\conj(\Delta(3N^2)))$ contains all conjugacy classes except for $B_1,B_2$. On the other hand, when $N/3$ is an integer, the conjugate pairs are given as follows:
\begin{align}
    (C^{N/3}_1,C^{2N/3}_1),\quad (C^{(k,l)}_3,C^{(-k,-l)}_3),\quad (B^{(1,p)}_{N^2/3},B^{(2,-p)}_{N^2/3}).
\end{align}
It follows that $\com(\conj(\Delta(3N^2)))$ contains the classes $C_1, C^{N/3}_1, C^{2N/3}_1,$ and $C^{(k,l)}_3$ satisfying $k+l=0\bmod3$. Moreover, the groupification of $\Delta(3N^2)$ can be determined from the above multiplication rules by examining the products between elements in $\com(\conj(D_N))$ and the remaining conjugacy classes. 
Thus, we find that when $N/3$ is not an integer,
\begin{align}
    \gr[\conj(\Delta(3N^2))] =  \{ [C_1],[B^1_{N^2}],
[B^2_{N^2}]\}\cong\mathbb{Z}_3 ,
\end{align}
where we define
\begin{gather}
    [C_1] = \com(\conj(\Delta(3N^2))),\quad [B^1_{N^2}]=\{ B^1_{N^2} \},\quad [B^2_{N^2}] = \{ B^2_{N^2} \}.
\end{gather}
Their $\mathbb{Z}_3$ charges are given by
\begin{align}
    &[C_1]: 0,\quad
    [B_{N^2}^1] : 1,\quad [B_{N^2}^2] : 2.
\end{align}

On the other hand, when $N/3$ is an integer, we find
\begin{align}
    \gr[\conj(\Delta(3N^2))] =  \{ [C_1],[C_3^{(n,0)}],[B^{(m,n)}] | n,m=0,1,2\}\cong \mathbb{Z}_3\times \mathbb{Z}_3,
\end{align}
where we define
\begin{align}
    [C_1] = \com(\conj(D_N)),\quad [C^{(n,0)}_3]=\{ C^{(n,0)}_3 \},\quad [B^{(m,n)}] = \{ B^{(m,n)} \}.
\end{align}
Here, their $\mathbb{Z}_3\times \mathbb{Z}_3'$ charges are given by
\begin{align}
    &[C_1]: (0,0),\quad
    [C^{(n,0)}_3] : (n,0),\quad [B^{(m,n)}_3] : (n,m),
\end{align}
where $n=1,2$ and $m=0,1,2$.

\subsubsection{Example: $A_4$}

Here, we consider a concrete example,\footnote{$A_4$ orbifolded heterotic models were studied, e.g., in Ref.~\cite{Fischer:2012qj}.} $A_4 \cong (\mathbb{Z}_2\times \mathbb{Z}'_2)\rtimes \mathbb{Z}_3$ conjugacy classes, which are listed as follows,
\begin{align}
    &C_1 = \{ e \},  \notag \\
    &C_3 = \{ a,a',aa' \},  \notag\\
    &B^1_4 = \{ b,ba,ba',baa' \},   \notag\\
    &B^2_4 = \{ b^2,b^2a,b^2a',b^2aa' \}.
\end{align}
They obey the multiplication rules shown in Table~\ref{tab:A4}. 
Note that the multiplication rules have the permutation symmetry $S_2$ between $C_7^1$ and $C_7^2$ as well as permutation between $B_4^1$ and $B_4^2$, corresponding to the outer automorphism.
\begin{table}[H]
    \caption{Multiplication rules for conjugacy classes of $A_4$.}
    \label{tab:A4}
    \centering
    \begin{tabular}{|c||c|c|c|c|}
    \hline
    &$C_1$&$C_3$&$B^1_4$&$B^2_4$\\
    \hline\hline
    $C_1$&$C_1$&$C_3$&$B^1_4$&$B^2_4$\\
    \hline
    $C_3$&$C_3$&$3C_1+2C_3$&$3B^1_4$&$3B^2_4$\\
    \hline
    $B^1_4$&$B^1_4$&$3B^1_4$&$4B^2_4$&$4C_1+4C_3$\\
    \hline
    $B^2_4$&$B^2_4$&$3B^2_4$&$4C_1+4C_3$&$4B^1_4$\\
    \hline
    \end{tabular}
\end{table}
Here, we simplify the notation by $C_3$, which corresponds to $C_3^{(1,0)}$ in the previous section, because there is a single conjugacy class including only $a$ and $a'$.
From the multiplication rules, we can obtain the subset $\com(\conj(A_4))$ as 
\begin{align}
    \com(\conj(A_4))= \{ C_1, C_3 \} .
\end{align}
We can also compute the groupification of $\conj(A_4)$ as,
\begin{align}
    \gr[\conj(A_4)] = \{ [C_1], [B^1_4], [B^2_4] \} \cong \mathbb{Z}_3,
\end{align}
where 
\begin{align}
    [C_1]= \{ C_1, C_3 \},\quad [B^1_4] = \{  B^1_4 \}, \quad
    [B^2_4] = \{  B^2_4 \}.
\end{align}

Now, let us examine the loop effects more explicitly.
For that purpose, we introduce the new notation:
\begin{align}
    &0 =C_1= \{ e \},  \notag \\
    &\dot{1} =  C_3= \{ a,a',aa' \},  \notag \\
    &2 = B_4^1 = \{ b,ba,ba',baa' \},  \notag\\
    &\bar{2} =B_4^2 = \{ b^2,b^2a,b^2a',b^2aa' \}.
\end{align}
Using this notation, we denote the corresponding fields, $\phi_0,\phi_{\dot{1}},\cdots$ and their couplings by, e.g., $\lambda_{00\dot{1}}$.
At tree level, the selection rules due to $\conj(A_4)$ allow the following 3-point couplings:
\begin{align}
    \lambda^{(0)}_{000}, \quad \lambda^{(0)}_{0\dot{1}\dot{1}}, \quad \lambda^{(0)}_{0 2 \bar 2}, \quad \lambda^{(0)}_{\dot{1} \dot{1} \dot{1}} \quad \lambda^{(0)}_{\dot{1} 2 \bar 2}, \quad \lambda^{(0)}_{222}, \quad \lambda^{(0)}_{\bar 2 \bar 2 \bar 2},
\end{align}
and the others are forbidden.
We can define an approximate $\mathbb{Z}_2'$ symmetry such that $\phi_{\dot{1}}$, corresponding to $C_3$, carries $\mathbb{Z}_2'$ charge odd, while the others are $\mathbb{Z}_2'$ even.
However, the couplings $\lambda^{(0)}_{\dot{1} \dot{1} \dot{1}}$ and $\lambda^{(0)}_{\dot{1} 2 \bar 2}$ violate the $\mathbb{Z}_2'$ symmetry.
Note that $C_3$ belongs to $\com(\conj(A_4))$.
In the spurion language, these couplings are $\mathbb{Z}_2'$ odd, while the other couplings are $\mathbb{Z}_2'$ even.

The loop diagrams induce the following coupling:
\begin{align}
\label{eq:loop-A4}
\lambda^{(1)}_{ 0 0 \dot{1}
}&\propto
\lambda^{(0)}_{ 0 \dot{1} \dot{1} }\lambda^{(0)}_{ 0 \dot{1} \dot{1} }\lambda^{(0)}_{ \dot{1} \dot{1} \dot{1} }
+
\lambda^{(0)}_{ 0 2 \bar{2} }\lambda^{(0)}_{ 0 2 \bar{2} }\lambda^{(0)}_{ \dot{1} 2 \bar{2} },
\end{align}
which violate the tree-level selection rules.
However, they are consistent with $\gr[\conj(A_4)]$, which forbids the other three-point couplings.
We can verify the correspondence of $\mathbb{Z}_2'$ charges between the left and right hand sides in Eq.~(\ref{eq:loop-A4}).
When both $\lambda^{(0)}_{\dot{1} \dot{1} \dot{1}}$ and $\lambda^{(0)}_{\dot{1} 2 \bar 2}$ are small, the induced coupling 
$\lambda^{(1)}_{ 0 0 \dot{1}}$ is also small.
In the limit $\lambda^{(0)}_{\dot{1} \dot{1} \dot{1}}, \lambda^{(0)}_{\dot{1} 2 \bar 2} \to 0$, the loop-induced coupling 
$\lambda^{(0)}_{\dot{1} 2 \bar 2}$ vanishes, and the tree-level selection rules are not violated. 
Indeed, in this limit, the approximate $\mathbb{Z}_2'$ becomes exact.
Table \ref{tab:3_point_coupling_A4} shows the 3-point couplings, which are allowed at tree level and 1-loop level.

Since the multiplication rules have the permutation symmetry $S_2$, let us impose this $S_2$ symmetry in a field theory by requiring $\lambda^{(0)}_{222}=\lambda^{(0)}_{\bar 2 \bar 2  \bar 2}$.
When we combine $S_2$ with $\gr[\conj(A_4)]\cong \mathbb{Z}_3$, the remaining symmetry can be enhanced to $S_3$.
When $\lambda^{(0)}_{222} \neq \lambda^{(0)}_{\bar 2 \bar 2  \bar 2}$, 
the symmetry is not enhanced from $\gr[\conj(A_4)]\cong \mathbb{Z}_3$.

\begin{longtable}[H]{|c|p{12cm}|c|}
\caption{Allowed $3$-point couplings obtained from $\conj(A_4)$.}
\label{tab:3_point_coupling_A4} \\
\hline
& \multicolumn{1}{c|}{3-point coupling}& types\\
\hline
\endhead
tree & 
$C_1C_1 C_1$,
$C_1C_3 C_3$,
$C_1B^1_4 B^2_4$,
$C_3C_3 C_3$,
$C_3B^1_4 B^2_4$,
$B^1_4B^1_4B^1_4$,
$B^2_4B^2_4B^2_4$,
&7\\
\hline
1-loop & 
$C_1C_1C_3$
&1(8)\\
\hline
\end{longtable}

\subsubsection{Example: $\Delta(27)$}

Let us consider a concrete example, $\Delta(27)\cong (\mathbb{Z}_3\times \mathbb{Z}'_3)\rtimes \mathbb{Z}_3$.
$\Delta(27)$ orbifolded heterotic models were studied, e.g., in Refs.~\cite{Fischer:2012qj,Fischer:2013qza}.
The conjugacy classes of $\Delta(27)$ are listed as follows:
\begin{align}
    &C_1 = \{e\},\notag\\
    &C^1_1 = \{aa'^2\},\notag\\
    &C^2_1 = \{a^2a'\},\notag\\
    &C^{(1, 0)}_3 = \{a, a', a^2a'^2\},\notag\\
    &C^{(2, 0)}_3 = \{a^2, a'^2, aa'\},\notag\\
    &B^{(1, 0)}_3 = \{b, baa'^2, ba^2a'\},\notag\\
    &B^{(1, 1)}_3 = \{ba, ba', ba^2a'^2\},\notag\\
    &B^{(1, 2)}_3 = \{ba^2, ba'^2, baa'\},\notag\\
    &B^{(2, 0)}_3 = \{b^2, b^2aa'^2, b^2a^2a'\},\notag\\
    &B^{(2, 1)}_3 = \{b^2a, b^2a', b^2a^2a'^2\},\notag\\
    &B^{(2, 2)}_3 = \{b^2a^2, b^2a'^2, b^2aa'\}.
\end{align}
They obey the multiplication rules shown in Table~\ref{tab:Delta27}. 
Note that the multiplication rules have the $((\mathbb{Z}_3\times \mathbb{Z}'_3) \rtimes Q_8)\rtimes S_3$ symmetry, 
where the action of $\mathrm{GL}(2,3)\cong Q_8\rtimes S_3$ is described by
\begin{align}
\label{eq:Delta27-Q8-S3}
    u_1 : \ &C^1_1 \to C^2_1,\quad C^2_1\to C^1_1,\quad C^{(1,0)}_3\to B^{(2,0)}_3,\quad C^{(2,0)}_3\to B^{(1,0)}_3,\nonumber\\
    &B^{(1,0)}_3\to C^{(2,0)}_3,\quad B^{(1,1)}_3\to B^{(2,2)}_3,\quad B^{(2,0)}_3\to C^{(1,0)}_3,\quad B^{(2,2)}_3\to B^{(1,1)}_3,
\nonumber\\
    u_2 : \ &C^{(1,0)}_3\to B^{(1,2)}_3,\quad C^{(2,0)}_3\to B^{(2,1)}_3,\quad B^{(1,0)}_3\to B^{(2,0)}_3,\quad B^{(1,1)}_3\to C^{(2,0)}_3,\nonumber\\
    &B^{(1,2)}_3\to B^{(1,1)}_3,\quad B^{(2,0)}_3\to B^{(1,0)}_3,\quad B^{(2,1)}_3\to B^{(2,2)}_3,\quad B^{(2,2)}_3\to C^{(1,0)}_3. 
\end{align}
and $\mathbb{Z}_3\times \mathbb{Z}'_3$ act as
\begin{align}
\label{eq:Delta27-Z3-Z3}
   \mathbb{Z}_3 &: B_3^{(n,p)} \to \omega^{n+p}B_3^{(n,p)},\nonumber\\
   \mathbb{Z}'_3 &: B_3^{(n,p)} \to \omega^n B_3^{(n,p)}.
\end{align}
with $n,p=1,2$. 
Here, $\mathrm{GL}(2,3)$ is the outer automorphism of $\Delta(27)$~\cite{Holthausen:2012dk}, and the $\mathbb{Z}'_3$ action corresponds to the transformation $b \to \omega b$.

\begin{table}[H]
    \centering
    \caption{Multiplication rules for conjugacy classes of $\Delta(27)$.}
\label{tab:Delta27}
\resizebox{\textwidth}{!}{
    \begin{tabular}{|c||c|c|c|c|c|}
      \hline
       & $C_1$ & $C^1_1$ & $C^2_1$ & $C^{(1, 0)}_3$ & $C^{(2, 0)}_3$ \\
       \hline
       \hline
       $C_1$& $C_1$ & $C^1_1$ & $C^2_1$ & $C^{(1, 0)}_3$ & $C^{(2, 0)}_3$ \\
       \hline
       $C^1_1$& $C^1_1$& $C^2_1$ & $C_1$ & $C^{(1, 0)}_3$ & $C^{(2, 0)}_3$ \\
       \hline
       $C^2_1$& $C^2_1$ &$C_1$  & $C^1_1$ & $C^{(1, 0)}_3$ & $C^{(2, 0)}_3$ \\
       \hline
       $C^{(1, 0)}_3$& $C^{(1, 0)}_3$ & $C^{(1, 0)}_3$ & $C^{(1, 0)}_3$ & $3 C^{(2, 0)}_3$ & $3 C_1 + 3 C^1_1 + 3 C^2_1$ \\
       \hline
       $C^{(2, 0)}_3$& $C^{(2, 0)}_3$ & $C^{(2, 0)}_3$ &$C^{(2, 0)}_3$  & $3 C_1 + 3 C^1_1 + 3 C^2_1$ & $3 C^{(1, 0)}_3$ \\
       \hline
       
       $B^{(1,0)}_3$& $B^{(1,0)}_3$ & $B^{(1,0)}_3$ &$B^{(1,0)}_3$  & $3 B^{(1,1)}_3$ & $3 B^{(1,2)}_3$ \\
       \hline  
       $B^{(1,1)}_3$& $B^{(1,1)}_3$ & $B^{(1,1)}_3$ &$B^{(1,1)}_3$  & $3 B^{(1,2)}_3$ & $3 B^{(1,0)}_3$ \\
       \hline  
       $B^{(1,2)}_3$& $B^{(1,2)}_3$ & $B^{(1,2)}_3$ &$B^{(1,2)}_3$  & $3 B^{(1,0)}_3$ & $3 B^{(1,1)}_3$ \\
       \hline  
       $B^{(2,0)}_3$& $B^{(2,0)}_3$ & $B^{(2,0)}_3$ &$B^{(2,0)}_3$  & $3 B^{(2,1)}_3$ & $3 B^{(2,2)}_3$ \\
       \hline  
       $B^{(2,1)}_3$& $B^{(2,1)}_3$ & $B^{(2,1)}_3$ &$B^{(2,1)}_3$  & $3 B^{(2,2)}_3$ & $3 B^{(2,0)}_3$ \\
       \hline  
       $B^{(2,2)}_3$& $B^{(2,2)}_3$ & $B^{(2,2)}_3$ &$B^{(2,2)}_3$  & $3 B^{(2,0)}_3$ & $3 B^{(2,1)}_3$ \\
       \hline  
       \end{tabular}}
       \resizebox{\textwidth}{!}{
    \begin{tabular}{|c||c|c|c|c|c|c|}
      \hline
      & $B^{(1,0)}_3$ & $B^{(1,1)}_3$ & $B^{(1,2)}_3$& $B^{(2,0)}_3$ & $B^{(2,1)}_3$ & $B^{(2,2)}_3$\\
      \hline
      \hline
      $C_1$&$B^{(1,0)}_3$ & $B^{(1,1)}_3$ & $B^{(1,2)}_3$& $B^{(2,0)}_3$ & $B^{(2,1)}_3$ & $B^{(2,2)}_3$\\
      \hline
      $C^1_1$&$B^{(1,0)}_3$&$B^{(1,1)}_3$&$B^{(1,2)}_3$&$B^{(2,0)}_3$&$B^{(2,1)}_3$&$B^{(2,2)}_3$\\
      \hline
      $C^2_1$&$B^{(1,0)}_3$&$B^{(1,1)}_3$&$B^{(1,2)}_3$&$B^{(2,0)}_3$&$B^{(2,1)}_3$&$B^{(2,2)}_3$\\
      \hline
      $C^{(1,0)}_3$&$3B^{(1,1)}_3$&$3B^{(1,2)}_3$&$3B^{(1,0)}_3$&$3B^{(2,1)}_3$&$3B^{(2,2)}_3$&$3B^{(2,0)}_3$\\
      \hline
      $C^{(2,0)}_3$&$3B^{(1,2)}_3$&$3B^{(1,0)}_3$&$3B^{(1,1)}_3$&$3B^{(2,2)}_3$&$3B^{(2,0)}_3$&$3B^{(2,1)}_3$\\
      \hline
      $B^{(1,0)}_3$ & $3 B^{(2,0)}_3$ & $3 B^{(2,1)}_3$ & $3 B^{(2,2)}_3$&$3 C_1 + 3 C^1_1 + 3 C^2_1$&$3C^{(1,0)}_3$&$3C^{(2,0)}_3$\\
      \hline
      $B^{(1,1)}_3$ & $3 B^{(2,1)}_3$ & $3 B^{(2,2)}_3$ & $3 B^{(2,0)}_3$&$3C^{(1,0)}_3$&$3C^{(2,0)}_3$&$3 C_1 + 3 C^1_1 + 3 C^2_1$\\
      \hline
      $B^{(1,2)}_3$ & $3 B^{(2,2)}_3$ & $3 B^{(2,0)}_3$ & $3 B^{(2,1)}_3$&$3C^{(2,0)}_3$&$3 C_1 + 3 C^1_1 + 3 C^2_1$&$3C^{(1,0)}_3$\\
      \hline
      $B^{(2,0)}_3$ & $3 C_1 + 3 C^1_1 + 3 C^2_1$ & $3 C^{(1, 0)}_3$ & $3 C^{(2, 0)}_3$&$3 B^{(1,0)}_3$ & $3 B^{(1,1)}_3$ & $3 B^{(1,2)}_3$\\
      \hline
      $B^{(2,1)}_3$ & $3 C^{(1, 0)}_3$ & $3 C^{(2, 0)}_3$ & $3 C_1 + 3 C^1_1 + 3 C^2_1$&$3 B^{(1,1)}_3$ & $3 B^{(1,2)}_3$ & $3 B^{(1,0)}_3$\\
      \hline
      $B^{(2,2)}_3$ & $3 C^{(2, 0)}_3$ & $3 C_1 + 3 C^1_1 + 3 C^2_1$ & $3 C^{(1, 0)}_3$&$3 B^{(1,2)}_3$ & $3 B^{(1,0)}_3$ & $3 B^{(1,1)}_3$\\
      \hline
    \end{tabular}}
  \end{table}
From these results, we can obtain the subset $\com(\conj(\Delta(27)))$ as
\begin{align}
  \com(\conj(\Delta(27))) = \{ C_1, C^1_1, C^2_1 \},
\end{align}
and we can also compute the groupification of $\conj(\Delta(27))$ as follows:
\begin{align}
  &\gr(\conj(\Delta(27))) \notag\\
  &= \{ [C_1], [C^{(1,0)}_3],[C^{(2,0)}_3],[B^{(1,0)}_3],[B^{(1,1)}_3],[B^{(1,2)}_3],[B^{(2,0)}_3],[B^{(2,1)}_3],[B^{(2,2)}_3]\}\cong \mathbb{Z}_3\times \mathbb{Z}_3',
\end{align}
where we define
\begin{align}
  &[C_1] = \{C_1,C^1_1,C^2_1\},\quad [C^{(n,0)}_3]=\{C^{(n,0)}_3\},\quad [B^{(m,n)}_3] = \{ B^{(m,n)}_3 \}.
\end{align}
Note that their $\mathbb{Z}_3\times \mathbb{Z}_3'$ charges are given by
\begin{align}
    &[C_1]: (0,0),\quad
    [C^{(n,0)}_3] : (n,0),\quad [B^{(m,n)}_3] : (n,m),
\end{align}
where $n=1,2$ and $m=0,1,2$.

We now turn to check the loop effects more explicitly. We introduce the following new notation:
\begin{gather}
    0 = C_1,\quad 
    1 = C^1_1,\quad
    \bar{1} = C^2_1,\quad 
    2 = C^{(1,0)}_3,\quad 
    \bar{2} = C^{(2,0)}_3,\notag\\
    3 = B^{(1,0)}_3,\quad 
    4 = B^{(1,1)}_3,\quad 
    5 = B^{(1,2)}_3,\quad\notag\\
    \bar{3} = B^{(2,0)}_3,\quad
    \bar{5}= B^{(2,1)}_3,\quad
    \bar{4} = B^{(2,2)}_3.
\end{gather}
By using this notation, we denote the corresponding fields by $\phi_0$, $\phi_{1}, \cdots$, and their couplings by, e.g., $\lambda_{001}$.
The selection rules imposed by $\conj(\Delta(27))$ allow the tree-level 3-point couplings shown in Table \ref{tab:3_point_coupling_Delta27}, which correspond to the following couplings:
\begin{align}
    &\lambda_{000}, \notag \\
    &\lambda_{kkk},\lambda_{\bar{k}\bar{k}\bar{k}},\quad{\rm for\ all}\ k, \notag \\
    &\lambda_{0k\bar{k}},\quad k\geq 1, \notag \\
    &\lambda_{1k\bar{k}},
    \lambda_{\bar{1}k\bar{k}},\quad
    k\geq 2 \notag \\
    &\lambda_{klm},\quad k,l,m\neq 0,1,\bar{1}, \quad {\rm s.t.}\ \mathbb{Z}_3\times\mathbb{Z}_3{\rm\ invariant.}
\end{align}
and the others are forbidden at tree level. We can define an approximate $\mathbb{Z}_3'$ symmetry such that 
$\phi_1$ and $\phi_{\bar{1}}$, corresponding to $C^1_1$ and $C^2_1$, carry $\mathbb{Z}_3'$ charges, 1 and 2, respectively, while the others have charge 0.
However, the following couplings:
\begin{align}
    \lambda_{1k\bar{k}},\quad
    \lambda_{\bar{1}k\bar{k}},\quad k\geq 2
\end{align}
violate the approximate $\mathbb{Z}_3'$ symmetry.
Note that $C^1_1$ and $C^2_1$ belong to $\com(\conj(\Delta(27)))$.
In the spurion language, the former coupling carries $\mathbb{Z}_3'$ charge 1, while the latter carries $\mathbb{Z}_3'$ charge 2.
The loop diagrams induce the following couplings:
\begin{align}
\label{eq:loop-Delta27}
\lambda^{(1)}_{ 0 0 1
}&\propto
\lambda^{(0)}_{ 0 2 \bar{2} }\lambda^{(0)}_{ 0 2 \bar{2} }\lambda^{(0)}_{ 1 2 \bar{2} }
+
\lambda^{(0)}_{ 0 3 \bar{3} }\lambda^{(0)}_{ 0 3 \bar{3} }\lambda^{(0)}_{ 1 3 \bar{3} }
+
\lambda^{(0)}_{ 0 4 \bar{4} }\lambda^{(0)}_{ 0 4 \bar{4} }\lambda^{(0)}_{ 1 4 \bar{4} }
+
\lambda^{(0)}_{ 0 5 \bar{5} }\lambda^{(0)}_{ 0 5 \bar{5} }\lambda^{(0)}_{ 1 5 \bar{5} },\notag\\
\lambda^{(1)}_{ 0 0 \bar{1}
}&\propto
\lambda^{(0)}_{ 0 2 \bar{2} }\lambda^{(0)}_{ 0 2 \bar{2} }\lambda^{(0)}_{ \bar{1} 2 \bar{2} }
+
\lambda^{(0)}_{ 0 3 \bar{3} }\lambda^{(0)}_{ 0 3 \bar{3} }\lambda^{(0)}_{ \bar{1} 3 \bar{3} }
+
\lambda^{(0)}_{ 0 4 \bar{4} }\lambda^{(0)}_{ 0 4 \bar{4} }\lambda^{(0)}_{ \bar{1} 4 \bar{4} }
+
\lambda^{(0)}_{ 0 5 \bar{5} }\lambda^{(0)}_{ 0 5 \bar{5} }\lambda^{(0)}_{ \bar{1} 5 \bar{5} },\notag\\
\lambda^{(1)}_{ 0 1 1
}&\propto
\lambda^{(0)}_{ 0 2 \bar{2} }\lambda^{(0)}_{ 1 2 \bar{2} }\lambda^{(0)}_{ 1 2 \bar{2} }
+
\lambda^{(0)}_{ 0 3 \bar{3} }\lambda^{(0)}_{ 1 3 \bar{3} }\lambda^{(0)}_{ 1 3 \bar{3} }
+
\lambda^{(0)}_{ 0 4 \bar{4} }\lambda^{(0)}_{ 1 4 \bar{4} }\lambda^{(0)}_{ 1 4 \bar{4} }
+
\lambda^{(0)}_{ 0 5 \bar{5} }\lambda^{(0)}_{ 1 5 \bar{5} }\lambda^{(0)}_{ 1 5 \bar{5} },\notag\\
\lambda^{(1)}_{ 0 \bar{1} \bar{1}
}&\propto
\lambda^{(0)}_{ 0 2 \bar{2} }\lambda^{(0)}_{ \bar{1} 2 \bar{2} }\lambda^{(0)}_{ \bar{1} 2 \bar{2} }
+
\lambda^{(0)}_{ 0 3 \bar{3} }\lambda^{(0)}_{ \bar{1} 3 \bar{3} }\lambda^{(0)}_{ \bar{1} 3 \bar{3} }
+
\lambda^{(0)}_{ 0 4 \bar{4} }\lambda^{(0)}_{ \bar{1} 4 \bar{4} }\lambda^{(0)}_{ \bar{1} 4 \bar{4} }
+
\lambda^{(0)}_{ 0 5 \bar{5} }\lambda^{(0)}_{ \bar{1} 5 \bar{5} }\lambda^{(0)}_{ \bar{1} 5 \bar{5} },\notag\\
\lambda^{(1)}_{ 1 1 \bar{1}
}&\propto
\lambda^{(0)}_{ 1 2 \bar{2} }\lambda^{(0)}_{ 1 2 \bar{2} }\lambda^{(0)}_{ \bar{1} 2 \bar{2} }
+
\lambda^{(0)}_{ 1 3 \bar{3} }\lambda^{(0)}_{ 1 3 \bar{3} }\lambda^{(0)}_{ \bar{1} 3 \bar{3} }
+
\lambda^{(0)}_{ 1 4 \bar{4} }\lambda^{(0)}_{ 1 4 \bar{4} }\lambda^{(0)}_{ \bar{1} 4 \bar{4} }
+
\lambda^{(0)}_{ 1 5 \bar{5} }\lambda^{(0)}_{ 1 5 \bar{5} }\lambda^{(0)}_{ \bar{1} 5 \bar{5} },\notag\\
\lambda^{(1)}_{ 1 \bar{1} \bar{1}
}&\propto
\lambda^{(0)}_{ 1 2 \bar{2} }\lambda^{(0)}_{ \bar{1} 2 \bar{2} }\lambda^{(0)}_{ \bar{1} 2 \bar{2} }
+
\lambda^{(0)}_{ 1 3 \bar{3} }\lambda^{(0)}_{ \bar{1} 3 \bar{3} }\lambda^{(0)}_{ \bar{1} 3 \bar{3} }
+
\lambda^{(0)}_{ 1 4 \bar{4} }\lambda^{(0)}_{ \bar{1} 4 \bar{4} }\lambda^{(0)}_{ \bar{1} 4 \bar{4} }
+
\lambda^{(0)}_{ 1 5 \bar{5} }\lambda^{(0)}_{ \bar{1} 5 \bar{5} }\lambda^{(0)}_{ \bar{1} 5 \bar{5} }.
\end{align}
These couplings are shown in Table \ref{tab:3_point_coupling_Delta27} and violate the tree-level selection rules. 
Nevertheless, all the couplings shown in Table \ref{tab:3_point_coupling_Delta27} are consistent with $\gr(\conj(\Delta(27)))=\mathbb{Z}_3\times\mathbb{Z}_3$, which forbids the other three-point couplings. 
We can verify the correspondence of $\mathbb{Z}'_3$ charges between left and right sides in Eq.~(\ref{eq:loop-Delta27}). 
In particular, when all spurionic couplings carrying $\mathbb{Z}'_3 $ charges 1 and 2 vanish, the tree-level selection rules are not violated.
Indeed, in this limit, the approximate $\mathbb{Z}'_3$ becomes exact.


\begin{longtable}[H]{|c|p{12cm}|c|}
\caption{Allowed $3$-point couplings obtained from $\conj(\Delta(27))$.}
\label{tab:3_point_coupling_Delta27} \\
\hline
& \multicolumn{1}{c|}{3-point coupling}& types\\
\hline
\endhead
tree & 
$C_1C_1C_1$,\quad\allowbreak
$C_1C^1_1C^2_1$,\quad\allowbreak
$C_1C^{(1,0)}_3C^{(2,0)}_3$,\quad\allowbreak
$C_1B^{(1,0)}_3B^{(2,0)}_3$,\quad\allowbreak
$C_1B^{(1,1)}_3B^{(2,2)}_3$,\quad\allowbreak
$C_1B^{(1,2)}_3B^{(2,1)}_3$,\quad\allowbreak
$C^1_1C^1_1C^1_1$,\quad\allowbreak
$C^1_1C^{(1,0)}_3C^{(2,0)}_3$,\quad\allowbreak
$C^1_1B^{(1,0)}_3B^{(2,0)}_3$,\quad\allowbreak
$C^1_1B^{(1,1)}_3B^{(2,2)}_3$,\quad\allowbreak
$C^1_1B^{(1,2)}_3B^{(2,1)}_3$,\quad\allowbreak
$C^2_1C^2_1C^2_1$,\quad\allowbreak
$C^2_1C^{(1,0)}_3C^{(2,0)}_3$,\quad\allowbreak
$C^2_1B^{(1,0)}_3B^{(2,0)}_3$,\quad\allowbreak
$C^2_1B^{(1,1)}_3B^{(2,2)}_3$,\quad\allowbreak
$C^2_1B^{(1,2)}_3B^{(2,1)}_3$,\quad\allowbreak
$C^{(1,0)}_3C^{(1,0)}_3C^{(1,0)}_3$,\quad\allowbreak
$C^{(1,0)}_3B^{(1,0)}_3B^{(2,2)}_3$.\quad\allowbreak
$C^{(1,0)}_3B^{(1,1)}_3B^{(2,1)}_3$,\quad\allowbreak
$C^{(1,0)}_3B^{(1,2)}_3B^{(2,0)}_3$,\quad\allowbreak
$C^{(2,0)}_3C^{(2,0)}_3C^{(2,0)}_3$,\quad\allowbreak
$C^{(2,0)}_3B^{(1,0)}_3B^{(2,1)}_3$,\quad\allowbreak
$C^{(2,0)}_3B^{(1,1)}_3B^{(2,0)}_3$,\quad\allowbreak
$C^{(2,0)}_3B^{(1,2)}_3B^{(2,2)}_3$,\quad\allowbreak
$B^{(1,0)}_3B^{(1,0)}_3B^{(1,0)}_3$,\quad\allowbreak
$B^{(1,0)}_3B^{(1,1)}_3B^{(1,2)}_3$,\quad\allowbreak
$B^{(1,1)}_3B^{(1,1)}_3B^{(1,1)}_3$,\quad\allowbreak
$B^{(1,2)}_3B^{(1,2)}_3B^{(1,2)}_3$,\quad\allowbreak
$B^{(2,0)}_3B^{(2,0)}_3B^{(2,0)}_3$,\quad\allowbreak
$B^{(2,0)}_3B^{(2,1)}_3B^{(2,2)}_3$,\quad\allowbreak
$B^{(2,1)}_3B^{(2,1)}_3B^{(2,1)}_3$,\quad\allowbreak
$B^{(2,2)}_3B^{(2,2)}_3B^{(2,2)}_3$.\quad\allowbreak
&32\\
\hline
1-loop & 
$C_1C_1C^1_1$,
$C_1C_1C^2_1$,
$C_1C^1_1C^1_1$,
$C_1C^2_1C^2_1$,
$C^1_1C^1_1C^2_1$,
$C^1_1C^2_1C^2_1$.
&6(38)\\
\hline
\end{longtable}

Note that the conjugacy classes of $\Delta(27)$ have quite large automorphism as shown in Eq.~(\ref{eq:Delta27-Q8-S3}) as well as Eq.~(\ref{eq:Delta27-Z3-Z3}).
We can impose this automorphism $Q_3 \rtimes S_3$ or its subsymmetries in the theory by setting proper relations among couplings.
When we combine it with $\gr[(\Delta(27))]$, the remaining symmetry can be enhanced to $((\mathbb{Z}_3 \times \mathbb{Z}_3')\rtimes Q_8) \rtimes S_3$ or its subgroups.

\subsection{Generic properties for residual symmetries}

The results of our analyses in the previous sections demonstrate the generic properties of groupification. 
So far, we have considered the discrete groups $G =H \rtimes \mathbb{Z}_N$ and examined the structure of $\conj(G)$. 
Here, we denote elements of $H$ by $h_i$ and 
the generator of $\mathbb{Z}_N$ by $b$, which satisfies $b^N=e$. 
The group structure defined by the semi-direct product means
\begin{align}
\label{eq:bhb=h}
    bh_ib^{-1}=h_j,
\end{align}
which leads to 
\begin{align}
\label{eq:b=bh}
    h_jbh_j^{-1}=bh_ih_j^{-1}.
\end{align}
That is, we find that 
\begin{align}
    gb^kg^{-1}=b^kh_\ell,
\end{align}
where $g \in G$.
Thus, the conjugacy classes of $G =H \rtimes \mathbb{Z}_N$ can be represented as
\begin{align}
    B^k=\{ b^k h_1, b^k h_2, \cdots \},
\end{align}
where $h_\ell$ denotes some elements in $H$. 
Each conjugacy class has a definite $\mathbb{Z}_N$ charge $k$, although two or more classes may have the same charge $k$ like $B'^k=\{  b^k h'_1, b^k h'_2, \cdots\}$. 
Consequently, the algebra $\conj(G)$ exhibits $\mathbb{Z}_N$ symmetry, and obviously $\gr[\conj(G)]$ includes $\mathbb{Z}_N$ symmetry.

Now, let us consider the conjugacy class $C_{ij}$ including $h_ih^{-1}_j$, where $h_i$ and $h_j$ satisfy Eq.~(\ref{eq:bhb=h}).
Eq.~(\ref{eq:b=bh}) leads to the following multiplication rule:
\begin{align}
    B^1C_{ij}=B^1,
\end{align}
where $B^1$ denotes the conjugacy class including $b$.
It follows that
\begin{align}
 \bar B^1 B^1= C_{ij} + \dots,
\end{align}
which implies $C_{ij} \in \com(\conj(G))$. 
If all the elements in $H$ can be expressed as $h_ih_j^{-1}$ satisfying Eq.~(\ref{eq:bhb=h}), all the conjugacy classes of $H$ correspond to $\com(\conj(G))$.

However, all the elements in $H$ cannot be written by $h_ih_j^{-1}$ in certain groups. 
A simple example is $D_N$ with $N=$ even, where $H$ corresponds to $\mathbb{Z}_N$ with generator $a$, i.e. $a^N=e$.
By using the algebraic relation $bab^{-1}=a^{-1}$ of $D_N$, we find that $h_ih^{-1}_j=a^{2k}$ in the notation introduced above. 
For $N=$ even, the elements $a^{2k+1}$ cannot be written by $h_ih^{-1}_j$ satisfying Eq.~(\ref{eq:bhb=h}). 
Here, we denote the conjugacy class including $a^k$ and $a^{-k}$ by $C^k$. 
Then, $\com(\conj(D_N))$ includes 
only the conjugacy classes $C^{2k}$, but not $C^{2k+1}$. 
Thus, a $\mathbb{Z}_2$ symmetry remains in $\gr[\conj(D_N)]$. 
Consequently, we find that $\gr[\conj(D_N)] = \mathbb{Z}_2 \times \mathbb{Z}_2$. 
Note that we also find $h_ih^{-1}_j=a^{2k}$ from $bab^{-1}=a^{-1}$ for $N=$ odd, too, but all elements of $\mathbb{Z}_N$ with $N=$ odd can be written by $a^{2k}$ with proper numbers $k$.
Then, all conjugacy classes including $a^n$ belong to $\com(\conj(D_N))$, yielding $\gr[\conj(D_N)] = \mathbb{Z}_2$.

Similarly, we can discuss $\Delta(3N^2)$ and other groups.
Consider $\Delta(3N^2)= (\mathbb{Z}_N\times \mathbb{Z}_N') \rtimes \mathbb{Z}_3$. 
When $N/3 \neq$ integer,  all elements of $\mathbb{Z}_N\times \mathbb{Z}_N'$ can be written by $h_ih^{-1}_j$, which satisfy Eq.~(\ref{eq:bhb=h}). 
Thus, the conjugacy classes including all elements of $\mathbb{Z}_N \times \mathbb{Z}_N'$ belong to $\com(\conj(\Delta(3N^2)))$.
However, when $N/3  =$ integer, 
all elements of $\mathbb{Z}_N\times \mathbb{Z}_N'$ cannot be written in the form $h_ih^{-1}_j$ satisfying Eq.~(\ref{eq:bhb=h}).
For example, by using the algebraic relations $bab^{-1}=a^{-1}a'^{-1}$ and $ba'b^{-1}=a$ for $N=3$, we find that $h_ih^{-1}_j$ includes only $aa'^2$ in $C_1^1$ and $a^2a'$ and $C_1^2$, but no other elements. 
Then, $\com(\conj(\Delta(27)))$ includes $C_1^1$ and $C_1^2$, but not $C_3^{(1,0)}$ or $C_3^{(2,0)}$. 
The $\mathbb{Z}_3$ symmetry remains without be contained in $\com(\conj(\Delta(3N^2)))$.
Overall, we find $\gr[\conj(\Delta(3N^2))] =\mathbb{Z}_3 \times \mathbb{Z}_3$. 
Similarly, we can investigate $\gr[\conj(\Delta(3N^2))]$ for general $N$.

We provide an overview of the residual group-like symmetries $\gr[A]$ associated with the conjugacy classes of the finite discrete groups discussed above. 
As summarized in Table~\ref{tab:Gr_summary}, the loop-induced group-like symmetries $\gr[A]$ are described by Abelian symmetries such as $\mathbb{Z}_N$ and $\mathbb{Z}_N\times \mathbb{Z}_M$.
See Appendix \ref{app:Deta-6N} for $\Delta(6N^2)$.
\begin{table}[H]
    \caption{Group-like symmetries $\mathrm{Gr}$ for conjugacy classes of finite groups.}
    \centering
    \begin{tabular}{|c||c|}\hline 
       $G$  & $\mathrm{Gr}[\mathrm{Conj}(G)]$  \\\hline\hline
       $D_{N=\mathrm{even}}$  & $\mathbb{Z}_2\times \mathbb{Z}_2$ \\ \hline 
       $D_{N=\mathrm{odd}}$  & $\mathbb{Z}_2$ \\\hline
       $T_{N=\mathrm{prime}}$  & $\mathbb{Z}_3$ \\\hline
       $\Delta(3N^2)|_{N\in 3\mathbb{Z}_{>0}}$  & $\mathbb{Z}_3\times \mathbb{Z}_3$ \\\hline
       $\Delta(3N^2)|_{N\notin 3\mathbb{Z}_{>0}}$  & $\mathbb{Z}_3$ \\\hline
       $S_4, ~\Delta(54)$  & $\mathbb{Z}_2$ \\\hline 
    \end{tabular}
    \label{tab:Gr_summary}
\end{table}

Furthermore, we can define an approximate symmetries, which are relevant to conjugacy classes belonging to $\com(\conj(G))$, 
e.g., $\mathbb{Z}_2'$ for $\conj(S_3)$ and $\conj(D_4)$, and $\mathbb{Z}_3'$ for $\conj(T_7)$.
The more conjugacy classes belong to $\com(\conj(G))$, the larger approximate symmetries become.
See Appendix \ref{app:Deta-6N} for $\Delta(6N^2)$.
Certain tree-level couplings violate these approximate symmetries and they induce further violating terms by loop effects.
In the spurion language, violating coupling constants carry charges under approximate symmetries.
We can find the correspondence of charges between tree-level and 1-loop level couplings.
When all the tree-level violating terms vanish, 
violating terms cannot be induced at loop levels.
The approximate symmetries become exact in this limit.
Thus, approximate symmetries control loop-induced couplings.

In general, the following couplings:
\begin{align}
\label{eq:Y-00k}
    \lambda_{00k},
\end{align}
with $k\neq 0$ are forbidden at tree level by the property of the unit.
However, if $\gr[\conj(G)]$ allows, such couplings can be induced by loop effects.
They may be suppressed by loop factors and depend on other tree-level couplings.
Similarly, the following couplings:
\begin{align}
\label{eq:Y-0kl}
    \lambda_{0 k\ell},
\end{align}
with $k \neq \bar \ell$ are forbidden at tree level.
However, if $\gr[\conj(G)]$ allows, such couplings can be induced by loop effects.
There are other couplings, which are forbidden at tree level, but induced by loop effects.
(See for such examples Appendix \ref{app:Deta-6N}.)
They may be suppressed by loop factors and depend on other tree-level couplings.
These small couplings would be phenomenologically useful to explain small values and some hierarchies, e.g. Yukawa hierarchies.

\subsection{Implications in string theory}

So far, we have examined the implications of $\conj(G)$ in field theory.
Similarly, we can study its implications in string theory.
In heterotic orbifold models, string states correspond to conjugacy classes \cite{Dixon:1986jc,Dixon:1986qv,Hamidi:1986vh}.
Coupling selection rules follow the multiplication rules of conjugacy classes.
In Ref.~\cite{Kobayashi:2025ocp},  these selection rules at tree level were studied in heterotic non-Abelian orbifold models.
One of the simple orbifold models is $S_3$ orbifolded heterotic models \cite{Inoue:1987ak,Inoue:1988ki,Inoue:1990ci,Konopka:2012gy,Fischer:2012qj,Fischer:2013qza,Funakoshi:2025lxs,Hernandez-Segura:2025sfr}, because $S_3$ is the smallest non-Abelian discrete group.
Here, we denote the twists of string boundary conditions by $\theta_a$ and $\theta_b$, which correspond to the generators $a$ and $b$ of $S_3$, i.e., $(\theta_a)^3=(\theta_b)^2=1$.
Since there are three conjugacy classes, there are three string sectors, the untwisted sector, 
the $\theta_a$ twisted sector including $\theta_a$ and $(\theta_a)^2$ boundary conditions, and 
the $\theta_b$ twisted sector including $\theta_b (\theta_a)^k$ boundary conditions with $k=0,1,2$.
At tree level, their coupling section rules are controlled by $\conj(S_3)$ as studied in Ref.~\cite{Kobayashi:2025ocp}.
However, the analysis in the previous section shows that tree-level selection rules are violated by loop effects, and only $\gr[\conj(S_3)]\cong \mathbb{Z}_2$ remains, where 
the untwisted and $\theta_a$ twisted sectors are $\mathbb{Z}_2$ even sectors, while 
the $\theta_b$ twisted sector is a $\mathbb{Z}_2$ odd sector.
This $\mathbb{Z}_2$ symmetry can play a role as a flavor symmetry in low energy effective field theory.\footnote{In heterotic Abelian orbifold models, interesting flavor symmetries were derived \cite{Dijkgraaf:1987vp,Kobayashi:2004ya,Kobayashi:2006wq,Beye:2014nxa}. Similar flavor symmetries were also derived in intersecting/magnetized D-brane models \cite{Abe:2009vi,Berasaluce-Gonzalez:2012abm,Marchesano:2013ega}.}
Allowed couplings should include an even number of $\theta_b$ twisted sector, while the numbers of untwisted sector and $\theta_a$ twisted sector are not limited at the loop level.
In addition, there is an approximate $\mathbb{Z}_2'$ symmetry, where $\theta_a$ sector is $\mathbb{Z}_2'$ odd, while the others are $\mathbb{Z}_2'$ even.
The 3-point couplings among three $\theta_a$ twisted sectors, and the couplings among a single $\theta_a$ and two $\theta_b$ twisted sectors violate this approximate $\mathbb{Z}_2'$ symmetry.
If these couplings are small, loop-induced $\mathbb{Z}_2'$ violating terms are also small.

Similarly, we can analyze $D_4$ orbifolded heterotic models.
The difference is as follows. 
$D_4$ orbifolded heterotic string theory has five string sectors, the untwisted sector, 
the $\theta_a$ and $(\theta_a)^2$ twisted sectors, and the $\theta_b$ and $\theta_a \theta_b$ twisted sectors, where the notation for the twists is the same as in the $S_3$ orbifolded heterotic models. 
Their coupling selection rules are controlled by $\conj(D_4)$.
However, loop effects violate them and only $\gr[\conj(D_4)]\cong \mathbb{Z}_2 \times \mathbb{Z}_2$ remains, where $\theta_a$, $\theta_b$, $\theta_a \theta_b$ twisted sectors correspond to 
(even,odd), (odd,even), and (odd, odd) charges under $ \mathbb{Z}_2 \times \mathbb{Z}_2$, respectively, while the others are correspond to (even,even) charges.
There is also an approximate $\mathbb{Z}_2'$ symmetry, where the $(\theta_a)^2$ twisted sector is $\mathbb{Z}_2'$ odd, while the others are $\mathbb{Z}_2'$ even.

Furthermore, we can examine the coupling selection rules in $T_7$ orbifolded heterotic models, which have five string sectors: the untwisted sector, the $\theta_a$ and $(\theta_a)^3$ twisted sectors, and the $\theta_b$ and $(\theta_b)^2$ twisted sectors. 
The $\theta_a$ and $(\theta_a)^3$ twisted sectors have $(\theta_a)^k$ boundary conditions with $k=1,2,4$ and $k=3,5,6$, respectively, while the $\theta_b$ and $(\theta_b)^2$ twisted sectors include $\theta_b(\theta_a)^k$ and $(\theta_b)^2(\theta_a)^k$ boundary conditions with $k=0,1,\cdots,6$. 
Their tree-level coupling selection rules are controlled by $\conj(T_7)$. 
However, loop effects violate them, and only $\gr[\conj(T_7)]\cong \mathbb{Z}_3$ remains, where 
the $\theta_b$ and $(\theta_b)^3$ twisted sectors carry $\mathbb{Z}_3$ charges 1 and 2, respectively, while the others are $\mathbb{Z}_3$ invariant.
There is also an approximate $\mathbb{Z}_3'$ symmetry, where the $\theta_a$ and $(\theta_a)^3$ twisted sectors carry $\mathbb{Z}_3$ charges  1 and 2, respectively, while the others are $\mathbb{Z}_3$ invariant.

Similarly, we can investigate the coupling selection rules in other heterotic non-Abelian orbifold models, e.g., $A_4$ and $\Delta(27)$ orbifolded models as well as $S_4$ and $\Delta(54)$ orbifolded models.
The results are summarized in Table \ref{tab:twisted}, where U denotes the untwisted sector.
Here, $\theta_c$ denotes the string twist corresponding to the element $c$. 
$S_4$ and $\Delta(54)$ orbifolded models have the $\mathbb{Z}_2$ symmetry, where 
the $\theta_c$ twisted sectors are $\mathbb{Z}_2$ odd, while the others are $\mathbb{Z}_2$ even.
There are two or more sectors that are invariant under the full $\mathbb{Z}_N$ and $\mathbb{Z}_N\times \mathbb{Z}_N$ symmetries.
In order to distinguish these sectors, we can define approximate $\mathbb{Z}_M'$ symmetries.
However, these approximate symmetries are violated by certain types of tree-level couplings, which in turn induce further violating terms via loop effects.

\begin{table}[ht]
    \centering
    \caption{Symmetries in heterotic non-Abelian orbifold models. Note that $D_4$ orbifolded models and $\Delta(27)$ orbifolded models have $\mathbb{Z}_2 \times \mathbb{Z}_2$ and $\mathbb{Z}_3 \times \mathbb{Z}_3$ symmetries. $k$ denotes $\mathbb{Z}_N$ charges.}
    \label{tab:twisted}
    \begin{tabular}{|c|c|c|c|c|}
    \hline
    Orbifold     &  $\mathbb{Z}_N$ & $k=0$  sector & $k=1$ sector & $k=2$ sector \\ \hline
      $S_3$   & $\mathbb{Z}_2$  & U, $\theta_a$ & $\theta_b$  &  ---- \\ \hline
            $D_4$   &  first $\mathbb{Z}_2$ & U, $(\theta_a)^\ell$ & $\theta_b,~\theta_a\theta_b$  &  ---- \\ 
         &  second $\mathbb{Z}_2$ & U, $(\theta_a)^2,\theta_b$ & $\theta_a,~\theta_a\theta_b$  &  ---- \\ \hline
         $T_7$ &  $\mathbb{Z}_3$ & U, $\theta_a,~(\theta_a)^3$ & $\theta_b$ & $(\theta_b)^2$ \\ \hline
         $A_4$ &  $\mathbb{Z}_3$ & U, $\theta_a$ & $\theta_b$ & $(\theta_b)^2$     \\ \hline
         $\Delta(27)$ & first $\mathbb{Z}_3$  & U, $(\theta_a)\ell(\theta_{a'})^m$ & 
         $\theta_b(\theta_a)^\ell$ & $(\theta_b)^2(\theta_a)^\ell$ \\ 
          & second $\mathbb{Z}_3$  & U, $\theta_a(\theta_{a'}^2),~(\theta_a)^2\theta_{a'}^2, (\theta_b)^\ell$ & $\theta_a,~\theta_a(\theta_b)^\ell$ & $(\theta_a)^2,~(\theta_a)^2(\theta_b)^\ell$ \\ \hline
          $S_4$ & $\mathbb{Z}_2$ & sectors without $\theta_c$ twist & $\theta_c$ twisted sectors & --- \\ \hline 
             $\Delta(54)$ & $\mathbb{Z}_2$ & sectors without $\theta_c$ twist & $\theta_c$ twisted sectors & --- \\ \hline        
    \end{tabular}
\end{table}

The couplings between two untwisted sectors and one twisted sector correspond to the couplings (\ref{eq:Y-00k}) and are forbidden at tree level.
However, if $\gr[\conj(G)]$ allows, such couplings can be induced by loop effects.
Similarly, the couplings (\ref{eq:Y-0kl}) correspond to the couplings including a untwisted sector and two twisted sectors, which are not conjugate to each other, and are not forbidden at tree level. 
However, if $\gr[\conj(G)]$ allows, such couplings can be induced by loop effects.
There are other couplings that are forbidden at tree level, but induced by loop effects.
(See for such examples Appendix \ref{app:Deta-6N}.)
These couplings are suppressed by loop factors and depend on tree-level couplings, which may be functions of moduli. 
Consequently, these couplings can be small compared to other couplings. 
For example, the 3-point couplings including only untwisted sectors originate from the ten-dimensional gauge coupling, and are of ${\cal O}(1)$, whereas the above loop-induced couplings can be smaller. 
Such hierarchical couplings would be phenomenologically useful, e.g. for explaining Yukawa hierarchies. 
Computations of couplings in heterotic orbifold models have been carried out mainly for Abelian orbifolds \cite{Hamidi:1986vh,Dixon:1986qv,Burwick:1990tu,Choi:2007nb}.
It is important to compute tree-level and loop-induced couplings in heterotic non-Abelian orbifold models as well.
That is beyond our scope and is left for future work.

\subsection{Emergent non-Abelian symmetries}
\label{sec:non-Abelian}

In this section, we explore a realization of non-Abelian symmetries from groupification. 
As previously noted, when we impose the symmetry $G^{(\mathrm{outer})}$ originating from the outer automorphism of groups, residual Abelian group-like symmetries can be enhanced to non-Abelian ones for certain cases, as shown in Table~\ref{tab:non_AbelianGr}. This is the case when $G^{(\mathrm{outer})}$ non-trivially acts on conjugacy classes of the groups. 
\begin{table}[H]
    \caption{Non-Abelian group-like symmetries for conjugacy classes of finite groups.}
    \centering
    \begin{tabular}{|c||c|}\hline 
       $G$  & $\mathrm{Gr}[\mathrm{Conj}(G)]\rtimes G^{(\mathrm{outer})}$  \\\hline\hline
       $D_{N=\mathrm{even}}$  & $\mathbb{Z}_2\times D_4$ \\ \hline 
       $T_{N=\mathrm{prime}}$  & $S_3$ \\\hline
       $A_4$  & $S_3$ \\\hline
       $\Delta(27)$ & $((\mathbb{Z}_3 \times \mathbb{Z}_3')\rtimes Q_8) \rtimes S_3$ \\ \hline
    \end{tabular}
    \label{tab:non_AbelianGr}
\end{table}

Another possibility is relevant to the CP symmetry.
When the fields in a theory are labeled by conjugacy classes of finite groups in which the charge conjugate field belongs to a different class, the residual discrete symmetry can be enlarged to a non-Abelian one.
That is the generalized CP symmetry~\cite{Holthausen:2012dk,Chen:2014tpa,Feruglio:2012cw}.\footnote{Recently, the generalized CP symmetry with non-invertible selection rules was discussed as concrete examples in Ref.~\cite{Kobayashi:2025wty}.}

For illustrative purposes, let us consider a theory where fields are labeled by the conjugacy classes of $T_7$. 
As discussed in Sec.~\ref{sec:TN}, the groupification of $\mathrm{Conj}(T_7)$ is described by 
\begin{align}
  \gr(\conj(T_7)) = \{ [C_1],[C^1_7],[C^2_7] \}\cong \mathbb{Z}_3.
\end{align}
Suppose that three scalar fields $\phi_1$, $\phi_2$ and $\phi_3$ are respectively labeled by $[C_1]$, $[C^1_7]$ and $[C^2_7]$. 
Under the $\mathbb{Z}_3$ symmetry, $\phi_1$ transforms as a trivial singlet, whereas $\phi_{2}$ and $\phi_3$ transform as
\begin{align}
    \begin{pmatrix}
        \phi_2\\
        \phi_3
    \end{pmatrix}
    \rightarrow
    \begin{pmatrix}
        \omega & 0 \\
        0 & \omega^2
    \end{pmatrix}
    \begin{pmatrix}
        \phi_2\\
        \phi_3
    \end{pmatrix}
    \equiv \rho
    \begin{pmatrix}
        \phi_2\\
        \phi_3
    \end{pmatrix},
\end{align}
with $\omega =e^{2\pi i/3}$. 
Furthermore, we further consider a CP transformation:
\begin{align}
    \begin{pmatrix}
        \phi_2(x)\\
        \phi_3(x)
    \end{pmatrix}
    \rightarrow
    \begin{pmatrix}
        \phi_3(x_P)\\
        \phi_2(x_P)
    \end{pmatrix}
    ,
\end{align}
with $x=(t,\mathbf{x})$ and $x_P=(t,-\mathbf{x})$. 
Here, we use the fact that the charge conjugate of $\phi_2$ is $\phi_3$, and vice versa. 
It originates from the $\mathbb{Z}_2^{\mathrm{CP}}$ transformation $[C_7^1]\leftrightarrow [C_7^2]$. 
Hence, combining the $\mathbb{Z}_3$ and CP transformations, one arrives at
\begin{align}
    \phi_i \xrightarrow{\mathrm{CP}} \bar{\phi_i} \xrightarrow{\mathbb{Z}_3} \rho^{\ast}\phi_i \xrightarrow{\mathrm{CP}^{-1}} \rho^\ast\phi_i.
\end{align}
Note that $\rho^\ast = \rho^{-1}$ also belongs to $\mathbb{Z}_3$ symmetry. 
Since the transformations of $\mathbb{Z}_3$ and $\mathbb{Z}_2^{\mathrm{CP}}$ do not commute, the residual symmetry is enhanced to the non-Abelian group $S_3\cong \mathbb{Z}_3\rtimes \mathbb{Z}_2^{\mathrm{CP}}$. 
Under this symmetry, $\phi_1$ corresponds to a trivial singlet under $S_3$, but $\phi_2$ and $\phi_3$ form an $S_3$ doublet. 
This provides one of the simplest example of realizing non-Abelian discrete symmetry through a procedure of groupification. 
The same $S_3$ symmetry can be realized for other $T_N$ and $\Delta(3N^2)|_{N\notin 3\mathbb{Z}_{>0}}$ groups in which the charge conjugate class differs from the original one. 
Furthermore, one can realize $\Sigma(18)\cong (\mathbb{Z}_3\times \mathbb{Z}'_3)\rtimes \mathbb{Z}_2^{\mathrm{CP}}$ symmetry for $\Delta(3N^2)|_{N\in 3\mathbb{Z}_{>0}}$.

\section{Anomalies of groupification}
\label{sec:anomaly}

We have found that the tree-level selection rules are violated by loop effects, 
while the symmetry $\gr[A]$ remains.
As shown explicitly in the previous sections, 
$\gr[A]$ is given by $\mathbb{Z}_N$ or $\mathbb{Z}_N \times \mathbb{Z}_M$.
Fermion fields have definite charges under these Abelian discrete symmetries and definite transformation behaviors.
Thus, we can define anomalies of $\gr[A]$ in the same manner as for conventional discrete symmetries \cite{Krauss:1988zc,Ibanez:1991hv,Araki:2008ek,Chen:2015aba,Kobayashi:2021xfs}.

We consider a gauge theory with a continuous gauge symmetry $G_g$ and a discrete $\mathbb{Z}_N$ symmetry. 
The theory includes fermion fields transforming in representations ${\bf R}^{(f)}$ of $G_g$ and the $\mathbb{Z}_N$ charge $q^{(f)}_N$.
There exists a mixed $\mathbb{Z}_N-G_g-G_g$ anomaly, and its anomaly coefficient is given by \cite{Araki:2008ek},
\begin{align}
    A_{\mathbb{Z}_N-G_g-G_g} = \frac{2}{N}\sum_{{\bf R}^{(f)}} q_N^{(f)}T_2({\bf R}^{(f)}),
\end{align}
where $T_2({\bf R}^{(f)})$ denotes the Dynkin index of the representation ${\bf R}^{(f)}$, e.g. ${\bf R}^{(f)}=1/2$ for the fundamental representation of $SU(N)$.
The anomaly-free condition requires 
\begin{align}
    \sum_{{\bf R}^{(f)}} q_N^{(f)}T_2({\bf R}^{(f)}) =0,~~~{\rm mod}~~N/2.
\end{align}
Similarly, the coefficient of the $\mathbb{Z}_N$-gravity-gravity anomaly is written by 
\begin{align}
    A_{\mathbb{Z}_N-{\rm gravity- gravity}} = \frac{2}{N}\sum_{{\bf R}^{(f)}} q_N^{(f)}{\dim}({\bf R}^{(f)}),
\end{align}
where ${\dim}({\bf R}^{(f)})$ denotes the dimension of the representation ${\bf R}^{(f)}$, and the anomaly-free condition requires
\begin{align}
  \sum_{{\bf R}^{(f)}} q_N^{(f)}{\dim}({\bf R}^{(f)})=0,~~~{\rm mod}~~N/2.
\end{align}

Here, we explicitly apply these anomaly-free conditions to the selection rules in the previous sections. 
First of all, for $D_N$ with $N=$odd,  we find that $\gr[\conj(D_N)]$ is $\mathbb{Z}_2$, where fields corresponding to $[C_1]$ and $[B_1]$ carry $\mathbb{Z}_2$ even and odd charges, respectively.
Hence, only fermion fields corresponding to $[B_1]$ contribute to the anomalies.
The anomaly-free condition for the $\mathbb{Z}_N-G_g-G_g$ anomaly is expressed as
\begin{align}
    \sum_{[B_1]} T_2({\bf R}^{(f)}) =0,~~~{\rm mod}~~1.
\end{align}
In all cases, $\gr[\conj(D_N)]$ with $N=$odd is free from the $\mathbb{Z}_N$-gravity-gravity anomaly.
Several models based on $\mathbb{Z}_2$ gauging $\mathbb{Z}_N$ with $N=3,5$ were studied in Refs.~\cite{Kobayashi:2024cvp,Kobayashi:2025znw,Kobayashi:2025ldi,Jiang:2025psz,Liang:2025dkm,Kobayashi:2025thd,Kobayashi:2025rpx}.
These models do not include fields corresponding to $[B_1]$, and they are anomaly-free.
For $S_3$ orbifolded heterotic models, only the $\theta_b$ twisted sector contributes to this anomaly.

For $D_N$ with $N=$even,  we find that $\gr[\conj(D_N)]$ is $\mathbb{Z}_2 \times \mathbb{Z}_2$, where fields corresponding to $[C_1]$, $[C_2^{\rm odd}]$, $[B_1]$ and $[B_2]$ carry (even,even), (even,odd), (odd,even), and (odd,odd) charges under 
$\mathbb{Z}_2 \times \mathbb{Z}_2$, respectively.
The anomaly-free conditions are written by 
\begin{align}
    \sum_{[B_1], [B_2]} T_2({\bf R}^{(f)}) =0,~~~{\rm mod}~~1,
\end{align}
for the first $\mathbb{Z}_2$,
\begin{align}
    \sum_{[C_2^{\rm odd}], [B_2]} T_2({\bf R}^{(f)}) =0,~~~{\rm mod}~~1,
\end{align}
for the second $\mathbb{Z}_2$.
For example, in Refs.~\cite{Kobayashi:2024cvp,Kobayashi:2025znw,Kobayashi:2025ldi}, Yukawa textures derived from $\mathbb{Z}_2$ gauging $\mathbb{Z}_4$ were studied. 
These models can be classified into anomalous and anomaly-free cases. 
In addition, we studied matter symmetries of supersymmetric standard model 
imposed by $\mathbb{Z}_2$ gauging $\mathbb{Z}_4$ in Ref.~\cite{Kobayashi:2025lar}.
These can be further constrained by anomaly-free conditions. 
For the $D_4$ orbifolded heterotic models, only the $\theta_b$ twisted sectors contribute to the anomalies of the first $\mathbb{Z}_2$, while $\theta_a$ and $\theta_a\theta_b$ twisted sectors contribute to the anomalies of the second $\mathbb{Z}_2$.

Similarly, we can study anomalies for $T_7$. 
We find that $\gr[\conj(T_7)]$ is $\mathbb{Z}_3$, where the fields corresponding to $[C_1]$, $[C_7^1]$, and $[C_7^2]$, carry $\mathbb{Z}_3$ charges, 0, 1, and 2, respectively. 
The anomaly-free conditions are given by
\begin{align}
    \sum_{[C^1_7]} T_2({\bf R}^{(f)}) + \sum_{[C^2_7]} 2T_2({\bf R}^{(f)}) =0,~~~{\rm mod}~~\frac{3}{2},
\end{align}
for the $\mathbb{Z}_N-G_g-G_g$ anomaly and 
\begin{align}
    \sum_{[C^1_7]} {\rm dim}({\bf R}^{(f)}) + \sum_{[C^2_7]} 2{\rm dim}({\bf R}^{(f)}) =0,~~~{\rm mod}~~\frac{3}{2},
\end{align}
for $\mathbb{Z}_N$-gravity-gravity anomaly.
For $T_7$ orbifolded heterotic models, only the $\theta_b$ and $(\theta_b)^2$ twisted sectors contribute to the anomalies.
Similarly, for other $\gr[\conj(G)]$, we can analyze anomalies.
Moreover, we can impose the symmetry $G^{\rm outer}$ originating from the outer automorphism of groups by setting proper relations among couplings as studied in the previous section. 
Its anomalies can also be analyzed in a similar way.

Thus, when we require anomaly-free conditions on $\gr[A]$, we can constrain models with non-invertible selection rules. 
Alternatively, if $\gr[A]$ is anomalous, it is important to study whether or not the Green-Schwarz mechanism can cancel the anomaly in the presence of the non-invertible selection rules within the framework of string-derived low energy effective field theory.

Finally, let us give a comment on approximate $\mathbb{Z}_N'$ symmetries. 
As an example, we consider the approximate $\mathbb{Z}_2'$ symmetry in 
$\conj(S_3)$. 
As mentioned in the previous section, when the tree-level couplings 
$\lambda^{(0)}_{\dot{1} \dot{1} \dot{1}}$ and $\lambda^{(0)}_{\dot{1}\dot{2  } \dot{2} }$ vanish, 
there appears the exact $\mathbb{Z}_2'$ symmetry at tree level.
However, suppose that this $\mathbb{Z}_2'$ symmetry is anomalous. 
In that case, non-perturbative effects may induce the couplings 
$\lambda^{({\rm non-per})}_{\dot{1} \dot{1} \dot{1}}$ and $\lambda^{({\rm non-per})}_{\dot{1}\dot{2  } \dot{2} }$ as well as 
$\lambda^{({\rm non-per})}_{0 0 \dot{1}}$. 
Such anomalous $\mathbb{Z}_2'$ may lead to selection rules similar to those obtained from $\conj(S_3)$.
However, our selection rules due to $\conj(S_3)$ imply a relation among $\lambda^{({\rm non-per})}_{\dot{1} \dot{1} \dot{1}}$,  $\lambda^{({\rm non-per})}_{\dot{1}\dot{2  } \dot{2} }$, and 
$\lambda^{({\rm non-per})}_{0 0 \dot{1}}$. 
In particular, the coupling $\lambda^{({\rm non-per})}_{0 0 \dot{1}}$ is expressed in terms of $\lambda^{({\rm non-per})}_{\dot{1} \dot{1} \dot{1}}$ and $\lambda^{({\rm non-per})}_{\dot{1}\dot{2  } \dot{2} }$, and are suppressed by a loop factor compared with the other two couplings. 

The same reasoning applies to other approximate symmetries. 
Anomalous approximate symmetries may lead to the same selection rules, but our selection rules derived from $\conj(G)$ further imply the relations among couplings such that loop induced couplings can be written in terms of tree level ones. 
These relations among couplings can lead to phenomenologically interesting hierarchies among couplings, and these aspects are important in particle physics, e.g., in the radiative generation of neutrino masses and other phenomena~\cite{Suzuki:2025oov,Kobayashi:2025cwx,Suzuki:2025bxg,Nomura:2025sod,Chen:2025awz,Okada:2025kfm}.

\section{Selection rules on Calabi-Yau manifolds}
\label{sec:CY}

We have studied loop aspects of $\conj(G)$, whose selection rules cannot be understood by group theory. 
These can be realized in heterotic non-Abelian orbifold models, although we mainly focus on field theory.
The selection rules arising from Calabi-Yau compactifications are determined not by group theory but by topology, and those are non-trivial. 
Here, let us compare their loop effects with the results in the previous sections. 

We consider heterotic Calabi-Yau compactifications with the standard embedding, where there is a one-to-one correspondence between moduli and chiral matter fields.
Recently, their coupling selection rules were classified for complete intersection Calabi-Yau threefolds with five or fewer moduli \cite{Dong:2025pah}, and their phenomenological implications were studied \cite{Dong:2026iwa}. 
Table \ref{tab:h11=3} shows such a classification for Calabi-Yau threefolds with three moduli and three corresponding fields $\phi_i$.
The table lists the forbidden couplings $\kappa_{ijk} \phi_i \phi_j\phi_k$.
Calabi-Yau compactifications yield four-dimensional $\mathcal{N}=1$ supersymmetric effective field theories at low energies. 
There is the non-renormalization theorem on superpotential in $\mathcal{N}=1$ supersymmetric theories. 
Since we are interested in generic loop effects on selection rules without assuming any other symmetries, we do not take supersymmetric cancellations into account.
Instead, we may assume that supersymmetry is broken just below the compactification scale. 
Under this assumption, we study loop effects on the selection rules originating from Calabi-Yau topology.
The result is as follows. 
Although certain 3-point couplings are forbidden at tree level as shown in Table \ref{tab:h11=3}, all 3-point couplings among possible fields are induced by loop effects. 
This result is quite different from what we have studied in the previous section.
For non-invertible selection rules arising from $\conj(G)$, certain groups $\gr[A]$ remain even after including loop effects. 
In addition, as another difference, we can show that there is no unit in the selection rules dictated by Calabi-Yau topology, while the fusion algebra $\conj(G)$ has the unit.
See Appendix~\ref{app:CY} for further discussion.

\begin{table}[H]
    \centering
    \caption{Types of prepotential for CICYs with $h^{1,1}=3$. Vanishing $\kappa_{abc}$ are shown.}
    \label{tab:h11=3}
    \begin{tabular}{|c||c|}
    \hline
    Type & $\kappa_{abc}=0$ \\ \hline \hline
      1 & $\kappa_{111}, \kappa_{222}, \kappa_{333}$ \\
      \hline
       2 & $\kappa_{111}, \kappa_{222}$ \\
      \hline
      3 & $\kappa_{111}, \kappa_{112}, \kappa_{113}$ \\
      \hline
       4 & $\kappa_{111}, \kappa_{112}, \kappa_{113}, \kappa_{222}$ \\
      \hline
       5 & $\kappa_{111}$ \\
      \hline
       6 & none \\
      \hline
       7 & $\kappa_{111}, \kappa_{112}, \kappa_{113} ,\kappa_{122} ,\kappa_{222} ,\kappa_{223}$ \\
      \hline
       8 & $\kappa_{111}, \kappa_{112}, \kappa_{113} ,\kappa_{122} ,\kappa_{222}$ \\
      \hline
       9 & $\kappa_{111}, \kappa_{112}, \kappa_{113}, \kappa_{222}, \kappa_{333}$ \\
      \hline
       10 & $\kappa_{111}, \kappa_{112}, \kappa_{113}, \kappa_{122} ,\kappa_{222}, \kappa_{333}$ \\
      \hline
       11 & $\kappa_{111}, \kappa_{112}, \kappa_{113}, \kappa_{122}, \kappa_{222}, \kappa_{223}, \kappa_{333}$ \\
      \hline
    \end{tabular}
  \end{table}

\section{Conclusions}
\label{sec:con}

We have studied non-invertible selection rules without group structures, in particular, focusing on the multiplication rules of conjugacy classes of finite discrete groups. 
Although these selection rules are expected to be violated via loop effects~\cite{Heckman:2024obe,Kaidi:2024wio,Funakoshi:2024uvy}, certain group-like symmetries remain exact at all loop orders through a procedure referred to as groupification.

In this paper, we revealed the residual group-like symmetries associated with selection rules based on the conjugacy classes of various finite groups realized in heterotic string theory on non-Abelian orbifolds. 
Our analysis show that these residual group-like symmetries $\gr[A]$ are described by $\mathbb{Z}_N$ or $\mathbb{Z}_N\times \mathbb{Z}_M$ symmetries, as summarized in Table~\ref{tab:Gr_summary}. 
We comment on the possibility of realizing non-Abelian symmetries through groupification. 
One possible avenue is to consider the 
symmetry originating from the outer automorphism of groups including CP symmetry, which enhances $\gr[A]$ to a non-Abelian symmetry, as shown in Table~\ref{tab:non_AbelianGr}. 
Furthermore, we found an approximate discrete symmetry that controls loop-induced couplings. 
Since this approximate symmetry becomes exact in the limit where the tree-level selection rules are preserved, most parameters in the non-invertible selection rules are natural in the sense of 't Hooft. 
This observation supports phenomenological studies in which couplings are radiatively generated, e.g., in the context of radiative seesaw models~\cite{Kobayashi:2025cwx}. 
Our obtained results are based on toroidal orbifolds, but we have also analyzed specific heterotic Calabi-Yau compactifications. 
In heterotic string theory with standard embedding, 
we found that all three-point couplings are induced by loop effects. While we have focused on complete intersection Calabi-Yau threefolds, it would be interesting to investigate the structure of non-invertible selection rules on more general Calabi-Yau threefolds. We leave a comprehensive analysis of such cases for future work.

Since the residual symmetry is described by the group-like symmetry $\gr[A]$, its anomalies can be analyzed in the same manner as conventional discrete symmetries. The anomaly cancellation conditions we proposed for $\gr[A]$ can provide constraints on models with non-invertible selection rules if $\gr[A]$ is anomalous. 

\acknowledgments

This work was supported by JSPS KAKENHI Grant Numbers JP23K03375 (T.K.) and JP25H01539 (H.O.).

\appendix
\section{Loop-induced 2-point couplings}
\label{app:2-point}

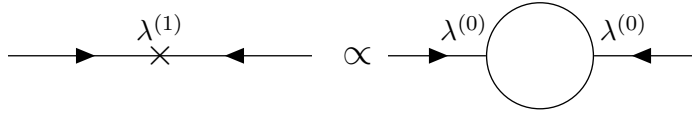
\begin{figure}[H]
  \centering
  \begin{tikzpicture}
  \begin{feynhand}  
    \vertex (i1) at (-2,0);
    \vertex (i2) at (2,0);
    \vertex [ringblob] (a) [draw, text=black, circle] at (0,0) {{}};
    \propag [fermion] (i1) to  (a);
    \propag [fermion] (i2) to  (a);
    \vertex (v_1) at (1.1,0) {};
\node[above=2pt] at (v_1) {$\lambda^{(0)}$};
\vertex (v_2) at (-1.0,0) {};
\node[above=2pt] at (v_2) {$\lambda^{(0)}$};
    \vertex (i3) at (-7,0);
    \vertex (i4) at (-3,0);
    \vertex (v_3) at (-5,0) ;
    \propag [fermion] (i3) to  (v_3);
    \propag [fermion] (i4) to  (v_3);
    \vertex (v_3) at (-5,0);
    \node[above=2pt] at (v_3) {$\lambda^{(1)}$};
        \node at (-2.4, 0) {\Large{$\propto$}};
        \node at (v_3) {\Large{$\times$}};
  \end{feynhand}
  \end{tikzpicture}
  \caption{Loop effects.}
    \label{fig:2-point_1_loop_diagram}
\end{figure}

We have examined loop effects due to the diagram of Figure \ref{fig:3-point_1_loop_diagram}.
Similarly, we can study other diagrams and find the same results on $\gr(\conj(G))$.
Here, we discuss 2-point couplings as shown in Figure \ref{fig:2-point_1_loop_diagram}.

$\bullet$~{\bf $S_3$}

At tree level, $\conj(S_3)$ allows the following 2-point couplings:
\begin{align}
    \lambda^{(0)}_{00}, \quad \lambda^{(0)}_{\dot{1} \dot{1}}, \quad \lambda^{(0)}_{\dot{2}\dot{2}}.
\end{align}
These are diagonal.
The loop effects of Figure \ref{fig:2-point_1_loop_diagram} induce the following new coupling:
\begin{align}
\label{eq:2-point-S3}
{
\lambda^{(1)}_{ 0 \dot{1}
}
}&\propto
\lambda^{(0)}_{ 0 \dot{1} \dot{1} }\lambda^{(0)}_{ \dot{1} \dot{1} \dot{1} }
+
\lambda^{(0)}_{ 0 \dot{2} \dot{2} }\lambda^{(0)}_{ \dot{1} \dot{2} \dot{2} }.
\end{align}
The other couplings are forbidden by $\gr[\conj(S_3)]=\mathbb{Z}_2$.
In the above relation (\ref{eq:2-point-S3}), 
there is the approximate $\mathbb{Z}'_2$ correspondence between the left and right hand sides.
In the limit $\lambda^{(0)}_{ \dot{1} \dot{1} \dot{1} },\lambda^{(0)}_{ \dot{1} \dot{2} \dot{2} } \to 0 $, the coupling $\lambda^{(1)}_{ 0 \dot{1}}$ is also vanishing, and the approximate $\mathbb{Z}'_2$ becomes exact.

\vspace{0.5cm}
$\bullet$~{\bf $D_4$}

At tree level, $\conj(D_4)$ allows the following 2-point couplings:
\begin{align}
    \lambda^{(0)}_{00}, \quad \lambda^{(0)}_{\dot{1} \dot{1}}, \quad \lambda^{(0)}_{\dot{2}\dot{2}}, \quad 
    \lambda^{(0)}_{\dot{3} \dot{3}}, \quad \lambda^{(0)}_{\dot{4}\dot{4}}.
\end{align}
These are diagonal.
The loop effects of Figure \ref{fig:2-point_1_loop_diagram} induce the following new coupling:
\begin{align}
\label{eq:2-point-D4}
{
\lambda^{(1)}_{ 0 \dot{2}
}
}&\propto
\lambda^{(0)}_{ 0 \dot{1} \dot{1} }\lambda^{(0)}_{ \dot{1} \dot{1} \dot{2} }
+
\lambda^{(0)}_{ 0 \dot{3} \dot{3} }\lambda^{(0)}_{ \dot{2} \dot{3} \dot{3} }
+
\lambda^{(0)}_{ 0 \dot{4} \dot{4} }\lambda^{(0)}_{ \dot{2} \dot{4} \dot{4} }.
\end{align}
The other couplings are forbidden by $\gr[\conj(D_4)]=\mathbb{Z}_2 \times \mathbb{Z}_2$.
In the above relation (\ref{eq:2-point-D4}), 
there is the approximate $\mathbb{Z}'_2$ correspondence between the left and right hand sides.
In the limit $\lambda^{(0)}_{ \dot{1} \dot{1} \dot{2} },\lambda^{(0)}_{ \dot{2} \dot{3} \dot{3} } ,\lambda^{(0)}_{ \dot{2} \dot{4} \dot{4} } \to 0 $, the coupling $\lambda^{(1)}_{ 0 \dot{2}}$ is also vanishing, and the approximate $\mathbb{Z}'_2$ becomes exact.

\vspace{0.5cm}
$\bullet$~{\bf $T_7$}

At tree level, $\conj(T_7)$ allows the following 2-point couplings:
\begin{align}
    \lambda^{(0)}_{00}, \quad \lambda^{(0)}_{{1} \bar{1}}, \quad \lambda^{(0)}_{{2}\bar{2}}.
\end{align}
These are diagonal.
The loop effects of Figure \ref{fig:2-point_1_loop_diagram} induce the following new coupling:
\begin{align}
\label{eq:2-point-T7}
{
\lambda^{(1)}_{ 0 1}
}&\propto
\lambda^{(0)}_{ 0 1 \bar{1} }\lambda^{(0)}_{ 1 1 \bar{1} }
+
\lambda^{(0)}_{ 0 2 \bar{2} }\lambda^{(0)}_{ 1 2 \bar{2} }, \notag \\
{
\lambda^{(1)}_{ 0 \bar{1}}
}&\propto
\lambda^{(0)}_{ 0 1 \bar{1} }\lambda^{(0)}_{ 1 \bar{1} \bar{1} }
+
\lambda^{(0)}_{ 0 2 \bar{2} }\lambda^{(0)}_{ \bar{1} 2 \bar{2} }, \notag \\
{
\lambda^{(1)}_{ 1 1}
}&\propto
\lambda^{(0)}_{ 1 1 1 }\lambda^{(0)}_{ 1 \bar{1} \bar{1} }
+
\lambda^{(0)}_{ 1 1 \bar{1} }\lambda^{(0)}_{ 1 1 \bar{1} }
+
\lambda^{(0)}_{ 1 2 \bar{2} }\lambda^{(0)}_{ 1 2 \bar{2} }, \notag \\
{
\lambda^{(1)}_{ \bar{1} \bar{1}}
}&\propto
\lambda^{(0)}_{ 1 1 \bar{1} }\lambda^{(0)}_{ \bar{1} \bar{1} \bar{1} }
+
\lambda^{(0)}_{ 1 \bar{1} \bar{1} }\lambda^{(0)}_{ 1 \bar{1} \bar{1} }
+
\lambda^{(0)}_{ \bar{1} 2 \bar{2} }\lambda^{(0)}_{ \bar{1} 2 \bar{2} }.
\end{align}
The other couplings are forbidden by $\gr[\conj(T_7)]=\mathbb{Z}_3 $.
In the above relation (\ref{eq:2-point-T7}), 
there is the approximate $\mathbb{Z}'_3$ correspondence between the left and right hand sides.
In the limit that the couplings with non-zero $\mathbb{Z}_3'$ charges vanish, all the couplings $\lambda^{(1)}_{ 0 1}, \lambda^{(1)}_{ 0 \bar{1}}, \lambda^{(1)}_{ 1 1}, \lambda^{(1)}_{ \bar{1} \bar{1}} $ are also vanishing, and the approximate $\mathbb{Z}'_2$ becomes exact.

\section{$\Delta(6N^2)$}
\label{app:Deta-6N}

In this section, we discuss $\Delta(6N^2) \cong (\mathbb{Z}_N\times \mathbb{Z}'_{N}) \rtimes S_3$, whose order is $6N^2$. Putting $a$ as the $\mathbb{Z}_N$ generator, $a'$ as the $\mathbb{Z}'_N$ generator, and $b,c$ as the $S_3$ generators, they satisfy the following algebraic relations:
\begin{gather}
    a^N = a'^N = b^3 = c^2 = (bc)^2=e,\quad aa' = a'a,\notag\\
  bab^{-1} = a^{-1}a'^{-1},\quad ba'b^{-1} = a,\notag\\
  cac^{-1} = a'^{-1},\quad ca'c^{-1}=a^{-1},
\end{gather}
where $e$ denotes the identity. 
Note that arbitrary elements in $\Delta(6N^2)$ can be expressed as $b^kc^la^ma'^n,\ k=0,1,2,\ l=0,1,m,n=0,1,...,N-1$. 

\subsection{Conjugacy classes}

We obtain the conjugacy classes of $\Delta(6N^2)$ as follows \cite{Ishimori:2010au,Kobayashi:2022moq}: 
\begin{itemize}
  \item When $N/3$ is not an integer,
  \begin{align}
    &C_1 = \{ e \},\notag\\
    &C^{(k)}_3 = \{ a^ka'^{-k},a^{-2k}a'^{-k},a^{k}a'^{2k} \},\ k=1,...,N-1,\notag\\
    &C^{(l,m)}_6 = \{ a^{l}a'^{m}, a^{m-l}a'^{-l},a^{-m}a'^{l-m},a^{-m}a'^{-l},a^{m-l}a'^{m},a^{l}a'^{l-m}\},\notag\\
    &C_{2N^2} = \{ b^ka^la'^m\ |\ k=1,2,\ l,m=0,1,...,N-1 \},\notag\\
    &C^{(k)}_{3N} = \{ ca^{k+n}a'^n, b^2ca^{-k}a'^{-k-n}, bca^{-n}a'^{k}\ |\ n= 0,1,...,N-1 \},\ k=0,1,...,N-1.
  \end{align}
  Note that $C^{(l,m)}_6$ is defined only for $(l,m)$ such that $l+m\neq 0\pmod N,\ 2l-m\neq 0\pmod N,\ l-2m\neq 0\pmod N$.
  \item When $N/3$ is an integer,
  \begin{align}
    &C_1 = \{ e \},\notag\\
    &C^{(p)}_1 = \{ a^pa'^{-p} \},\quad p = \frac{N}{3},\frac{2N}{3},\notag\\
    &C^{(k)}_3 = \{ a^ka'^{=k},a^{-2k}a'^{-k},a^{k}a'^{2k} \},\quad k = 1,...,N-1,k\neq \frac{N}{3},\frac{2N}{3},\notag\\
    &C^{(l,m)}_6 = \{ a^la'^m,a^{m-l}a'^{-l},a^{-m}a'^{l-m},a^{-m}a'^{-l},a^{m-l}a'^{m},a^{l}a'^{l-m} \},\notag\\
    &C^{(q)}_{2N^2/3} = \{ ba^{q-n-3m}a'^n, b^2a^{-n}a'^{n+3m-q}\ |\ n=0,...,N-1,\ m=0,...,\frac{N-3}{3} \},\notag\\
    &C^{(l)}_{3N} = \{ ca^{l+n}a'^n, b^2ca^{-l}a'^{-l-n}, bca^{-n}a'^l\ |\ n=0,...,N-1 \},\quad l=0,...,N-1. 
  \end{align}
  Note that $C^{(l,m)}_6$ is defined only for pairs $(l,m)$ satisfying $l+m\neq 0\pmod N,\ 2l-m\neq 0\pmod N,\ l-2m\neq 0\pmod N$ and excluding the special cases $(0,0),(N/3,2N/3)$, and $(2N/3,N/3)$.
\end{itemize}

\subsection{Example: $S_4$}
Let us consider a concrete example in the case of $N=2$, i.e., $\Delta(24)\cong (\mathbb{Z}_2\times \mathbb{Z}'_2)\rtimes S_3\cong S_4$ conjugacy classes, which are given as follows:
\begin{align}
    &C_1 = \{ e \}, \notag \\
    &C^{(1,0)}_3 = \{ a,a',aa' \}, \notag \\
    &C_8 = \{ b,ba,ba',baa',b^2,b^2a,b^2a',b^2aa' \}, \notag \\
    &C^{(0)}_6 = \{ ca^na'^n,b^2ca^{-n},bca^{-n}|n=0,1 \},\notag \\
    &C^{(1)}_6 = \{ ca^{n+1}a'^n,b^2ca^{-1}a'^{-n-1},bca^{-n}a' |n=0,1\},
\end{align}
They obey the multiplication rules shown in Table~\ref{tab:Delta24}.
\begin{table}[H]
    \centering
        \caption{Multiplication rules for conjugacy classes of $S_4$.}
    \label{tab:Delta24}
    \resizebox{\textwidth}{!}{
    \begin{tabular}{|c||c|c|c|c|c|}
    \hline
    &$C_1$&$C^{(1,0)}_3$&$C_8$&$C^{(0)}_6$&$C^{(1)}_6$\\
    \hline\hline
    $C_1$&$C_1$&$C^{(1,0)}_3$&$C_8$&$C^{(0)}_6$&$C^{(1)}_6$\\
    \hline
    $C^{(1,0)}_3$&$C^{(1,0)}_3$&$3C_1+2C^{(1,0)}_3$&$3C_8$&$C^{(0)}_6+2C^{(1)}_6$&$2C^{(0)}_6+C^{(1)}_6$\\
    \hline
    $C_8$&$C_8$&$3C_8$&$8C_1+8C^{(1,0)}_3+4C_8$&$4C^{(0)}_6+4C^{(1)}_6$&$4C^{(0)}_6+4C^{(1)}_6$\\
    \hline
    $C^{(0)}_6$&$C^{(0)}_6$&$C^{(0)}_6+2C^{(1)}_6$&$4C^{(0)}_6+4C^{(1)}_6$&$6C_1+2C^{(1,0)}_3+3C_8$&$4C^{(1,0)}_3+3C_8$\\
    \hline
    $C^{(1)}_6$&$C^{(1)}_6$&$2C^{(0)}_6+C^{(1)}_6$&$4C^{(0)}_6+4C^{(1)}_6$&$4C^{(1,0)}_3+3C_8$&$6C_1+2C^{(1,0)}_3+3C_8$\\
    \hline
    \end{tabular}}
\end{table}
From these results, we can obtain the subset $\com(\conj(S_4))$ as
\begin{align}
  \com(\conj(S_4)) = \{ C_1, C^{(1,0)}_3, C_8 \},
\end{align}
and we can also compute the groupification of $\conj(\Delta(24))$ as follows.
\begin{align}
  \gr(\conj(S_4)) = \{ [C_1], [C^{(0)}_6]\}\cong \mathbb{Z}_2,
\end{align}
where we define
\begin{align}
  &[C_1] = \{ C_1, C^{(1,0)}_3, C_8 \},\quad  [C^{(0)}_6] = \{ C^{(0)}_6,C^{(1)}_6 \}.
\end{align}
Their $\mathbb{Z}_2$ charges are given by
\begin{align}
    &[C_1]: 0,\quad
    [C_{6}^{(0)}] : 1.
\end{align}

Let us check the loop effects more explicitly. We introduce the following new notation:
\begin{gather}
    0 = C_1,\quad
    \dot{1} = C^{(1,0)}_3,\quad
    \dot{2} = C_8,\notag\\
    \dot{3} = C^{(0)}_6,\quad
    \dot{4} = C^{(1)}_6.
\end{gather}
By using this notation, we denote the corresponding fields by $\phi_0$, $\phi_{1}, \cdots$, and their couplings by, e.g., $\lambda_{001}$.
The selection rules imposed by $\conj(S_4)$ allow the tee-level 3-point couplings shown in the second row of Table~\ref{tab:3_point_coupling_Delta24} , which correspond to the following couplings:
\begin{gather}
    \lambda_{0kk},\quad \lambda_{kkk},\quad \lambda_{\dot{1}kk},\quad \lambda_{\dot{2}kk},\notag\\
    \lambda_{\dot{1}\dot{3}\dot{4}},\quad \lambda_{\dot{2}\dot{3}\dot{4}},
\end{gather}
and the others are forbidden at  tree level.

We can define an approximate $\mathbb{Z}_2' \times \mathbb{Z}_2'$ symmetry, where $C_3^{(1,0)}$, $C_8$, and $C_6^{(1)}$ carry 
(odd, even), (even, odd), and (odd, odd) charges, while 
$C_1$ and $C_6^{(0)}$ carry (even,even) charges.
In Table~\ref{tab:3_point_coupling_Delta24}, the tree-level couplings except $C_1C_kC_k$ violate this approximate $\mathbb{Z}_2' \times \mathbb{Z}_2'$ symmetry, where $C_k=C_1, C_3^{(1,0)}, C_8, C_6^{(0)}, C_6^{(1)}$.
These violating couplings induce other violating couplings as 
\begin{align}
\lambda^{(1)}_{ 0 0 \dot{1}
}&\propto
\lambda^{(0)}_{ 0 \dot{1} \dot{1} }\lambda^{(0)}_{ 0 \dot{1} \dot{1} }\lambda^{(0)}_{ \dot{1} \dot{1} \dot{1} }
+
\lambda^{(0)}_{ 0 \dot{2} \dot{2} }\lambda^{(0)}_{ 0 \dot{2} \dot{2} }\lambda^{(0)}_{ \dot{1} \dot{2} \dot{2} }
+
\lambda^{(0)}_{ 0 \dot{3} \dot{3} }\lambda^{(0)}_{ 0 \dot{3} \dot{3} }\lambda^{(0)}_{ \dot{1} \dot{3} \dot{3} }
+
\lambda^{(0)}_{ 0 \dot{4} \dot{4} }\lambda^{(0)}_{ 0 \dot{4} \dot{4} }\lambda^{(0)}_{ \dot{1} \dot{4} \dot{4} }, \notag \\
\lambda^{(1)}_{ 0 0 \dot{2}
}&\propto
\lambda^{(0)}_{ 0 \dot{2} \dot{2} }\lambda^{(0)}_{ 0 \dot{2} \dot{2} }\lambda^{(0)}_{ \dot{2} \dot{2} \dot{2} }
+
\lambda^{(0)}_{ 0 \dot{3} \dot{3} }\lambda^{(0)}_{ 0 \dot{3} \dot{3} }\lambda^{(0)}_{ \dot{2} \dot{3} \dot{3} }
+
\lambda^{(0)}_{ 0 \dot{4} \dot{4} }\lambda^{(0)}_{ 0 \dot{4} \dot{4} }\lambda^{(0)}_{ \dot{2} \dot{4} \dot{4} }, \notag \\
\lambda^{(1)}_{ 0 \dot{1} \dot{2}
}&\propto
\lambda^{(0)}_{ 0 \dot{2} \dot{2} }\lambda^{(0)}_{ \dot{1} \dot{2} \dot{2} }\lambda^{(0)}_{ \dot{2} \dot{2} \dot{2} }
+
\lambda^{(0)}_{ 0 \dot{3} \dot{3} }\lambda^{(0)}_{ \dot{1} \dot{3} \dot{3} }\lambda^{(0)}_{ \dot{2} \dot{3} \dot{3} }
+
\lambda^{(0)}_{ 0 \dot{3} \dot{3} }\lambda^{(0)}_{ \dot{1} \dot{3} \dot{4} }\lambda^{(0)}_{ \dot{2} \dot{3} \dot{4} }
+
\lambda^{(0)}_{ 0 \dot{4} \dot{4} }\lambda^{(0)}_{ \dot{1} \dot{3} \dot{4} }\lambda^{(0)}_{ \dot{2} \dot{3} \dot{4} }
+
\lambda^{(0)}_{ 0 \dot{4} \dot{4} }\lambda^{(0)}_{ \dot{1} \dot{4} \dot{4} }\lambda^{(0)}_{ \dot{2} \dot{4} \dot{4} }, 
\notag \\
\lambda^{(1)}_{ 0 \dot{3} \dot{4}
}&\propto
\lambda^{(0)}_{ 0 \dot{1} \dot{1} }\lambda^{(0)}_{ \dot{1} \dot{3} \dot{3} }\lambda^{(0)}_{ \dot{1} \dot{3} \dot{4} }
+
\lambda^{(0)}_{ 0 \dot{1} \dot{1} }\lambda^{(0)}_{ \dot{1} \dot{3} \dot{4} }\lambda^{(0)}_{ \dot{1} \dot{4} \dot{4} }
+
\lambda^{(0)}_{ 0 \dot{2} \dot{2} }\lambda^{(0)}_{ \dot{2} \dot{3} \dot{3} }\lambda^{(0)}_{ \dot{2} \dot{3} \dot{4} }
+
\lambda^{(0)}_{ 0 \dot{2} \dot{2} }\lambda^{(0)}_{ \dot{2} \dot{3} \dot{4} }\lambda^{(0)}_{ \dot{2} \dot{4} \dot{4} }
+
\lambda^{(0)}_{ 0 \dot{3} \dot{3} }\lambda^{(0)}_{ \dot{1} \dot{3} \dot{3} }\lambda^{(0)}_{ \dot{1} \dot{3} \dot{4} }
\notag\\
&
+
\lambda^{(0)}_{ 0 \dot{3} \dot{3} }\lambda^{(0)}_{ \dot{2} \dot{3} \dot{3} }\lambda^{(0)}_{ \dot{2} \dot{3} \dot{4} }
+
\lambda^{(0)}_{ 0 \dot{4} \dot{4} }\lambda^{(0)}_{ \dot{1} \dot{3} \dot{4} }\lambda^{(0)}_{ \dot{1} \dot{4} \dot{4} }
+
\lambda^{(0)}_{ 0 \dot{4} \dot{4} }\lambda^{(0)}_{ \dot{2} \dot{3} \dot{4} }\lambda^{(0)}_{ \dot{2} \dot{4} \dot{4} },
\notag \\
\lambda^{(1)}_{ \dot{1} \dot{1} \dot{2}
}&\propto
\lambda^{(0)}_{ \dot{1} \dot{2} \dot{2} }\lambda^{(0)}_{ \dot{1} \dot{2} \dot{2} }\lambda^{(0)}_{ \dot{2} \dot{2} \dot{2} }
+
\lambda^{(0)}_{ \dot{1} \dot{3} \dot{3} }\lambda^{(0)}_{ \dot{1} \dot{3} \dot{3} }\lambda^{(0)}_{ \dot{2} \dot{3} \dot{3} }
+
\lambda^{(0)}_{ \dot{1} \dot{3} \dot{3} }\lambda^{(0)}_{ \dot{1} \dot{3} \dot{4} }\lambda^{(0)}_{ \dot{2} \dot{3} \dot{4} }
+
\lambda^{(0)}_{ \dot{1} \dot{3} \dot{4} }\lambda^{(0)}_{ \dot{1} \dot{3} \dot{4} }\lambda^{(0)}_{ \dot{2} \dot{3} \dot{3} }
+
\lambda^{(0)}_{ \dot{1} \dot{3} \dot{4} }\lambda^{(0)}_{ \dot{1} \dot{3} \dot{4} }\lambda^{(0)}_{ \dot{2} \dot{4} \dot{4} }
\notag\\
&
+
\lambda^{(0)}_{ \dot{1} \dot{3} \dot{4} }\lambda^{(0)}_{ \dot{1} \dot{4} \dot{4} }\lambda^{(0)}_{ \dot{2} \dot{3} \dot{4} }
+
\lambda^{(0)}_{ \dot{1} \dot{4} \dot{4} }\lambda^{(0)}_{ \dot{1} \dot{4} \dot{4} }\lambda^{(0)}_{ \dot{2} \dot{4} \dot{4} }
\end{align}
which are shown in the third row of Table \ref{tab:3_point_coupling_Delta24}.
This approximate symmetry becomes exact when these tree-level violating terms vanish.

\begin{longtable}[H]{|c|p{12cm}|c|}
\caption{Allowed $3$-point couplings obtained from $\conj(S_4)$.}
\label{tab:3_point_coupling_Delta24} \\
\hline
& \multicolumn{1}{c|}{3-point coupling}& types\\
\hline
\endhead
tree & 
$C_1C_1 C_1$, \quad\allowbreak
$C_1 C^{(1,0)}_3C^{(1,0)}_3$, \quad\allowbreak
$C_1C_8C_8$,\quad\allowbreak
$C_1C^{(0)}_6C^{(0)}_6$,\quad\allowbreak
$C_1 C^{(1)}_6C^{(1)}_6$,\quad\allowbreak
$C^{(1,0)}_3C^{(1,0)}_3C^{(1,0)}_3$,\quad\allowbreak
$C^{(1,0)}_3C_8C_8$,\quad\allowbreak
$C^{(1,0)}_3C^{(0)}_6C^{(0)}_6$,\quad\allowbreak
$C^{(1,0)}_3C^{(0)}_6C^{(1)}_6$, \quad\allowbreak
$C^{(1,0)}_3C^{(1)}_6C^{(1)}_6$,\quad\allowbreak
$C_8C_8C_8$,\quad\allowbreak
$C_8C^{(0)}_6C^{(0)}_6$,\quad\allowbreak
$C_8C^{(0)}_6 C^{(1)}_6$, \quad\allowbreak
$C_8C^{(1)}_6C^{(1)}_6$.\quad\allowbreak
&14\\
\hline
1-loop & 
$C_1C_1C^{(1,0)}_3$,\quad\allowbreak
$C_1C_1 C_8$,\quad\allowbreak
$C_1 C^{(1,0)}_3C_8$,\quad\allowbreak
$C_1 C^{(0)}_6 C^{(1)}_6$, \quad\allowbreak
$C^{(1,0)}_3C^{(1,0)}_3C_8$.\quad\allowbreak
&5(19)\\
\hline
\end{longtable}

\subsection{Example: $\Delta(54)$}

As a concrete example with $N\in3\mathbb{Z}$, we examine $\Delta(54)\cong (\mathbb{Z}_3\times \mathbb{Z}'_3)\rtimes S_3$ conjugacy classes, which are given as follows:
\begin{align}
&C_1=\{ e \},\notag\\
&C^{(1)}_1=\{ aa'^2 \},\notag\\
&C^{(2)}_1=\{ a^2a' \},\notag\\
&C^{(1,0)}_6=\{ a,a',a^2a'^2,a^2,a'^2,aa' \},\notag\\
&C^{(0)}_6 = \{ b,ba^2a',baa'^2,b^2,b^2a^2a',b^2aa'^2 \},\notag\\
&C^{(1)}_6 = \{ ba,ba',ba^2a'^2,b^2a^2,b^2a'^2,b^2aa' \},\notag\\
&C^{(2)}_6 = \{ ba^2,baa',ba'^2,b^2a,b^2a',b^2a^2a'^2 \},\notag\\
&C^{(l)}_9 = \{ ca^{l+n}a'^n, b^2ca^{-l}a'^{-l-n},bca^{-n}a'^l \ |\  n=0,1,2\},\quad  (l= 0,1,2).
\end{align}
They obey the multiplication rules shown in Table~\ref{tab:Delta54}.
\begin{table}[H]
  \centering
  \caption{Multiplication rules for conjugacy classes of $\Delta(54)$.}
  \label{tab:Delta54}
  \resizebox{\textwidth}{!}{
  \begin{tabular}{|c||c|c|c|c|c|c|c|}
  \hline
     &$C_1$ &$C^{(1)}_1$ &$C^{(2)}_1$ &$C^{(1,0)}_6$&$C^{(0)}_6$&$C^{(1)}_6$&$C^{(2)}_6$ \\
     \hline
     \hline
     $C_1$&$C_1$ &$C^{(1)}_1$ &$C^{(2)}_1$ &$C^{(1,0)}_6$&$C^{(0)}_6$&$C^{(1)}_6$&$C^{(2)}_6$ \\
     \hline
     $C^{(1)}_1$& $C^{(1)}_1$&$C^{(2)}_1$ & $C_1$& $C^{(1,0)}_6$&$C^{(0)}_6$&$C^{(1)}_6$&$C^{(2)}_6$\\
     \hline
     $C^{(2)}_1$& $C^{(2)}_1$&$C_1$ &$C^{(1)}_1$ & $C^{(1,0)}_6$&$C^{(0)}_6$&$C^{(1)}_6$&$C^{(2)}_6$\\
     \hline
     $C^{(1,0)}_6$&$C^{(1,0)}_6$ & $C^{(1,0)}_6$&$C^{(1,0)}_6$ &$6C_1+6C^{(1)}_1+6C^{(2)}_1+3C^{(1,0)}_6$&$3C^{(1)}_6+3C^{(2)}_6$&$3C^{(0)}_6+3C^{(2)}_6$&$3C^{(0)}_6+3C^{(1)}_6$\\
     \hline
     $C^{(0)}_6$&$C^{(0)}_6$&$C^{(0)}_6$&$C^{(0)}_6$&$3C^{(1)}_6+3C^{(2)}_6$&$3C^{(0)}_6+6C_1+6C^{(1)}_1+6C^{(2)}_1$&$3B^{(2,0)}_{6}+3C^{(1,0)}_6$&$3B^{(1,0)}_{6}+3C^{(1,0)}_6$\\
     \hline
     $C^{(1)}_6$&$C^{(1)}_6$&$C^{(1)}_6$&$C^{(1)}_6$&$3C^{(0)}_6+3C^{(2)}_6$&$3C^{(2)}_6+3C^{(1,0)}_6$&$3C^{(1)}_6+6C_1+6C^{(1)}_1+6C^{(2)}_1$&$3C^{(0)}_6+3C^{(1,0)}_6$\\
     \hline
     $C^{(2)}_6$&$C^{(2)}_6$&$C^{(2)}_6$&$C^{(2)}_6$&$3C^{(0)}_6+3C^{(1)}_6$&$3B^{(1,0)}_{6}+3C^{(1,0)}_6$&$3C^{(0)}_6+3C^{(1,0)}_6$&$3C^{(2)}_6+6C_1+6C^{(1)}_1+6C^{(2)}_1$\\
     \hline
     $C^{(0)}_9$&$C^{(0)}_9$&$C^{(2)}_9$&$C^{(1)}_9$&$2C^{(0)}_9+2C^{(1)}_9+2C^{(2)}_9$&$2C^{(0)}_9+2C^{(1)}_9+2C^{(2)}_9$&$2C^{(0)}_9+2C^{(1)}_9+2C^{(2)}_9$&$2C^{(0)}_9+2C^{(1)}_9+2C^{(2)}_9$\\
     \hline
     $C^{(1)}_9$&$C^{(1)}_9$&$C^{(0)}_9$&$C^{(2)}_9$&$2C^{(0)}_9+2C^{(1)}_9+2C^{(2)}_9$&$2C^{(0)}_9+2C^{(1)}_9+2C^{(2)}_9$&$2C^{(0)}_9+2C^{(1)}_9+2C^{(2)}_9$&$2C^{(0)}_9+2C^{(1)}_9+2C^{(2)}_9$\\
     \hline
     $C^{(2)}_9$&$C^{(2)}_9$&$C^{(1)}_9$&$C^{(0)}_9$&$2C^{(0)}_9+2C^{(1)}_9+2C^{(2)}_9$&$2C^{(0)}_9+2C^{(1)}_9+2C^{(2)}_9$&$2C^{(0)}_9+2C^{(1)}_9+2C^{(2)}_9$&$2C^{(0)}_9+2C^{(1)}_9+2C^{(2)}_9$\\
     \hline
  \end{tabular}
  }
    \resizebox{\textwidth}{!}{
  \begin{tabular}{|c||c|c|c|}
  \hline
     &$C^{(0)}_9$&$C^{(1)}_9$&$C^{(2)}_9$ \\
     \hline
     \hline
     $C_1$&$C^{(0)}_9$&$C^{(1)}_9$&$C^{(2)}_9$ \\
     \hline
     $C^{(1)}_1$&$C^{(2)}_9$&$C^{(0)}_9$&$C^{(1)}_9$\\
     \hline
     $C^{(2)}_1$&$C^{(1)}_9$&$C^{(2)}_9$&$C^{(0)}_9$\\
     \hline
     $C^{(1,0)}_6$&$2C^{(0)}_9+2C^{(1)}_9+2C^{(2)}_9$&$2C^{(0)}_9+2C^{(1)}_9+2C^{(2)}_9$&$2C^{(0)}_9+2C^{(1)}_9+2C^{(2)}_9$ \\
     \hline
     $C^{(0)}_6$&$2C^{(0)}_9+2C^{(1)}_9+2C^{(2)}_9$&$2C^{(0)}_9+2C^{(1)}_9+2C^{(2)}_9$&$2C^{(0)}_9+2C^{(1)}_9+2C^{(2)}_9$\\
     \hline
     $C^{(1)}_6$&$2C^{(0)}_9+2C^{(1)}_9+2C^{(2)}_9$&$2C^{(0)}_9+2C^{(1)}_9+2C^{(2)}_9$&$2C^{(0)}_9+2C^{(1)}_9+2C^{(2)}_9$\\
     \hline
     $C^{(2)}_6$&$2C^{(0)}_9+2C^{(1)}_9+2C^{(2)}_9$&$2C^{(0)}_9+2C^{(1)}_9+2C^{(2)}_9$&$2C^{(0)}_9+2C^{(1)}_9+2C^{(2)}_9$\\
     \hline
     $C^{(0)}_9$&$9C_1+3C^{(1,0)}_6+3C^{(0)}_6+3C^{(1)}_6+3C^{(2)}_6$&$9C^{(2)}_1+3C^{(1,0)}_6+3C^{(0)}_6+3C^{(1)}_6+3C^{(2)}_6$&$9C^{(1)}_1+3C^{(1,0)}_6+3C^{(0)}_6+3C^{(1)}_6+3C^{(2)}_6$\\
     \hline
     $C^{(1)}_9$&$9C^{(2)}_1+3C^{(1,0)}_6+3C^{(0)}_6+3C^{(1)}_6+3C^{(2)}_6$&$9C^{(1)}_1+3C^{(1,0)}_6+3C^{(0)}_6+3C^{(1)}_6+3C^{(2)}_6$&$9C_1+3C^{(1,0)}_6+3C^{(0)}_6+3C^{(1)}_6+3C^{(2)}_6$\\
     \hline
     $C^{(2)}_9$&$9C^{(1)}_1+3C^{(1,0)}_6+3C^{(0)}_6+3C^{(1)}_6+3C^{(2)}_6$&$9C_1+3C^{(1,0)}_6+3C^{(0)}_6+3C^{(1)}_6+3C^{(2)}_6$&$9C^{(2)}_1+3C^{(1,0)}_6+3C^{(0)}_6+3C^{(1)}_6+3C^{(2)}_6$\\
     \hline
  \end{tabular}
  }
\end{table}
From these results, we can obtain the subset $\com(\conj(\Delta(54)))$ as
\begin{align}
  \com(\conj(\Delta(54))) = \{ C_1,C^{(1)}_1,C^{(2)}_1,C^{(1,0)}_6,C^{(0)}_6,C^{(1)}_6,C^{(2)}_6 \},
\end{align}
and we can also compute the groupification of $\conj(\Delta(54))$ as follows.
\begin{align}
  \gr(\conj(\Delta(54))) = \{ [C_1], [C^{(0)}_9]\}\cong \mathbb{Z}_2,
\end{align}
where we define
\begin{align}
  &[C_1] = \{C_1,C^{(1)}_1,C^{(2)}_1,C^{(1,0)}_6,C^{(0)}_6,C^{(1)}_6,C^{(2)}_6\}, \quad [C^{(0)}_9] = \{C^{(0)}_9,C^{(1)}_9,C^{(2)}_9\}.
\end{align}
Their $\mathbb{Z}_2$ charges are given by
\begin{align}
    &[C_1]: 0,\quad
    [C_{9}^{(0)}] : 1.
\end{align}

Let us check the loop effects more explicitly. We introduce the following new notation:
\begin{gather}
0=C_1,\quad
1=C^{(1)}_1,\quad
\bar{1}=C^{(2)}_1,\quad
\dot{2}=C^{(1,0)}_6,\notag\\
\dot{3} = C^{(0)}_6,\quad
\dot{4} = C^{(1)}_6,\quad
\dot{5} = C^{(2)}_6,\notag\\
\dot{6} = C^{(0)}_9,\quad\quad
7 = C^{(1)}_9,\quad
\bar{7} =C^{(2)}_9.
\end{gather}
By using this notation, we denote the corresponding fields by $\phi_0$, $\phi_{1}, \cdots$, and their couplings by, e.g., $\lambda_{001}$.
The selection rules imposed by $\conj(\Delta(54))$ allow the 3-point tree-level couplings shown in  Table \ref{tab:3_point_coupling_Delta54}, which correspond to the following couplings:
\begin{align}
&\lambda_{kkk},\quad k\neq \dot{6},7,\bar{7}, \notag \\
&\lambda_{0k\bar{k}},\quad\lambda_{0\dot{k}\dot{k}}, \notag \\
&\lambda_{1\dot{k}\dot{k}},\quad \lambda_{\bar{1}\dot{k}\dot{k}},\quad k\neq 6, \notag \\
&\lambda_{1\bar{7}\bar{7}},\quad 
\lambda_{1\dot{6}7},\quad 
\lambda_{\bar{1}\dot{6}\bar{7}},\quad 
\lambda_{\bar{1}77}, \notag \\
&\lambda_{\dot{2}\dot{k}\dot{l}},\quad k\neq l,\quad k,l = 3,4,5, \notag \\
&\lambda_{\dot{2}kl},\quad k,l=\dot{6},7,\bar{7}, \notag \\
&\lambda_{\dot{3}\dot{4}\dot{5}},\notag \\
&\lambda_{\dot{k}lm},\quad k=3,4,5,\quad l,m=\dot{6},7,\bar{7},
\end{align}
and the others are forbidden at tree level.

We can define approximate $\mathbb{Z}_3'$ symmetry.
The conjugacy classes $C_1^{(1)}$ and $C_9^{(1)}$ carry $\mathbb{Z}_3'$ charge 1 and their conjugate $C_1^{(2)}$ and $C_9^{(2)}$ carry $\mathbb{Z}_3'$ charge 2, while the others are $\mathbb{Z}_3'$ invariant.
Similarly, we have an independent approximate $\mathbb{Z}_2'$ symmetry corresponding to 
each of self-conjugate classes $C_6^{(1,0)}, C_6^{(0)}, C_6^{(1)}, C_6^{2}, C_9^{(0)}$, where the corresponding class is $\mathbb{Z}_2'$ odd, while the others are $\mathbb{Z}_2'$ even.
Totally, we an define approximate $\mathbb{Z}_3'\times(\mathbb{Z}_2')^5$ symmetry.
Certain tree-level couplings violate this approximate 
$\mathbb{Z}_3'\times(\mathbb{Z}_2')^5$ symmetry, and they induce further violating couplings at loop levels as
\begin{align}
\lambda^{(1)}_{ 0 0 k
}&\propto
\lambda^{(0)}_{ 0 \dot{2} \dot{2} }\lambda^{(0)}_{ 0 \dot{2} \dot{2} }\lambda^{(0)}_{ k \dot{2} \dot{2} }
+
\lambda^{(0)}_{ 0 \dot{3} \dot{3} }\lambda^{(0)}_{ 0 \dot{3} \dot{3} }\lambda^{(0)}_{ k \dot{3} \dot{3} }
+
\lambda^{(0)}_{ 0 \dot{4} \dot{4} }\lambda^{(0)}_{ 0 \dot{4} \dot{4} }\lambda^{(0)}_{ k \dot{4} \dot{4} }
+
\lambda^{(0)}_{ 0 \dot{5} \dot{5} }\lambda^{(0)}_{ 0 \dot{5} \dot{5} }\lambda^{(0)}_{ k \dot{5} \dot{5} },
\end{align}
with $k=1, \bar 1$, 
\begin{align}
\lambda^{(1)}_{ 0 0 \dot{k}
}&\propto
\lambda^{(0)}_{ 0 \dot{k} \dot{k} }\lambda^{(0)}_{ 0 \dot{k} \dot{k} }\lambda^{(0)}_{ \dot{k} \dot{k} \dot{k} }
+
\lambda^{(0)}_{ 0 \dot{6} \dot{6} }\lambda^{(0)}_{ 0 \dot{6} \dot{6} }\lambda^{(0)}_{ \dot{k} \dot{6} \dot{6} }
+
\lambda^{(0)}_{ 0 7 \bar{7} }\lambda^{(0)}_{ 0 7 \bar{7} }\lambda^{(0)}_{ \dot{k} 7 \bar{7} },
\end{align}
with $k=2,3,4,5$,
\begin{align}
\lambda^{(1)}_{ 0 k k
}&\propto
\lambda^{(0)}_{ 0 \dot{2} \dot{2} }\lambda^{(0)}_{ k \dot{2} \dot{2} }\lambda^{(0)}_{ k \dot{2} \dot{2} }
+
\lambda^{(0)}_{ 0 \dot{3} \dot{3} }\lambda^{(0)}_{ k \dot{3} \dot{3} }\lambda^{(0)}_{ k \dot{3} \dot{3} }
+
\lambda^{(0)}_{ 0 \dot{4} \dot{4} }\lambda^{(0)}_{ k \dot{4} \dot{4} }\lambda^{(0)}_{ k \dot{4} \dot{4} }
+
\lambda^{(0)}_{ 0 \dot{5} \dot{5} }\lambda^{(0)}_{ k \dot{5} \dot{5} }\lambda^{(0)}_{ k \dot{5} \dot{5} },
\end{align}
with $k=1, \bar 1$,  
\begin{align}
\lambda^{(1)}_{ 0 k \dot{l}
}&\propto
\lambda^{(0)}_{ 0 \dot{l}\dot{l} }\lambda^{(0)}_{ k \dot{l}\dot{l} }\lambda^{(0)}_{ \dot{l}\dot{l}\dot{l}}
+
\lambda^{(0)}_{ 0 \dot{6} \dot{6} }\lambda^{(0)}_{ k \dot{6} \bar{7} }\lambda^{(0)}_{ \dot{l} \dot{6} 7 }
+
\lambda^{(0)}_{ 0 7 \bar{7} }\lambda^{(0)}_{ k \dot{6} \bar{7} }\lambda^{(0)}_{ \dot{l} \dot{6} 7 }
+
\lambda^{(0)}_{ 0 7 \bar{7} }\lambda^{(0)}_{k77 }\lambda^{(0)}_{ \dot{l} \bar{7} \bar{7} },
\end{align}
with $k=1,\bar{1}$ and $l=2,3,4,5$, 
\begin{align}
\lambda^{(1)}_{ 0 \dot{k} \dot{l}
}&\propto
\lambda^{(0)}_{ 0 \dot{m} \dot{m} }\lambda^{(0)}_{ \dot{k} \dot{m} \dot{n} }\lambda^{(0)}_{ \dot{l} \dot{m} \dot{n} }
+
\lambda^{(0)}_{ 0 \dot{n} \dot{n} }\lambda^{(0)}_{ \dot{k} \dot{m} \dot{n} }\lambda^{(0)}_{ \dot{l} \dot{m} \dot{n} }
+
\lambda^{(0)}_{ 0 \dot{p} \dot{p} }\lambda^{(0)}_{ \dot{k} \dot{p} \dot{p} }\lambda^{(0)}_{ \dot{l} \dot{p} \dot{p} }
+
\lambda^{(0)}_{ 0 \dot{6} \dot{6} }\lambda^{(0)}_{ \dot{k} \dot{6} 7 }\lambda^{(0)}_{ \dot{l} \dot{6} \bar{7} }
+
\lambda^{(0)}_{ 0 \dot{6} \dot{6} }\lambda^{(0)}_{ \dot{k} \dot{6} \bar{7} }\lambda^{(0)}_{ \dot{l} \dot{6} 7 } \notag \\
&
+
\lambda^{(0)}_{ 0 7 \bar{7} }\lambda^{(0)}_{ \dot{k} \dot{6} 7 }\lambda^{(0)}_{ \dot{l} \dot{6} \bar{7} }
+
\lambda^{(0)}_{ 0 7 \bar{7} }\lambda^{(0)}_{ \dot{k} \dot{6} \bar{7} }\lambda^{(0)}_{ \dot{l} \dot{6} 7 }
+
\lambda^{(0)}_{ 0 7 \bar{7} }\lambda^{(0)}_{ \dot{k} 7 7 }\lambda^{(0)}_{ \dot{l} \bar{7} \bar{7} }
+
\lambda^{(0)}_{ 0 7 \bar{7} }\lambda^{(0)}_{ \dot{k} 7 \bar{7} }\lambda^{(0)}_{ \dot{l} 7 \bar{7} }
+
\lambda^{(0)}_{ 0 7 \bar{7} }\lambda^{(0)}_{ \dot{k} \bar{7} \bar{7} }\lambda^{(0)}_{ \dot{l} 7 7 },
\end{align}
with $(k,l,m,n,p)=(2,3,4,5.6)$ and its possible permutations, 
\begin{align}
\lambda^{(1)}_{ 0 \dot{6} k
}\propto 
\sum_{\ell=2,3,4,5} &\lambda^{(0)}_{ 0 \dot{\ell} \dot{\ell} }\lambda^{(0)}_{ \dot{\ell} \dot{6} \dot{6} }\lambda^{(0)}_{ \dot{\ell} \dot{6} k }
+
\lambda^{(0)}_{ 0 \dot{\ell} \dot{\ell} }\lambda^{(0)}_{ \dot{\ell} \dot{6} k }\lambda^{(0)}_{ \dot{\ell} k \bar{k} }
+
\lambda^{(0)}_{ 0 \dot{\ell} \dot{\ell} }\lambda^{(0)}_{ \dot{\ell} \dot{6} \bar{k} }\lambda^{(0)}_{ \dot{\ell} k k }
\notag\\
&
+
\lambda^{(0)}_{ 0 \dot{6} \dot{6} }\lambda^{(0)}_{ \dot{\ell} \dot{6} \dot{6} }\lambda^{(0)}_{ \dot{\ell} \dot{6} k }
+
\lambda^{(0)}_{ 0 k \bar{k} }\lambda^{(0)}_{ \dot{\ell} \dot{6} k }\lambda^{(0)}_{ \dot{\ell} k \bar{k} }
+
\lambda^{(0)}_{ 0 k \bar{k} }\lambda^{(0)}_{ \dot{\ell} \dot{6} \bar{k} }\lambda^{(0)}_{ \dot{\ell} k k },
\end{align}
with $k=7,\bar{7}$,
\begin{align}
\lambda^{(1)}_{ 0 k k
}\propto  \sum_{\ell=2,3,4,5} &
\lambda^{(0)}_{ 0 \dot{\ell} \dot{\ell} }\lambda^{(0)}_{ \dot{\ell} \dot{6} k }\lambda^{(0)}_{ \dot{\ell} \dot{6} k }
+
\lambda^{(0)}_{ 0 \dot{\ell} \dot{\ell} }\lambda^{(0)}_{ \dot{\ell} k k }\lambda^{(0)}_{ \dot{\ell} k \bar{k} }
+
\lambda^{(0)}_{ 0 \dot{6} \dot{6} }\lambda^{(0)}_{ \dot{\ell} \dot{6} k }\lambda^{(0)}_{ \dot{\ell} \dot{6} k }
+
\lambda^{(0)}_{ 0 k \bar{k} }\lambda^{(0)}_{ \dot{\ell} k k }\lambda^{(0)}_{ \dot{\ell} k \bar{k} },
\end{align}
with $k=7,\bar{7}$, 
\begin{align}
\lambda^{(1)}_{ k k \bar{k}
}&\propto
\lambda^{(0)}_{ k \dot{2} \dot{2} }\lambda^{(0)}_{ k \dot{2} \dot{2} }\lambda^{(0)}_{ \bar{k} \dot{2} \dot{2} }
+
\lambda^{(0)}_{ k \dot{3} \dot{3} }\lambda^{(0)}_{ k \dot{3} \dot{3} }\lambda^{(0)}_{ \bar{k} \dot{3} \dot{3} }
+
\lambda^{(0)}_{ k \dot{4} \dot{4} }\lambda^{(0)}_{ k \dot{4} \dot{4} }\lambda^{(0)}_{ \bar{k} \dot{4} \dot{4} }
+
\lambda^{(0)}_{ k \dot{5} \dot{5} }\lambda^{(0)}_{ k \dot{5} \dot{5} }\lambda^{(0)}_{ \bar{k} \dot{5} \dot{5} },
\end{align}
with $k=1,\bar{1}$, 
\begin{align}
\lambda^{(1)}_{ k k \dot{2}
}&\propto
\lambda^{(0)}_{ k \dot{2} \dot{2} }\lambda^{(0)}_{ k \dot{2} \dot{2} }\lambda^{(0)}_{ \dot{2} \dot{2} \dot{2} }
+
\lambda^{(0)}_{ k \dot{6} 7 }\lambda^{(0)}_{ k \dot{6} 7 }\lambda^{(0)}_{ \dot{2} \bar{7} \bar{7} }
+
\lambda^{(0)}_{ k \dot{6} 7 }\lambda^{(0)}_{ k \bar{7} \bar{7} }\lambda^{(0)}_{ \dot{2} \dot{6} 7 },
\end{align}
with $k=1,\bar{1}$, 
\begin{align}
\lambda^{(1)}_{ 1 \bar{1} \dot{k}
}&\propto
\lambda^{(0)}_{ 1 \dot{k} \dot{k} }\lambda^{(0)}_{ \bar{1} \dot{k} \dot{k} }\lambda^{(0)}_{ \dot{k} \dot{k} \dot{k} }
+
\lambda^{(0)}_{ 1 \dot{6} 7 }\lambda^{(0)}_{ \bar{1} \dot{6} \bar{7} }\lambda^{(0)}_{ \dot{k} \dot{6} \dot{6} }
+
\lambda^{(0)}_{ 1 \dot{6} 7 }\lambda^{(0)}_{ \bar{1} \dot{6} \bar{7} }\lambda^{(0)}_{ \dot{k} 7 \bar{7} }
+
\lambda^{(0)}_{ 1 \bar{7} \bar{7} }\lambda^{(0)}_{ \bar{1} 7 7 }\lambda^{(0)}_{ \dot{k} 7 \bar{7} },
\end{align}
with $k=2,3,4,5$, 
\begin{align}
\lambda^{(1)}_{ k \dot{l} \dot{m}
}&\propto
\lambda^{(0)}_{ k \dot{n} \dot{n} }\lambda^{(0)}_{ \dot{l} \dot{n} \dot{p} }\lambda^{(0)}_{ \dot{m} \dot{n} \dot{p} }
+
\lambda^{(0)}_{ k \dot{p} \dot{p} }\lambda^{(0)}_{ \dot{l} \dot{n} \dot{p} }\lambda^{(0)}_{ \dot{m} \dot{n} \dot{p} }
+
\lambda^{(0)}_{ k \dot{6} 7 }\lambda^{(0)}_{ \dot{l} \dot{6} \dot{6} }\lambda^{(0)}_{ \dot{m} \dot{6} \bar{7} }
+
\lambda^{(0)}_{ k \dot{6} 7 }\lambda^{(0)}_{ \dot{l} \dot{6} 7 }\lambda^{(0)}_{ \dot{m} \bar{7} \bar{7} }
+
\lambda^{(0)}_{ k \dot{6} 7 }\lambda^{(0)}_{ \dot{l} \dot{6} \bar{7} }\lambda^{(0)}_{ \dot{m} \dot{6} \dot{6} }
\notag\\
&
+
\lambda^{(0)}_{ k \dot{6} 7 }\lambda^{(0)}_{ \dot{l} \dot{6} \bar{7} }\lambda^{(0)}_{ \dot{m} 7 \bar{7} }
+
\lambda^{(0)}_{ k \dot{6} 7 }\lambda^{(0)}_{ \dot{l} 7 \bar{7} }\lambda^{(0)}_{ \dot{m} \dot{6} \bar{7} }
+
\lambda^{(0)}_{ k \dot{6} 7 }\lambda^{(0)}_{ \dot{l} \bar{7} \bar{7} }\lambda^{(0)}_{ \dot{m} \dot{6} 7 }
+
\lambda^{(0)}_{ k \bar{7} \bar{7} }\lambda^{(0)}_{ \dot{l} \dot{6} 7 }\lambda^{(0)}_{ \dot{m} \dot{6} 7 }
+
\lambda^{(0)}_{ k \bar{7} \bar{7} }\lambda^{(0)}_{ \dot{l} 7 7 }\lambda^{(0)}_{ \dot{m} 7 \bar{7} }
\notag\\
&
+
\lambda^{(0)}_{ k \bar{7} \bar{7} }\lambda^{(0)}_{ \dot{l} 7 \bar{7} }\lambda^{(0)}_{ \dot{m} 7 7 },
\end{align}
with $k=1,\bar{1}$, $(l,m,n,p) = (2,3,4,5)$ and its possible permutations,
\begin{align}
\lambda^{(1)}_{ k l l
}\propto \sum_{\ell=2,3,4,5}  &
\lambda^{(0)}_{ k \dot{\ell} \dot{\ell} }\lambda^{(0)}_{ \dot{\ell} \dot{6} l}\lambda^{(0)}_{ \dot{\ell} \dot{6} l }
+
\lambda^{(0)}_{ k \dot{\ell} \dot{\ell} }\lambda^{(0)}_{ \dot{\ell} k 7 }\lambda^{(0)}_{ \dot{\ell} k \bar{7} }
+
\lambda^{(0)}_{ k \dot{6} 7 }\lambda^{(0)}_{ \dot{\ell} \dot{6} l }\lambda^{(0)}_{ \dot{\ell} l \bar{7} }
+
\lambda^{(0)}_{ k \bar{7} \bar{7} }\lambda^{(0)}_{ \dot{\ell} l 7 }\lambda^{(0)}_{ \dot{\ell} l 7 },
\end{align}
    with $(k,l)=(1,7),(1,\dot{6}),(\bar{1},\dot{6}),(\bar{1},\bar{7})$, 
\begin{align}
\lambda^{(1)}_{ k l m
}\propto \sum_{\ell=2,3,4,,5} &
\lambda^{(0)}_{ k \dot{\ell} \dot{\ell} }\lambda^{(0)}_{ \dot{\ell} l \dot{6} }\lambda^{(0)}_{ \dot{\ell} \dot{6} m }
+
\lambda^{(0)}_{ k \dot{\ell} \dot{\ell} }\lambda^{(0)}_{ \dot{\ell} l 7 }\lambda^{(0)}_{ \dot{\ell} \bar{7} m }
+
\lambda^{(0)}_{ k \dot{\ell} \dot{\ell} }\lambda^{(0)}_{ \dot{\ell} l \bar{7} }\lambda^{(0)}_{ \dot{\ell} 7 m }
\notag\\
&
+
\lambda^{(0)}_{ k \dot{6} 7 }\lambda^{(0)}_{ \dot{\ell} l \dot{6} }\lambda^{(0)}_{ \dot{\ell} \bar{7} m }
+
\lambda^{(0)}_{ k \dot{6} 7 }\lambda^{(0)}_{ \dot{\ell} l \bar{7} }\lambda^{(0)}_{ \dot{\ell} \dot{6} m }
+
\lambda^{(0)}_{ k \bar{7} \bar{7} }\lambda^{(0)}_{ \dot{\ell} l 7 }\lambda^{(0)}_{ \dot{\ell} 7 m },
\end{align}
with $k=1,\bar{1}$ and $l,m=\dot{6},7,\bar{7}, \ l\neq m$,  
\begin{align}
\lambda^{(1)}_{ \dot{k} \dot{k} \dot{\ell}
}&\propto
\lambda^{(0)}_{ \dot{k} \dot{\ell} \dot{m} }\lambda^{(0)}_{ \dot{k} \dot{\ell} \dot{m} }\lambda^{(0)}_{ \dot{\ell} \dot{\ell} \dot{\ell} }
+
\lambda^{(0)}_{ \dot{k} \dot{\ell} \dot{n} }\lambda^{(0)}_{ \dot{k} \dot{\ell} \dot{n} }\lambda^{(0)}_{ \dot{\ell} \dot{\ell} \dot{\ell} }
+
\lambda^{(0)}_{ \dot{k} \dot{\ell} \dot{m} }\lambda^{(0)}_{ \dot{k} \dot{3\ell} \dot{n} }\lambda^{(0)}_{ \dot{\ell} \dot{m} \dot{n} }
+
\lambda^{(0)}_{ \dot{k} \dot{6} \dot{6} }\lambda^{(0)}_{ \dot{k} \dot{6} \dot{6} }\lambda^{(0)}_{ \dot{\ell} \dot{6} \dot{6} }
+
\lambda^{(0)}_{ \dot{k} \dot{6} \dot{6} }\lambda^{(0)}_{ \dot{k} \dot{6} 7 }\lambda^{(0)}_{ \dot{\ell} \dot{6} \bar{7} }
\notag\\
&
+
\lambda^{(0)}_{ \dot{k} \dot{6} \dot{6} }\lambda^{(0)}_{ \dot{k} \dot{6} \bar{7} }\lambda^{(0)}_{ \dot{\ell} \dot{6} 7 }
+
\lambda^{(0)}_{ \dot{k} \dot{6} 7 }\lambda^{(0)}_{ \dot{k} \dot{6} 7 }\lambda^{(0)}_{ \dot{\ell} \bar{7} \bar{7} }
+
\lambda^{(0)}_{ \dot{k} \dot{6} 7 }\lambda^{(0)}_{ \dot{k} \dot{6} \bar{7} }\lambda^{(0)}_{ \dot{\ell} \dot{6} \dot{6} }
+
\lambda^{(0)}_{ \dot{k} \dot{6} 7 }\lambda^{(0)}_{ \dot{k} \dot{6} \bar{7} }\lambda^{(0)}_{ \dot{ell} 7 \bar{7} }
+
\lambda^{(0)}_{ \dot{k} \dot{6} 7 }\lambda^{(0)}_{ \dot{k} 7 \bar{7} }\lambda^{(0)}_{ \dot{\ell} \dot{6} \bar{7} }
\notag\\
&
+
\lambda^{(0)}_{ \dot{k} \dot{6} 7 }\lambda^{(0)}_{ \dot{k} \bar{7} \bar{7} }\lambda^{(0)}_{ \dot{\ell} \dot{6} 7 }
+
\lambda^{(0)}_{ \dot{k} \dot{6} \bar{7} }\lambda^{(0)}_{ \dot{k} \dot{6} \bar{7} }\lambda^{(0)}_{ \dot{\ell} 7 7 }
+
\lambda^{(0)}_{ \dot{k} \dot{6} \bar{7} }\lambda^{(0)}_{ \dot{k} 7 7 }\lambda^{(0)}_{ \dot{\ell} \dot{6} \bar{7} }
+
\lambda^{(0)}_{ \dot{k} \dot{6} \bar{7} }\lambda^{(0)}_{ \dot{k} 7 \bar{7} }\lambda^{(0)}_{ \dot{\ell} \dot{6} 7 }
+
\lambda^{(0)}_{ \dot{k} 7 7 }\lambda^{(0)}_{ \dot{k} 7 \bar{7} }\lambda^{(0)}_{ \dot{\ell} \bar{7} \bar{7} }
\notag\\
&
+
\lambda^{(0)}_{ \dot{k} 7 7 }\lambda^{(0)}_{ \dot{k} \bar{7} \bar{7} }\lambda^{(0)}_{ \dot{\ell} 7 \bar{7} }
+
\lambda^{(0)}_{ \dot{k} 7 \bar{7} }\lambda^{(0)}_{ \dot{k} 7 \bar{7} }\lambda^{(0)}_{ \dot{\ell} 7 \bar{7} }
+
\lambda^{(0)}_{ \dot{k} 7 \bar{7} }\lambda^{(0)}_{ \dot{k} \bar{7} \bar{7} }\lambda^{(0)}_{ \dot{\ell} 7 7 },
\end{align}
where $k,\ell,m,n=2,3,4,5$ and all of them are different from each other.
These loop-induce couplings are shown in Table \ref{tab:3_point_coupling_Delta54}.
This $\mathbb{Z}_3'\times(\mathbb{Z}_2')^5$  approximate symmetry becomes exact when the tree-level violating couplings vanish.

\begin{longtable}[H]{|c|p{12cm}|c|}
\caption{Allowed $3$-point couplings obtained from $\conj(\Delta(54))$.} \label{tab:3_point_coupling_Delta54} \\
\hline
& \multicolumn{1}{c|}{3-point coupling}& types\\
\hline
\endhead
tree & 
$C_1 C_1 C_1$,\quad $C_1 C^{(1)}_1 C^{(2)}_1$,\quad\allowbreak
$C_1 C^{(1,0)}_6 C^{(1,0)}_6$,\quad\allowbreak
$C_1 C^{(0)}_6 C^{(0)}_6$,\quad\allowbreak
$C_1 C^{(1)}_6 C^{(1)}_6$,\quad\allowbreak
$C_1 C^{(2)}_6 C^{(2)}_6$,\quad\allowbreak
$C_1 C^{(0)}_9 C^{(0)}_9$,\quad\allowbreak
$C_1 C^{(1)}_9 C^{(2)}_9$,\quad\allowbreak
$C^{(1)}_1 C^{(1)}_1 C^{(1)}_1$,\quad\allowbreak
$C^{(1)}_1 C^{(1,0)}_6 C^{(1,0)}_6$,\quad\allowbreak
$C^{(1)}_1 C^{(0)}_6 C^{(0)}_6$,\quad\allowbreak
$C^{(1)}_1 C^{(1)}_6 C^{(1)}_6$,\quad\allowbreak
$C^{(1)}_1 C^{(2)}_6 C^{(2)}_6$,\quad\allowbreak
$C^{(1)}_1 C^{(0)}_9 C^{(1)}_9$,\quad\allowbreak
$C^{(1)}_1 C^{(2)}_9 C^{(2)}_9$,\quad\allowbreak
$C^{(2)}_1 C^{(2)}_1 C^{(2)}_1$,\quad\allowbreak
$C^{(2)}_1 C^{(1,0)}_6 C^{(1,0)}_6$,\quad\allowbreak
$C^{(2)}_1 C^{(0)}_6 C^{(0)}_6$,\quad\allowbreak
$C^{(2)}_1 C^{(1)}_6 C^{(1)}_6$,\quad\allowbreak
$C^{(2)}_1 C^{(2)}_6 C^{(2)}_6$,\quad\allowbreak
$C^{(2)}_1 C^{(0)}_9 C^{(2)}_9$,\quad\allowbreak
$C^{(2)}_1 C^{(1)}_9 C^{(1)}_9$,\quad\allowbreak
$C^{(1,0)}_6 C^{(1,0)}_6 C^{(1,0)}_6$,\quad\allowbreak
$C^{(1,0)}_6 C^{(0)}_6 C^{(1)}_6$,\quad\allowbreak
$C^{(1,0)}_6 C^{(0)}_6 C^{(2)}_6$,\quad\allowbreak
$C^{(1,0)}_6 C^{(1)}_6 C^{(2)}_6$,\quad\allowbreak
$C^{(1,0)}_6 C^{(0)}_9 C^{(0)}_9$,\quad\allowbreak
$C^{(1,0)}_6 C^{(0)}_9 C^{(1)}_9$,\quad\allowbreak
$C^{(1,0)}_6 C^{(0)}_9 C^{(2)}_9$,\quad\allowbreak
$C^{(1,0)}_6 C^{(1)}_9 C^{(1)}_9$,\quad\allowbreak
$C^{(1,0)}_6 C^{(1)}_9 C^{(2)}_9$,\quad\allowbreak
$C^{(1,0)}_6 C^{(2)}_9 C^{(2)}_9$,\quad\allowbreak
$C^{(0)}_6 C^{(0)}_6 C^{(0)}_6$,\quad\allowbreak
$C^{(0)}_6 C^{(1)}_6 C^{(2)}_6$,\quad\allowbreak
$C^{(0)}_6 C^{(0)}_9 C^{(0)}_9$,\quad\allowbreak
$C^{(0)}_6 C^{(0)}_9 C^{(1)}_9$,\quad\allowbreak
$C^{(0)}_6 C^{(0)}_9 C^{(2)}_9$,\quad\allowbreak
$C^{(0)}_6 C^{(1)}_9 C^{(1)}_9$,\quad\allowbreak
$C^{(0)}_6 C^{(1)}_9 C^{(2)}_9$,\quad\allowbreak
$C^{(0)}_6 C^{(2)}_9 C^{(2)}_9$,\quad\allowbreak
$C^{(1)}_6 C^{(1)}_6 C^{(1)}_6$,\quad\allowbreak
$C^{(1)}_6 C^{(0)}_9 C^{(0)}_9$,\quad\allowbreak
$C^{(1)}_6 C^{(0)}_9 C^{(1)}_9$,\quad\allowbreak
$C^{(1)}_6 C^{(0)}_9 C^{(2)}_9$,\quad\allowbreak
$C^{(1)}_6 C^{(1)}_9 C^{(1)}_9$,\quad\allowbreak
$C^{(1)}_6 C^{(1)}_9 C^{(2)}_9$,\quad\allowbreak
$C^{(1)}_6 C^{(2)}_9 C^{(2)}_9$,\quad\allowbreak
$C^{(2)}_6 C^{(2)}_6 C^{(2)}_6$,\quad\allowbreak
$C^{(2)}_6 C^{(0)}_9 C^{(0)}_9$,\quad\allowbreak
$C^{(2)}_6 C^{(0)}_9 C^{(1)}_9$,\quad\allowbreak
$C^{(2)}_6 C^{(0)}_9 C^{(2)}_9$,\quad\allowbreak
$C^{(2)}_6 C^{(1)}_9 C^{(1)}_9$,\quad\allowbreak
$C^{(2)}_6 C^{(1)}_9 C^{(2)}_9$,\quad\allowbreak
$C^{(2)}_6 C^{(2)}_9 C^{(2)}_9$.\quad\allowbreak
&54\\
\hline
1-loop & 
 $C_1 C_1 C^{(1)}_1$,\quad\allowbreak
 $C_1 C_1 C^{(2)}_1$,\quad\allowbreak
 $C_1 C_1 C^{(1,0)}_6$,\quad\allowbreak
 $C_1 C_1 C^{(0)}_6$,\quad\allowbreak
 $C_1 C_1 C^{(1)}_6$,\quad\allowbreak
 $C_1 C_1 C^{(2)}_6$,\quad\allowbreak
 $C_1 C^{(1)}_1 C^{(1)}_1$,\quad\allowbreak
 $C_1 C^{(1)}_1 C^{(1,0)}_6$,\quad\allowbreak
 $C_1 C^{(1)}_1 C^{(0)}_6$,\quad\allowbreak
 $C_1 C^{(1)}_1 C^{(1)}_6$,\quad\allowbreak
 $C_1 C^{(1)}_1 C^{(2)}_6$,\quad\allowbreak
 $C_1 C^{(2)}_1 C^{(2)}_1$,\quad\allowbreak
 $C_1 C^{(2)}_1 C^{(1,0)}_6$,\quad\allowbreak
 $C_1 C^{(2)}_1 C^{(0)}_6$,\quad\allowbreak
 $C_1 C^{(2)}_1 C^{(1)}_6$,\quad\allowbreak
 $C_1 C^{(2)}_1 C^{(2)}_6$,\quad\allowbreak
 $C_1 C^{(1,0)}_6 C^{(0)}_6$,\quad\allowbreak
 $C_1 C^{(1,0)}_6 C^{(1)}_6$,\quad\allowbreak
 $C_1 C^{(1,0)}_6 C^{(2)}_6$,\quad\allowbreak
 $C_1 C^{(0)}_6 C^{(1)}_6$,\quad\allowbreak
 $C_1 C^{(0)}_6 C^{(2)}_6$,\quad\allowbreak
 $C_1 C^{(1)}_6 C^{(2)}_6$,\quad\allowbreak
 $C_1 C^{(0)}_9 C^{(1)}_9$,\quad\allowbreak
 $C_1 C^{(0)}_9 C^{(2)}_9$,\quad\allowbreak
 $C_1 C^{(1)}_9 C^{(1)}_9$,\quad\allowbreak
 $C_1 C^{(2)}_9 C^{(2)}_9$,\quad\allowbreak
 $C^{(1)}_1 C^{(1)}_1 C^{(2)}_1$,\quad\allowbreak
 $C^{(1)}_1 C^{(1)}_1 C^{(1,0)}_6$,\quad\allowbreak
 $C^{(1)}_1 C^{(1)}_1 C^{(0)}_6$,\quad\allowbreak
 $C^{(1)}_1 C^{(1)}_1 C^{(1)}_6$,\quad\allowbreak
 $C^{(1)}_1 C^{(1)}_1 C^{(2)}_6$,\quad\allowbreak
 $C^{(1)}_1 C^{(2)}_1 C^{(2)}_1$,\quad\allowbreak
 $C^{(1)}_1 C^{(2)}_1 C^{(1,0)}_6$,\quad\allowbreak
 $C^{(1)}_1 C^{(2)}_1 C^{(0)}_6$,\quad\allowbreak
 $C^{(1)}_1 C^{(2)}_1 C^{(1)}_6$,\quad\allowbreak
 $C^{(1)}_1 C^{(2)}_1 C^{(2)}_6$,\quad\allowbreak
 $C^{(1)}_1 C^{(1,0)}_6 C^{(0)}_6$,\quad\allowbreak
 $C^{(1)}_1 C^{(1,0)}_6 C^{(1)}_6$,\quad\allowbreak
 $C^{(1)}_1 C^{(1,0)}_6 C^{(2)}_6$,\quad\allowbreak
 $C^{(1)}_1 C^{(0)}_6 C^{(1)}_6$,\quad\allowbreak
 $C^{(1)}_1 C^{(0)}_6 C^{(2)}_6$,\quad\allowbreak
 $C^{(1)}_1 C^{(1)}_6 C^{(2)}_6$,\quad\allowbreak
 $C^{(1)}_1 C^{(0)}_9 C^{(0)}_9$,\quad\allowbreak
 $C^{(1)}_1 C^{(0)}_9 C^{(2)}_9$,\quad\allowbreak
 $C^{(1)}_1 C^{(1)}_9 C^{(1)}_9$,\quad\allowbreak
 $C^{(1)}_1 C^{(1)}_9 C^{(2)}_9$,\quad\allowbreak
 $C^{(2)}_1 C^{(2)}_1 C^{(1,0)}_6$,\quad\allowbreak
 $C^{(2)}_1 C^{(2)}_1 C^{(0)}_6$,\quad\allowbreak
 $C^{(2)}_1 C^{(2)}_1 C^{(1)}_6$,\quad\allowbreak
 $C^{(2)}_1 C^{(2)}_1 C^{(2)}_6$,\quad\allowbreak
 $C^{(2)}_1 C^{(1,0)}_6 C^{(0)}_6$,\quad\allowbreak
 $C^{(2)}_1 C^{(1,0)}_6 C^{(1)}_6$,\quad\allowbreak
 $C^{(2)}_1 C^{(1,0)}_6 C^{(2)}_6$,\quad\allowbreak
 $C^{(2)}_1 C^{(0)}_6 C^{(1)}_6$,\quad\allowbreak
 $C^{(2)}_1 C^{(0)}_6 C^{(2)}_6$,\quad\allowbreak
 $C^{(2)}_1 C^{(1)}_6 C^{(2)}_6$,\quad\allowbreak
 $C^{(2)}_1 C^{(0)}_9 C^{(0)}_9$,\quad\allowbreak
 $C^{(2)}_1 C^{(0)}_9 C^{(1)}_9$,\quad\allowbreak
 $C^{(2)}_1 C^{(1)}_9 C^{(2)}_9$,\quad\allowbreak
 $C^{(2)}_1 C^{(2)}_9 C^{(2)}_9$,\quad\allowbreak
 $C^{(1,0)}_6 C^{(1,0)}_6 C^{(0)}_6$,\quad\allowbreak
 $C^{(1,0)}_6 C^{(1,0)}_6 C^{(1)}_6$,\quad\allowbreak
 $C^{(1,0)}_6 C^{(1,0)}_6 C^{(2)}_6$,\quad\allowbreak
 $C^{(1,0)}_6 C^{(0)}_6 C^{(0)}_6$,\quad\allowbreak
 $C^{(1,0)}_6 C^{(1)}_6 C^{(1)}_6$,\quad\allowbreak
 $C^{(1,0)}_6 C^{(2)}_6 C^{(2)}_6$,\quad\allowbreak
 $C^{(0)}_6 C^{(0)}_6 C^{(1)}_6$,\quad\allowbreak
 $C^{(0)}_6 C^{(0)}_6 C^{(2)}_6$,\quad\allowbreak
 $C^{(0)}_6 C^{(1)}_6 C^{(1)}_6$,\quad\allowbreak
 $C^{(0)}_6 C^{(2)}_6 C^{(2)}_6$,\quad\allowbreak
 $C^{(1)}_6 C^{(1)}_6 C^{(2)}_6$,\quad\allowbreak
 $C^{(1)}_6 C^{(2)}_6 C^{(2)}_6$.\quad\allowbreak
&72(126)\\
\hline
\end{longtable}

\section{Calabi-Yau selection rules}
\label{app:CY}
By assigning moduli of CICYs to each field in Yukawa couplings, we obtained abundant new Yukawa textures which have not appeared in group-like selection rules. We once speculated that there exist some relations between Calabi-Yau selection rules and non-invertible selection rules, and it is able to make them same by mixing the moduli-basis. It, however, turned out that it is impossible since we cannot create an unit element of moduli which is equivalent to the one in non-invertible selection rules. \par
To show this, let us begin by stating necessary conditions that the unit element of moduli must satisfy. In $h^{1, 1} = n$ case, we have a set of moduli $T = \left\lbrace \tau_i\ |\ i = 1, 2, ..., n \right\rbrace$ and they have following Calabi-Yau selection rules.
\begin{align}
    \tau_i \tau_j = \kappa_{i j k} \tau_k,
\end{align}
where $\kappa_{i j k}$ is the intersection number among $\tau_i, \tau_j$ and $\tau_k$. If there is a linear combination $e = E_i \tau_i$ of $\tau_i \in T$ satisfies
\begin{align}
    e \tau_j = \tau_j,
\end{align}
$e$ is the unit element we are finding since $e M_{i j} \tau_j = M_{i j} \tau_j$ for arbitrary mixing matrices $M_{i j}$. Let us focus on the number of equations about $E_i$ by changing the form of (E.2) to
\begin{align}
    E_i \kappa_{i j k} \tau_k = \tau_j.
\end{align}
It is, then, obvious that $E_i \kappa_{i j k} = \delta_{j k}$. Therefore, there are, considering the indices of $\kappa_{i j k}$ are totally symmetric, $n(n + 1)/2$ equations about $E_i$ while we only need $n$ equations to fix $E_i$. Thus, $E_i$ is overdetermined if we cannot find $n(n - 1)/2$ equations which are linearly dependent to others and there is no solution of $E_i$, i.e., we cannot create a unit element of moduli. We investigated the cases for $h^{1,1} = 2,\ldots,10$ and confirmed that such a solution does not exist.

\bibliography{references}{}
\bibliographystyle{JHEP}

\end{document}